\documentclass[journal]{IEEEtran}

\ifCLASSINFOpdf

\usepackage{cite}
\usepackage{amsmath,amssymb,amsthm,bm,mathrsfs,amsfonts,dsfont}
\usepackage{textcomp}
\usepackage{xcolor}
\usepackage{multirow}
\usepackage{tabularx}
\usepackage{tcolorbox}
\usepackage{caption}
\usepackage{subcaption}
\usepackage[colorlinks,bookmarksopen,bookmarksnumbered,citecolor=blue,urlcolor=blue]{hyperref}
\usepackage{graphicx}

\usepackage{lipsum}

\usepackage[linesnumbered,ruled,vlined]{algorithm2e}
\SetKwInput{KwInput}{Input}
\SetKwInput{KwOutput}{Output}
\allowdisplaybreaks

\def\BibTeX{{\rm B\kern-.05em{\sc i\kern-.025em b}\kern-.08em
    T\kern-.1667em\lower.7ex\hbox{E}\kern-.125emX}}
\usepackage[font=footnotesize]{caption}

\begin{document}
\title{Online Admission Control and Resource Allocation in Network Slicing under Demand Uncertainties}
\author{Sajjad~Gholamipour,~
        Behzad~Akbari,~
        Nader~Mokari,~
        Mohammad~Mahdi~Tajiki,~
        Eduard~Axel~Jorswieck% <-this % stops a space
\thanks{S. Gholamipour is with the Department
of Electrical and Computer Engineering, Tarbiat Modares University, Tehran, Iran (e-mail: sajjadgholamipour@modares.ac.ir).}% <-this % stops a space
\thanks{B. Akbari is with the Department
of Electrical and Computer Engineering, Tarbiat Modares University, Tehran, Iran (e-mail: b.akbari@modares.ac.ir).}% <-this % stops a space
\thanks{N. Mokari is with the Department
of Electrical and Computer Engineering, Tarbiat Modares University, Tehran, Iran (e-mail: nader.mokari@modares.ac.ir).}% <-this % stops a space
\thanks{M. M. Tajiki is with School of Electronic Engineering and Computer Science, Queen Mary University of London, London, UK (e-mail: m.tajiki@qmul.ac.uk).}% <-this % stops a space
\thanks{E. A. Jorswieck is with the Information Theory and Communication Systems Department, TU Braunschweig, Brunswick, Germany (e-mail: jorswieck@ifn.ing.tu-bs.de).}
}
\markboth{}%
{Shell \MakeLowercase{\textit{et al.}}: Bare Demo of IEEEtran.cls for IEEE Journals}
\maketitle
\begin{abstract}
One of the most important aspects of moving forward to the next generation networks like 5G/6G, is to enable network slicing in an efficient manner. The most challenging issues are the uncertainties in computation and communication demand. Because the slices' arrive to the network in different times and their lifespans vary, the solution should dynamically react to online slice requests. The joint problem of online admission control and resource allocation considering the energy consumption is formulated mathematically. It is based on Binary Linear Programming (BLP), where, the $\Gamma$-Robustness concept is exploited to overcome Virtual Links (VL) bandwidths' and Virtual Network Functions (VNF) workloads' uncertainties. Then, an optimal algorithm is proposed. This optimal algorithm cannot be solved in a reasonable amount of time for real-world and large-scale networks. To find near-optimal solution efficiently, a new heuristic algorithm is developed. The assessments’ results indicate that the efficiency of heuristic is vital in increasing the accepted requests’ count, decreasing power consumption and providing adjustable tolerance vs. the VNFs workloads’ and VLs traffics’ uncertainties, separately. Considering the acceptance ratio and power consumption that constitute the two important components of the objective function, heuristic has about 7\% and 10\% optimality gaps, respectively, while being about 30X faster than that of optimal algorithm. 
% The results obtained from apply exact optimization and heuristic algorithm on the Geant 2001 network show the efficiency of the proposed method in accepting requests, allocating resources and Providing separate adjustable tolerance against the uncertainty of VNFs workloads and virtual link traffic, as well as the heuristic produces results that are very close to those obtained from the exact optimization model. 
\end{abstract}


\IEEEpeerreviewmaketitle

\section{Introduction}
%%\IEEEPARstart{N}{ext}-
\textbf{Background.} Next generation network slicing is a concept introduced to meet the requirements of different services provided in such networks. By applying this concept, it is possible to deploy multiple logical networks on a common infrastructure network to provide different services that may have conflicting requirements, which includes Radio Access Network (RAN), transport network, and Core Network (CN) virtualization. A network slice consists of Virtual Network Functions (VNF) and Virtual Links (VL). Network resources are limited, therefore, they should be optimally allocated to the slices ~\cite{MSU-CSE-06-2, alliance2016description, MSU-CSE-06-1}. There exists many studies on the efficient allocation of network resources. In these studies, it is assumed that the requested resources’ volumes for VNFs and VLs are fix and known~\cite{ebrahimi2019joint,fendt2018network,farkiani2019fast,chen2020network,bhamare2017optimal,halabian2019distributed}, while, in real-world problems, these volumes change dynamically and are usually unknown ~\cite{marotta2017energy,bauschert2014network},  that is, the resource requirements of VNFs and VLs of each slice are uncertain for different reasons (e.g., changes in the users’ count of each slice). If the uncertainty conditions are of no concern in the resources allocation, it may lead to slice violation with a slight change in demand. In network slicing, the slice requests arrive in the network in a continuous manner, named online slice requests, which should be checked if the network has enough resources for newly arrived slices, their requested resources should be allocated. In this case, each slice has a lifespan and is active temporarily or permanently in the network. At the end of a slice lifespan, its resources are taken back.\\ 

\textbf{Contributions.} The focus of this article is on allocating resources to online slice requests where there exist uncertainties in the resources’ volumes required by VNFs and VLs. The two approaches in managing the uncertainty are: 1) reactive, and 2) proactive. The 
reactive approach leads to unpredictable and often significant delays in operations of slices as they need to resupply the resources to active slices on-demand for instance, through the Virtual Machines (VM) migrations. As to the proactive approach in resource allocation the uncertain demand information is applied to avoid the need for resource reallocation on the spot. In this article the proactive approach is of concern. The objective is to increase the accepted slices’ count and decrease servers’ and switches’ power consumption. In this article, the problem is considered as Energy-Aware Online Network Slicing under Uncertainties (EA-ONSU) and a system model is proposed for it. In this system model, each slice is considered as a Virtual Network (VN) which includes VMs and VLs. The tenants, the slice owners, know their requirements and are allowed to manage the VNFs needed on their VNs and serve their customers thereupon \cite{ebrahimi2019joint}. The admission control and proactive resource allocation in this proposed system model are formulated as being Binary Linear Programming (BLP) when the concept of $\Gamma$–Robustness is applied to overcome uncertainties.
The contributions of this article are summarized:
\begin{itemize}
\item Due to the practical limitations like fluctuations in resources demand (e.g., unusual changes in resources demand due to increasing slice’s customers' count), we model the  slices' VNFs' and VLs' requested resources' uncertainties.
\item The joint admission control and proactive resource allocation optimization problem is formulated to slice requests under demand uncertainties as BLP, by applying the $\Gamma$-Robustness concept, named the Robust Infrastructure Network Slicing (ROBINS). The objective is to increase the accepted slices’ count and decrease the physical servers’ and switches’ power consumption.
\item A new optimal algorithm named Optimal Energy-Aware Online Network Slicing under Uncertainties (OEA-ONSU) is proposed for accepting and allocating resources to online slice requests by applying the ROBINS. The OEA-ONSU is a three step algorithm: 1) the previous slices’ expiration are checked, 2) admission control and resource allocation to newly arrived slices are made, and 3) the status of infrastructure network is updated. 
In order to apply the proposed method to real-world and large-scale networks and a find near-optimal solution in reasonable time, a novel heuristic algorithm, named Near-optimal Energy-Aware Online Network Slicing under Uncertainties (NEA-ONSU), is developed.
\item The performance results of OEA-ONSU and NEA-ONSU algorithms on the Abilene network \cite{orlowski2010sndlib} are presented, where, the results reveal the efficiency of these proposed algorithms in accepting requests, allocating resources with considering energy efficiency, and providing adjustable tolerance  vs. the VNFs workloads’ and VLs traffics’ uncertainties, separately. Considering the acceptance ratio and power consumption, the two important components of the objective function, NEA-ONSU has about 7\% and 10\% optimality gaps, respectively, while being about 30X faster than that of OEA-ONSU.
% INJA
\end{itemize}
\textbf{Paper organization.} The rest of this article is organized as follows: the related works are reviewed in Sec. \ref{Related works}; the system is modeled in Sec.~\ref{System Model}; the online robust admission control and resource allocation: mathematical formulation is discussed in Sec.~\ref{online robust admission control and resource allocation: Mathematical Formulation}; the online robust admission control and resource allocation: solution methodology is introduced in Sec.~\ref{online robust admission control and resource allocation: Solution Methodology}; the numerical results are presented in Sec.~\ref{Numerical Results} and the article is concluded in Sec.~\ref{Conclusion}.
\newcolumntype{P}[1]{>{\centering\arraybackslash}p{#1}}
\begin{table*}[htpb]
\centering
\caption{Summary of Related Researches}
\label{Related-Research}
\small
\resizebox{\textwidth}{!}{
\begin{tabular}{||c|c|c|c|c|c|c|c|c|c|c|c|c||}
\hline
\textbf{ref.}& \begin{tabular}{c}\textbf{CPU} \\\textbf{Robustness} \end{tabular}& \begin{tabular}{c}\textbf{RAM} \\\textbf{Robustness} \end{tabular}& \begin{tabular}{c}\textbf{Storage} \\\textbf{Robustness}\end{tabular}& \begin{tabular}{c}\textbf{Bandwidth}\\ \textbf{Robustness}\end{tabular}& \begin{tabular}{c}\textbf{VNF} \\\textbf{Placement}\end{tabular}& \begin{tabular}{c}\textbf{VLE}\end{tabular}& \begin{tabular}{c}\textbf{Delay}\end{tabular}& \textbf{Online} & \begin{tabular}{c}\textbf{Applicable} \\\textbf{on Large}\\ \textbf{Networks}\end{tabular}& \begin{tabular}{c}\textbf{Decreasing} \\\textbf{Power} \\\textbf{Consumption} \\\textbf{of Servers}\end{tabular}& \begin{tabular}{c}\textbf{Decreasing}\\ \textbf{Power}\\ \textbf{Consumption}\\ \textbf{of Switches}\end{tabular}& \begin{tabular}{c}\textbf{Admission} \\\textbf{Control}\end{tabular} \\  
\hline
\cite{ebrahimi2019joint} &  &  &  &  & \checkmark & \checkmark & \checkmark &  &  & \checkmark & \checkmark & \checkmark\\
\hline
\cite{farkiani2019fast} &  &  &  &  & \checkmark & \checkmark &  &  & \checkmark & \checkmark & \checkmark & \checkmark\\
\hline
\cite{chen2020network} &  &  &  &  & \checkmark & \checkmark & \checkmark &  &  &  \checkmark & & \\ 
\hline
\cite{bhamare2017optimal} &  &  &  &  & \checkmark & \checkmark & \checkmark &  & \checkmark &  &  & \checkmark\\
\hline
\cite{sun2019energy} &  &  &  &  & \checkmark & \checkmark & & \checkmark & \checkmark & \checkmark & \checkmark & \checkmark\\
\hline
\cite{soualah2019online} &  &  &  &  & \checkmark & \checkmark & \checkmark & \checkmark & \checkmark & \checkmark &  & \checkmark \\
\hline
%\cite{bari2019esso} &  &  &  &  & \checkmark & \checkmark & \checkmark & \checkmark & \checkmark & \checkmark & \checkmark & \checkmark\\
%\hline
\cite{ghazizadeh2019joint} &  &  &  &  & \checkmark & \checkmark &  \checkmark &  & \checkmark &  &  & \\ 
\hline
\cite{varasteh2021holu} &  &  &  &  & \checkmark & \checkmark &  \checkmark & \checkmark & \checkmark & \checkmark & \checkmark & \checkmark\\ 
\hline
\cite{chen2021optimal} &  &  &  &  & \checkmark & \checkmark &  \checkmark & \checkmark  & \checkmark & \checkmark & \checkmark & \checkmark\\ 
\hline
%\cite{tajiki2018cect, tajiki2016qrtp, tajiki2017joint,tajiki2017optimal} &  &  &  &  &  & \checkmark & \checkmark &  & \checkmark &  &  & \\ 
%\hline
\cite{marotta2017energy} & \checkmark & \checkmark & \checkmark &  & \checkmark & \checkmark & \checkmark &  & \checkmark & \checkmark & \checkmark &  \\
\hline
\cite{hosseini2019probabilistic} & \checkmark &  &  & \checkmark & \checkmark & \checkmark &  & \checkmark & \checkmark &  &  & \\
\hline
\cite{marotta2017fast} & \checkmark & \checkmark &  &  & \checkmark & \checkmark & \checkmark &  & \checkmark & \checkmark & \checkmark & \\
\hline
\cite{nguyen2019proactive} & \checkmark &  &  & \checkmark & \checkmark & \checkmark & \checkmark &  & \checkmark & \checkmark & \checkmark &  \\
\hline
\cite{wen2018robustness} &  &  &  & \checkmark & \checkmark & \checkmark &  & \checkmark &  &  &  & \\
\hline
\cite{reddy2016robust} & \checkmark &  &  & \checkmark & \checkmark & \checkmark & \checkmark &  & \checkmark & \checkmark & \checkmark & \checkmark \\
\hline
\cite{baumgartner2017network} & \checkmark &  &  & \checkmark & \checkmark & \checkmark & \checkmark &  & \checkmark & \checkmark & \checkmark &  \\
\hline
\cite{wen2017robust} &  &  &  & \checkmark & \checkmark & \checkmark &  & \checkmark &  &  &  & \\
\hline
%\cite{tajiki2019software} & \checkmark &  &  & \checkmark & \checkmark & \checkmark & \checkmark &  & \checkmark &  &  & \\ \hline
%\cite{tajiki2019joint,tajiki2018joint} & \checkmark &  &  & \checkmark & \checkmark & \checkmark & \checkmark &  & \checkmark & \checkmark & \checkmark  & \\ \hline
%\cite{tajiki2018energy} & \checkmark &  &  & \checkmark & \checkmark & \checkmark & \checkmark &  & \checkmark & \checkmark &  & \\ \hline
\cite{bauschert2020fast} &  &  &  & \checkmark & \checkmark & \checkmark &  & \checkmark & \checkmark & \checkmark & \checkmark & \checkmark\\
\hline
\cite{nguyen2020deadline} & \checkmark &  &  & \checkmark & \checkmark & \checkmark & \checkmark &  & \checkmark & \checkmark & \checkmark & \\
\hline
\cite{luu2021uncertainty} & \checkmark & \checkmark &  & \checkmark & \checkmark & \checkmark & \checkmark &  & \checkmark & \checkmark & \checkmark & \\
\hline
\cite{luu2022admission} & \checkmark & \checkmark &  & \checkmark & \checkmark & \checkmark &  & \checkmark & \checkmark & \checkmark & \checkmark & \checkmark\\
\hline
This work & \checkmark & \checkmark & \checkmark & \checkmark & \checkmark & \checkmark & \checkmark & \checkmark & \checkmark & \checkmark & \checkmark & \checkmark\\ \hline
\end{tabular}
}
\end{table*}
\section{Related works}
\label{Related works}
%In next-generation networks (5G, 6G, etc.) slicing, each slice request can be considered as a Service Function Chain (SFC) which has its own requirements, for example in terms of bandwidth and latency.
%In next-generation networks (5G, 6G, etc.) slicing, resource allocation to each slice request is equivalent to placement of VNFs and embedding of VLs between them, with taking into account their requirements, for example, in terms of computational capacity, bandwidth and tolerable delay. This problem has received a lot of attention in recent years. 
The related works are categorized in terms of uncertainties: 1) Resource allocation with fixed resource demand \cite{ebrahimi2019joint,farkiani2019fast,sun2019energy,soualah2019online,chen2020network,bhamare2017optimal,ghazizadeh2019joint,varasteh2021holu,chen2021optimal}, 2) Resource allocation under demand uncertainties \cite{hosseini2019probabilistic,marotta2017energy,marotta2017fast,nguyen2019proactive,wen2018robustness,reddy2016robust,baumgartner2017network,wen2017robust,bauschert2020fast,nguyen2020deadline,luu2021uncertainty,luu2022admission}.
\subsection{Resource allocation with fixed resource demand}The energy consumption of cloud nodes and the cost of bandwidth consumption is decreased in \cite{ebrahimi2019joint}, where the provided framework allows tenants to manage their slices and serve their customers. For this purpose, an Integer Linear Programming (ILP) formulation is designed for resource allocation and an ILP formulation is designed for admission control.
The issue of energy-aware service deployment is studied in \cite{farkiani2019fast}, where, an ILP is formulated by considering limited VNF traffic processing capacity and management issues. By applying  the Benders decomposition, feasibility pump, and primal-dual algorithms, a fast and scalable algorithm with polynomial execution time is devised to compute a near-optimal solution.
The problem of resource allocation in network slicing is studied in \cite{chen2020network}, where, flexible routing, End-to-End (E2E) latency, and coordination overhead are of concern. The problem is formulated as a mixed binary linear programming, with the objective to reduce the energy consumption of cloud nodes.
The problem of placing VNFs to form Service Function Chains (SFCs) on geographically distributed clouds is addressed in \cite{bhamare2017optimal}. This problem is formulated as ILP to reduce inter-cloud traffic and response time in a multi-cloud scenario. The total implementation cost and Service Level Agreements (SLAs) are of concern. 
The energy efficiency optimization for orchestrating online SFC requests in multi-domain networks is assessed in \cite{sun2019energy}, where, the problem is formulated as ILP, followed by a heuristic algorithm, which next to meeting the needs of online SFCs, assures the privacy of each cloud efficiently.
An ILP formulation is presented in \cite{soualah2019online} to solve the Virtualized Network Function Forwarding Graph (VNF-FG) placement and chaining problem. VNFs are shared between tenants to optimize resource consumption and increase the infrastructure provider revenue. 
%In \cite{bari2019esso} a solution for allocating resources to SFCs with the goal of minimizing carbon footprint is provided. The goal of the ILP formulation is to minimize the carbon footprint in the initial placement of the new SFC request, consolidation of resources when leaving one or more SFCs, and migrate existing VNFs based on available resources and renewable energy, while SFCs E2E latency should not be violated with respect to their maximum allowed delays.
A mathematical formulation for reliability-aware VNF placement and routing by considering Quality of Service (QoS) parameters is presented in \cite{ghazizadeh2019joint}, where, a resource allocation algorithm, applying a shared protection scheme with Active-Standby redundancy, is proposed to optimize the redundant VNFs without affecting the QoS parameters.  The problem is formulated as Mixed Integer Linear Programming (MILP) and a meta-heuristic algorithm is proposed to make the solution scalable in large-scale networks. In \cite{varasteh2021holu}, the power-aware
and delay-constrained joint VNF placement and routing (PD-VPR) problem is formulated as an ILP. Then, a fast online heuristic named Holu is developed to overcome high computational complexity of ILP. In \cite{chen2021optimal}, the problem of resource allocation to SFCs by considering the E2E latency constraints of all services and all cloud and communication resource budget and energy efficiency constraints, is considered. The problem is formulated as Mixed Binary Linear Program (MBLP). Then, an alternative MBLP formulation is developed which shows same optimal solution and it is more computationally efficient when the dimension of the corresponding network becomes large.
\subsection{Resource allocation under demand uncertainties}
The manner where the robust strategies that place VNFs in virtual data centers impact on energy saving and the level of protection vs. uncertainties in demand is assessed in \cite{marotta2017energy}. Therefore, a robust optimization model with a heuristic algorithm is proposed for reducing the energy consumption of computational and network infrastructure that is robust to fluctuations in resources demand. To reduce energy consumption, the unused servers and switchs’ ports are turned off. 
%An optimal VNF placement scheme with taking into account the service chains' delay limits is devised.
The issue of mapping VLs into physical paths, known as Virtual Link Embedding (VLE), where the bandwidth requirements of VLs are uncertain, is assessed in \cite{hosseini2019probabilistic}. To have VLs with predictable performance, the mapping must assure the required E2E congestion probability of physical paths with no dependency on the characteristics of the paths where VLs are mapped. Accordingly, a general uncertainty model is proposed where the bandwidth requirements of VLs are uncertain. There exists a model for uncertainty in VNFs’ demand. The VLE problem is formulated as a nonlinear optimization one. Then, a model for large-scale networks is provided by applying the approximate formulation.
The robust network embedding problem under resource demand uncertainties is assessed in \cite{marotta2017fast}, where the limitations of delay are considered by applying the robust mixed integer optimization techniques.
A model for proactive SFC is proposed with the objective to prevent resource reallocation to SFCs  when demand fluctuate, prior to which, most of the proposed models are reactive, where, resources are reallocated during  the demand fluctuations, which has a negative effect on the delay-sensitive SFCs’ performance, \cite{nguyen2019proactive}. Consequently, first, a SFC orchestration is formulated with a predefined deadline limitation as the Mixed Integer Non-Linear Programming (MINLP) by applying the $\Gamma$-Robustness concept and next, an approximate algorithm is devised to solve the large-scale problems. 
The problem of robust and E2E network slicing, where the slices are considered as a set of VNFs and links between them is discussed in \cite{wen2018robustness}. Bugs may occur in some VNFs in a random manner, making some slices to lose their validity, hence, triggering a slice recovery process. Beacause traffic demand in each slice is stochastic, making drastic changes in traffic demand can lead to slice reconfiguration. As to slice recovery in network slicing under bandwidth uncertainty, a solution is presented in \cite{wen2018robustness} and a heuristic algorithm  is devised based on Variable Neighborhood Search (VNS) to accelerate the problem solving time in large-scale networks.
Researchers in \cite{reddy2016robust} applied  the $\Gamma$-Robustness concept in resource allocation to VNF chains by considering their tolerable delays and uncertain state in traffic demand, which leads to uncertainty in the requirements of VNFs resources, formulated as MILP. An admission control method is adopted and the objective function is to reduce the energy consumption of servers and links next to reducing the penalty due to the admission control process. To improve the scalability of the model, the MILP-based VNS algorithm is presented as well. 
To overcome the scalability issues in \cite{reddy2016robust}, a model when the concept of light robustness is applied presented in \cite{baumgartner2017network}.
The problem of failure recovery in network slicing under uncertainty in traffic demand is assessed in \cite{wen2017robust}, where, first, the problem is formulated as MILP, next, a robust optimization is applied to fulfill the stochastic traffic requests. %\cite{tajiki2019software} focuses on traffic engineering, failure recovery, fault prevention, and SFC with reliability considerations in the SDN-based networks. %\cite{tajiki2019software} proposes a novel failure recovery scheme that provides SFC in SDN-based networks. In the problem assumptions, in each timeslot, the required resources of each flow in terms of bandwidth and computational resources are variable. The problem is modeled as ILP and a heuristic algorithm is designed to make the solution scalable for large and real networks. \cite{tajiki2019joint} presents a fault-aware routing architecture for the SFC problem with energy consideration. This architecture is provided for SDN-based networks and also supports fog nodes. In the proposed scheme, the probability of failure in networking devices is optimized and in case of failure in fog nodes or switches, the network is reconfigured in real-time. Finally, in order to solve the scalability issues of the optimization problem, the sub-optimal heuristic algorithm is presented.
%\cite{tajiki2018joint} provides a resource allocation architecture that enables Energy-aware SFC in SDN-based networks with constraints on delay, link utilization, and server utilization. Hence, the focus is on the placement of VNFs, the allocation of VNFs to flows, and the flow routing as resource allocation, taking into account the initial resource allocation to flows and general resource reallocation. Then, the problem is formulated as ILP and finally, a set of heuristics is designed to find a near-optimal solution at a suitable time scale for practical applications.
%\cite{tajiki2018energy} investigates the SFC problem with the aim of reducing server power consumption and controls traffic congestion through network reconfiguration.
A Mixed Integer Programming (MIP) formulation is presented in \cite{bauschert2020fast} for the network slicing problem under traffic uncertainty and to reduce the computational complexity of mathematical optimization, a meta-heuristic algorithm based on ant colony optimization algorithms is devised for the robust network slice design problem. 
To avoid frequent resource re-provisioning, the deadline-aware, co-located, and geo-distributed SFC orchestration with demand uncertainties as robust optimization problem is formulated in \cite{nguyen2020deadline}, where, the exact and approximate algorithms are devised to solve it. Uncertain demand knowledge is used in computing the proactive SFC orchestration that can withstand fluctuations in dynamic service demand.
A resource allocation approach in network slicing that is robust to partly unknown users’ count with random usage of slice resources is proposed in \cite{luu2021uncertainty}, where, the objective is to increase the total earnings of the infrastructure provider (IP). The resource allocation to slices is made as to limit its impact on low-priority background services, which may coexist next to the slices in the infrastructure network. In this context, the probabilistic assurance is that the volume of allocated network resources to the slices will meet its requirements. In \cite{luu2022admission}, a prioritized admission control mechanism for concurrent slices based on an infrastructure resource reservation approach is proposed. The reservation accounts for the dynamic nature of slice requests while being robust to slice resource demands uncertainties. Adopting the perspective of an IP, reservation schemes are proposed that maximize the number of slices for which infrastructure resources can be granted while minimizing the costs charged to the Mobile Network Operators (MNOs).

The findings of this article are compared with that of the reviewed articles in Table~\ref{Related-Research}. In practice, there exist different types of resources in the infrastructure network: the CPU, RAM, storage, and bandwidth that a slice needs to operate. To confront the demand uncertainties, each resource demand uncertainty must be of concern. In this article, a comprehensive uncertainty formulation is devised and due to the business model introduced by 5GPPP \cite{queseth20175g} for network slicing, the online slice requests with their requirements including propagation delay of VLs is of concern. The joint admission control and resource allocation with the main objective of increasing the accepted slices’ count and decreasing the power consumption of the infrastructure network is of concern. Because of this comprehensive formulation of the problem, a large-scale problem that needs to be solved very fast as to be applicable on large-scale networks, a new and rapid near-optimal algorithm is devised. As observed, none of the available studies meets all the conditions addressed in this article.
\begin{table}[!ht]
	\renewcommand{\arraystretch}{1.5}
	\centering
	\scriptsize
	\caption{{Main Notations}}
	\label{table-notation}
	\resizebox{\columnwidth}{!}{
	\begin{tabular}{>{\color{black}}c |>{\color{black}}c}
	%\begin{tabular}{| c| c| }	
	    \hline
		\textbf{Notation}& \textbf{Definition}\\\hline
        \multicolumn{2}{c}{\textbf{Input Parameters}}\\\hline
        $\mathcal{N},\mathcal{L}$ &Sets of IP's physical nodes and links\\\hline
        $\mathcal{N}_\text{Used},\mathcal{N}_\text{Unused}$ &Sets of previously used and unused nodes\\\hline
        ${R}_{n}$, ${R'}_{n}$ &\begin{tabular}{c}Vector of node $n$'s total and available resources\\ that includes CPU, RAM, and storage\end{tabular}\\ \hline
        ${P}^\text{Max}_{n}$ , ${P}^\text{Idle}_{n}$ & Maximum and idle power consumption of node $n$\\ \hline
        ${S_n}, {S^\text{Port}_n}$& Switch $n$ and its one port power consumption\\ \hline
        ${S'_n}$& Number of connected ports of switch $n$ to other switches\\ \hline
		$B_{l_{n,n'}}$, $B'_{l_{n,n'}}$ &Total and available bandwidth of the physical link $l_{n,n'}$\\\hline
		${B}_\text{Total}$& Sum of all links' bandwidths\\ \hline
		$\tau_{l_{n,n'}}$ &Propagation delay of the physical link $l_{n,n'}$\\\hline
		${\zeta}_{l_{n,n'}}$&\begin{tabular}{c}
		Includes switch $n$ and $n'$ idle power and power of ports used in link ${l}_{n,n'}$\end{tabular}\\ \hline
		${\gamma}_{l_{n,n'}}$&\begin{tabular}{c}
		To specify used link $l_{n,n'}$, it is 1 if link $l_{n,n'}$ was used\end{tabular}\\ \hline
		$I^{l_{u,u'}}_{p_{n,n'}^b}$&\begin{tabular}{c}Indicator that determines physical link $l_{u,u'}$\\contributes in the $b^{\text{th}}$ path between $n$ and $n'$, if it has value 1\end{tabular}\\\hline
		$\mathcal{L}_{p_{n,n'}^b}$&\begin{tabular}{c}Set of all physical links $l_{u,u'}$ contribute in the $b^{\text{th}}$ path between $n$ and $n'$\end{tabular}\\\hline
		$\mathcal{P}_{n,n'}$ &Set of paths between $n$ and $n'$\\\hline
		$\mathcal{T}, \mathcal{T}_\text{c}$&Sets of all tenants that their slices are accepted and current tenants\\ \hline
        $\mathcal{D}_{t}, \mathcal{D}_{t\_\text{c}}$& \begin{tabular}{c} Sets of tenant $t$'s slices and current tenant $t$'s arrived slices \end{tabular} \\ \hline
        $\mathcal{S}, \mathcal{S}_\text{c}$ &Sets of total accepted slices and current time slot arrived slices\\ \hline
        $\phi_{t,d}$ &lifespan of the $d^\text{th}$ slice of tenant $t$\\ \hline
        $\mathcal{M}_{t,d}, \mathcal{E}_{t,d}$ &Sets of $s_{t,d}$'s requested VMs and VLs\\ \hline
        $\nu_{m_{t,d}}$
		&\begin{tabular}{c}Vector of the requested capacities for VM $m_{t,d}$ that\\ includes $\nu^\text{CPU}_{m_{t,d}}$, $\nu^\text{RAM}_{m_{t,d}}$ and, $\nu^\text{STOR}_{m_{t,d}}$\end{tabular}\\ \hline
		$\Upsilon_{e_ {m, m'}}$ &\begin{tabular}{c}Vector of specifications of requested VL between the two VMs $m$ and $m'$ \\in slice $s_{t, d}$ that includes $\omega_ {e_ {m, m'}}$ and $\tau_\text{max}^{e_ {m, m'}}$ \end{tabular}\\\hline
		$\omega_{e_{m,m'}}$ &Requested data rate between two VMs $m_{t,d}$ and $m'_{t,d}$\\ \hline
		$\tau^{e_{m,m'}}_{\max}$ &\begin{tabular}{c}Maximum tolerable propagation delay between two VMs $m_{t,d}$ and $m'_{t,d}$\end{tabular}\\ \hline
		$\alpha_n$&To specify previously turned-on node $n$, it is 1 if node $n$ is turned-on\\\hline
		${N}_\text{Total}$& Sum of All nodes power consumption\\ \hline
		${N}_\text{Used}$& Nodes power consumption in previous time slots\\ \hline
		${S}_\text{Total}$& Sum of All switches power consumption\\ \hline
		${S}_\text{Used}$& Switches power consumption in previous time slots\\ \hline
		$\Gamma_1$,$\Gamma_2$& To specify protection levels for VMs and VLs\\ \hline
		${\Delta}_1,{\Delta}_2$& To specify relative deviations for requested resources of VMs and VLs\\ \hline
		\multicolumn{2}{c}{\textbf{Decision Variables}}\\\hline
        ${\beta}_{n}$& 1 if node $n$ is turned on in current time slot\\\hline
        ${\delta}_{t,d}$& 1 if slice $d$ of tenant $t$ is accepted\\\hline
		$\pi_{n}^m$&\begin{tabular}{c} 1 if VM $m_{t,d}$ is placed on node $n$\end{tabular}\\\hline
		$\xi_{p_{n,n'}^b}^{e_{m,m'}}$&\begin{tabular}{c}1 if VL $e_{m,m'}$ is mapped on the $b^{\text{th}}$ path between $n$ and $n'$\end{tabular}\\\hline
        ${\vartheta}_{l_{n,n'}}$&\begin{tabular}{c}
		1 if link $l_{n,n'}$ used in current time slot\end{tabular}\\ \hline
%		$\epsilon_m$, $\epsilon'_{e_{m,m'}}$ & Auxiliary variables\\\hline
        \multicolumn{2}{c}{\textbf{Auxiliary Variables}}\\\hline
        ${\eta}$& Number of rejected slices in current time slot\\\hline
		${N}_\text{c}$&\begin{tabular}{c} Sum of used nodes power consumption with respect to\\ arrived slices in current time slot\end{tabular}\\ \hline
		${U}_{n}$&\begin{tabular}{c} Amount of node $n$'s used resources (CPU,RAM,Storage)\\ in current time slot \end{tabular}\\ \hline
		${O}^{t}_{n}$&\begin{tabular}{c} Amount of node $n$'s used resources (CPU,RAM,Storage) for robustness\\ in time slot $t$ \end{tabular}\\ \hline
		${S}_\text{c}$& Sum of the used switches power consumption in current time slot\\ \hline
		${U'}_{l_{n,n'}}$ &Amount of link $l_{n,n'}$'s used bandwidth in current time slot\\\hline
		${O'}^{t}_{l_{n,n'}}$ &Amount of link $l_{n,n'}$'s used bandwidth for robustness in time slot $t$\\\hline
%		$\epsilon_m$, $\epsilon'_{e_{m,m'}}$ & Auxiliary variables\\\hline
		\begin{tabular}{c}${\rho_{1}}^{m}_{n}$, ${z_{1}}_{n}$,\\ ${\rho_2}^{e_{m,m'}}_{p^b_{n,n'}}$, ${z_2}_{l_{u,u'}}$\end{tabular} & Robustness variables\\\hline
	\end{tabular}
	}
\end{table}
\section{System Model}\label{System Model}
This system includes:
1) Infrastructure network model and 2) Slice requests model. The main notations are summarized in Table \ref{table-notation}.
\begin{figure}[t]
\center{\includegraphics[width=\columnwidth]
{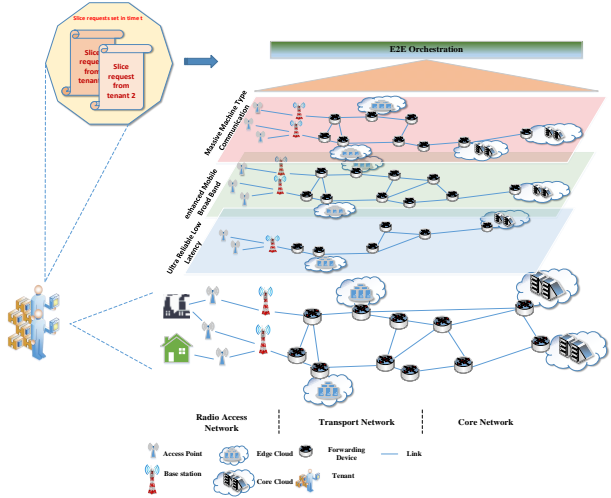}}
\caption{\label{Architecture} Infrastructure Provider's Network Architecture}
\end{figure}
\begin{figure}[t]
\centering
\includegraphics[width=\columnwidth]{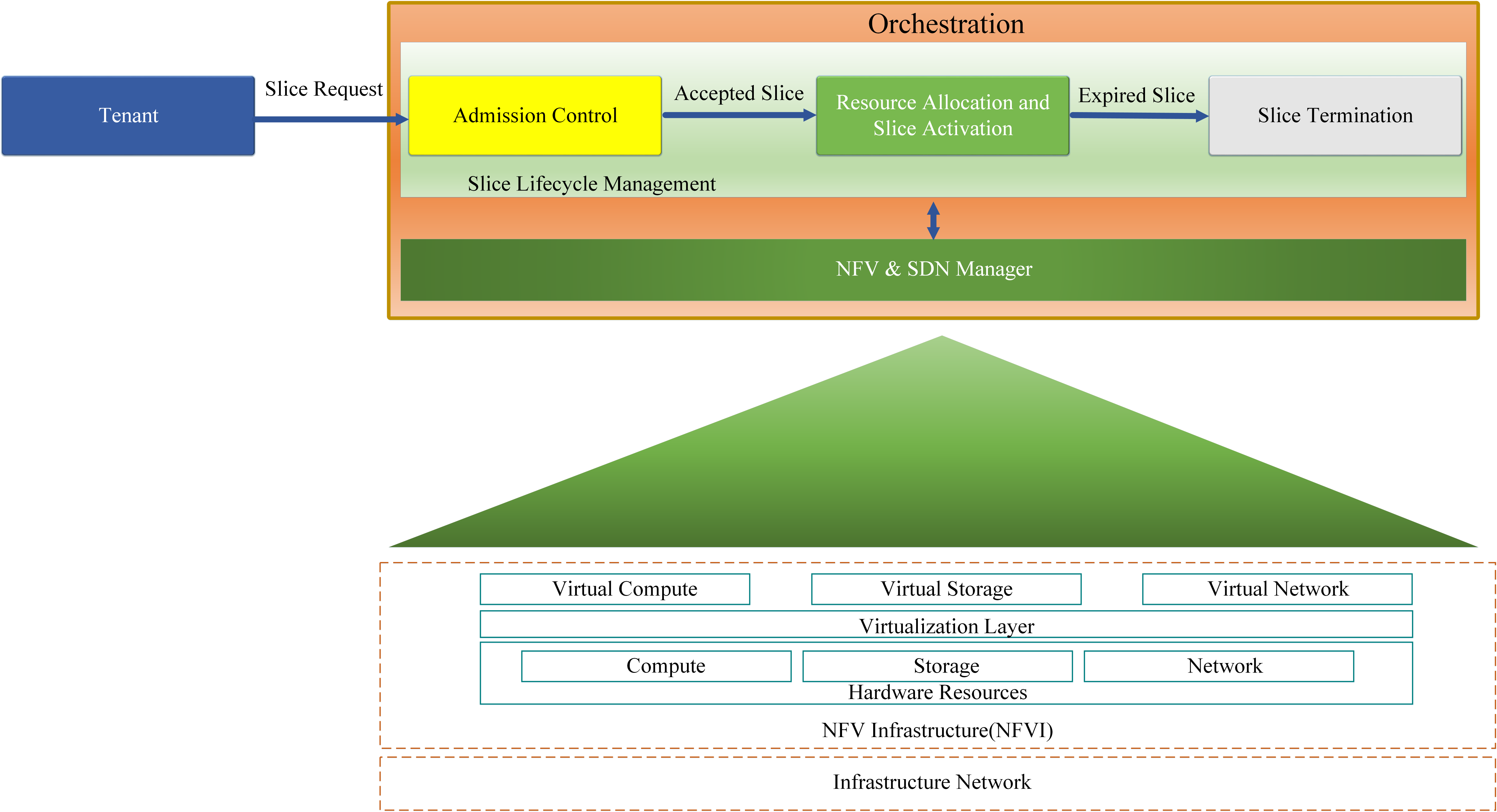}
\caption{ The Process of Processing Each Tenant's Slice Request }\label{Slice-request-processing}
\end{figure}

\subsection{Infrastructure Network Model}
An IP is assumed that provides network slices for several tenants on a shared network. The infrastructure network is considered as an undirected graph $G(\mathcal{N}; \mathcal{L})$, where $\mathcal{N}$ is the set of cloud nodes\footnote{Each cloud node includes a forwarding device and a cluster of servers.}  and base stations (BS), and $\mathcal{L} = [l_{n,n'} ]$ is the set of links which connect the nodes\footnote{We use the terms node and server interchangeably.} (the cloud nodes and BSs) of this graph, thus, if $ l_{n,n'}=1$, then the two nodes $n$ and $n'$ are connected, otherwise the opposite holds. As observed in Fig. (\ref{Architecture}), the distributed cloud nodes are considered in both the core and transport layers, and the abstract RAN \cite{wang2019reconfiguration} layer includes only BSs, where the nodes form a network together. It is worth noting that the abstract RAN implies that the RAN details are not considered. Only the computation and storage capacities of BSs and their forwarding devices \cite{wang2019reconfiguration}, which connect them to transport layer, are considered. It is assumed that all nodes support the Network Function Virtualization (NFV) and as observed in Fig. (\ref{Slice-request-processing}), there is an orchestrator that includes an NFV and Software Defined Networking (SDN) manager which, first, receives the tenants’ slice requests and next, if acceptable, creates them by allocating the resources thereupon. The resources of each node include the CPU, RAM, and storage, considered as a vector $ R_{n} = (R_{n}^\text{CPU}, R_{n}^\text{RAM}, R_{n}^\text{STOR})$ for each node $n\in \mathcal{N}$. The physical links of the network have fixed bandwidths, therefore, each link $l_{n,n'}\in \mathcal{L}$ has a limited bandwidth of $B_{l_{n,n'}}$. Because the nodes are distributed, there exists a considerable propagation delay, that is,  $\tau_{l_{n,n'}}$ is the propagation delay between nodes $n$ and $n'\in \mathcal{N}$.
\subsection{Slice Requests Model}
There exist some tenants who request slices. In this article, it is assumed that $\mathcal{T} = \{{1,…, T}\}$ is the set of all tenants, and set $\mathcal{D}_{t} = \{{1,…, d_{t}}\}$ is the requested slices of tenant $t$. Because each tenant can request different slices, in this set, $d_{t}$ is the number of the slices requested by tenant $t$. In this process, an online resource allocation is of concern. It is possible that, some tenants who may or may not already be in the set $\mathcal{T}$ transmit new slice requests to the IP at any given time, consequently, another set, named $\mathcal{T}_\text{c}$, is considered to include tenants who send slice requests and must decide whether to accept and allocate the resources thereupon. The set $\mathcal{D}_{t\_\text{c}}$ is considered as well, which includes the current slice requests of tenant $t\in \mathcal{T}_\text{c}$. A set named $\mathcal{S}_\text{c}$ is considered to include the current slice requests, that is, each $\mathcal{S}_\text{c}$’s member is represented by $s_{t,d}$ which is the $d^{\text{th}}$ slice request of tenant $t$ in current time. The set named $\mathcal{S}$ is considered to include all the accepted slices. As observed in Fig. (\ref{Slice-request-processing}), when a tenant transmits a slice request to the orchestrator, first, the possibility of accepting is checked, and if the slice is accepted, the required resources are allocated and the slice is activated. Whenever the tenant does not need the slice, the slice lifespan is expired, the slice termination process is run and its resources are taken back, consequently, the three sets $\mathcal{T}$, $\mathcal{D}_{t}$, and $\mathcal{S}$ are updated after both slice resource allocation and expiration. Each slice request specifies the set of VMs and their associated VLs represented by undirected graph $s_{t, d} = (\mathcal{M}_{t, d},\mathcal{E}_{t, d})$, where, $\mathcal{M}_{t, d}$ and $\mathcal{E}_{t, d}$ are the sets of the requested VMs and VLs by tenant $t$ for slice $d$, respectively. For each VM $m_ {t, d}\in \mathcal{M}_{t, d}$, a vector $\nu _ {m_ {t, d}} = [\nu _ {m_ {t, d}} ^ \text{CPU}, \nu _ {m_ {t, d}} ^ \text{RAM}, \nu _ {m_ {t , d}} ^ \text{STOR}]$ that indicates the volume of the resources that are needed by VM $m_ {t, d}$ is of concern. From now on $t$ and $d$ indices are removed from $m$ and $m'$ for readability. For each VL $e_{m,m'} \in \mathcal{E}_{t,d}$, $m$ and $m'\in \mathcal{M}_ {t, d}$, the specifications are expressed by $\Upsilon_{e_ {m, m'}} = [\omega_ {e_ {m, m'}}, \tau_\text{max}^{e_ {m, m'}}]$, where, $\omega_ {e_ {m, m'}}$ is the link data rate and $\tau_\text{max} ^ {e_ {m, m'}}$ is the maximum tolerable delay.
\section{online robust admission control and resource allocation: Mathematical Formulation}
\label{online robust admission control and resource allocation: Mathematical Formulation}
Based on the business model introduced by 5GPPP \cite{queseth20175g} for the relationship between the IP and the tenants, after a tenant transmits its slice request to the IP orchestrator, if the available resources are sufficient to accept the slice, the resources will be allocated. In the next generation networks slicing, the slice requests with different lifespans arrive at the orchestrator at different times, and the IP must be able to allocate resources to new slices in an online manner and release the resources containing expired slices. High power consumption of IT infrastructures due to environmental and economic reasons has become a major issue among researchers \cite{farkiani2019fast}. The total IT infrastructure power consumption consists of power consumed by the turned-on switches with their applied ports, and the power consumed by the turned-on servers. A robust optimization model is proposed for admission control and allocation of network resources to the online slice requests of tenants to reduce the IT infrastructure's total power consumption next to increasing the accepted slices’ count. Because formation of each slice takes time, in this study, the time slots are considered that at beginning of each time slot, the newly arrived slices during the previous time slot are processed and the resources of slices with expired lifespans are released. In this process ($\phi_{t,d}$) determines the slice lifespan.

\begin{figure}
	\centering
	\includegraphics[width=\columnwidth]{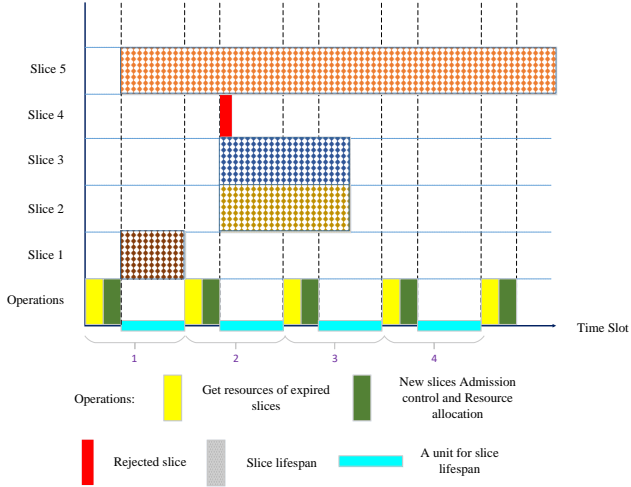}
	\caption{Online Slice Requests and Each Time Slot's Operations }\label{Time-slot-operations}
\end{figure}

An example of the admission control and resource allocation process to 5 slices that arrive at the network during 4 time slots is drawn in Fig. (\ref{Time-slot-operations}), where, at the beginning of each time slot, the process of checking the existence of expired slices and resource allocation to newly accepted slices are evident. A predefined unit of lifespan is considered. Slice 1 with lifespan 1 is accepted in time slot 1, therefore, at the beginning of time slot 2 that slices 2 and 3 are accepted, slice 1 is expired, and its resources are taken back. Because the lifespans of slices 2 and 3 are equal to 2, they are expired at the middle of time slot 3. Slice 4 that arrives at the network in time slot 2, is rejected, because its requested resources are not available in the infrastructure network. slice 5 which is arrived at network in time slot 1, is accepted and because it is a permanent slice, it is not expired. Due to the online allocation, when a slice request arrives, the admission control and the resource allocation are made concerning the remaining resources of the network. Because the admission control and resource allocation operations are made at the beginning of each time slot, the following optimal model makes the optimal allocation only in each time slot, thus, no global optimization. 
\subsection{Slice Acceptance}
To separate the accepted slices from the rejected slices in the slices’ admission control, the variable $\delta_{t,d}$ is defined, which in case of accepting a slice $d$ of tenant $t$, its volume is set to 1, that is, if it is not possible to meet the requirements of a slice in terms of the resources required by its VMs and VLs, and tolerable delays of VLs, this volume is set to 0. The following constraint is defined to specify the rejected slices’ count which should be decreased in the objective function.
\begin{align}\label{Slice Acceptance}
%% 8
&\text{C1: }\sum\limits_{t\in \mathcal{T}_\text{c}} \sum\limits_{d\in\mathcal{\mathcal{D}_\text{$t\_\text{c}$}}}(1-\mathcal{\delta}_{t,d}) = {\eta},
\end{align}
where, ${\eta}$ is an integer variable that specifies the count of rejected slices in current time slot.
\subsection{VNF Placement}
Each slice is considered as a VN that includes VMs and VLs and after the slice is accepted, the tenant as a slice owner manages the required VNFs on the VN. In this article, VNF placement means allocating resources to a VM. The binary variable $\pi^m_n$ is set to 1, if VM $m$ is placed on node $n$. To assure that a VM $m$ is placed on a single node and to prevent a rejected slice’s VM from being placed on the nodes, the following constraint is defined:
\begin{align}\label{VNF Placement}
%% 9
&\text{C2: }\sum\limits_{n \in \mathcal{N}} {\pi}_{n}^{m} = {\delta}_{t,d}, \forall m \in \mathcal{M}_{t,d}, \forall t \in \mathcal{T}_\text{c}, \forall d \in \mathcal{D}_{t\_\text{c}}.
%% 11
%&\text{C3: }{\pi}_{n}^{m} \le {\delta}_{t,d},\forall n\in \mathcal{N},\forall m \in \mathcal{M}_{t,d},\forall{t}\in \mathcal{T}_\text{c},\\
%&\forall {d} \in \mathcal{D}_{t\_\text{c}}.
\end{align}
In this constraint, ${\delta}_{t,d}$ is applied on the right side to assure that decisions are made only for the placement of the accepted slices’ VMs.
\subsection{VL Embedding}
There may be several paths between the two nodes $n$ and $n'$, as $\mathcal{P} _ {n, n'} = \{1, .., b, .., P _ {n, n'}\}$, such that $P _ {n, n'}$ is the physical paths’ count between the two nodes $n$ and $n'$, consequently, the $b^\text{th}$ path is symbolized as $p_ {n, n'} ^ {b}$. The binary decision variable $\xi _ {p^ {b}_{n, n'}} ^ {e_ {m, m'}}$ is defined to determine whether VL $e_ {m , m'}$ is embedded over path $p_ {n, n'} ^ {b}$ or not and if so, $\xi _ {p^ {b}_{n, n'}} ^ {e_ {m, m'}}$ is set to 1, otherwise, 0. The following constraints assure that the VLs of the accepted slices are located on only one path between the nodes that contain the associated VMs of the VLs:
\begin{align}\label{VL Embedding}\nonumber
%% 13
&\text{C3: }\sum\limits_{n\in \mathcal{N}} \sum\limits_{n' \in \mathcal{N}} \sum\limits_{b\in\mathcal{\text{$\mathcal{P}$}}_{n,n'}} \xi^{e_{m,m'}}_{p^b_{n,n'}} = \delta_{t,d},\,\forall e_{m,m'} \in \mathcal{E}_{t,d},\\
&\forall{t}\in \mathcal{T}_\text{c},\, 
\forall {d} \in \mathcal{D}_{t\_\text{c}}.\\\nonumber
%% 14
%&\text{C5: }\xi^{e_{m,m'}}_{p^b_{n,n'}} \le \delta_{t,d},\forall e_{m,m'}\in \mathcal{E}_{t,d},\,\forall{t}\in \mathcal{T}_\text{c},\,\\
%&\forall {d} \in \mathcal{D}_{t\_\text{c}}.\\\nonumber
%% 15
&\text{C4: }\xi_{p^{b}_{n,n'}}^{e_{m,m'}} = \pi_n^{m}\times \pi_{n'}^{m'}, \forall n,n' \in \mathcal{N},\forall b\in\mathcal{\text{$\mathcal{P}$}}_{n,n'}, \\
&\forall m,m' \in \mathcal{M}_{t,d},
\forall e_{m,m'} \in \mathcal{E}_{t,d},\forall t \in {\mathcal{T}_\text{c}},\forall{d}\in \mathcal{D}_{t\_\text{c}}.
\end{align}
By applying the nonlinear constraint C4, we assure that if $\xi_{p^{b}_{n,n'}}^{e_{m,m'}}$ is 1, $\pi_n^{m}$ and $\pi_{n'}^{m'}$ should be 1, consequently, to have a BLP model, the following transformation is advised: 
\begin{align}
%% 1
%&\text{C4-1: }\xi^{e_{m,m'}}_{p^b_{n,n'}} = \theta_{n,n'}^{m,m'}, \\
%% 16
&\text{C4-1: }\xi^{e_{m,m'}}_{p^b_{n,n'}} \le \pi_{n}^{m} + 1 - \pi_{n'}^{m'},\\
%% 17
&\text{C4-2: }\pi_{n}^{m} \le \xi^{e_{m,m'}}_{p^b_{n,n'}} + 1 - \pi_{n'}^{m'},\\\nonumber
%% 18
&\text{C4-3: }\xi^{e_{m,m'}}_{p^b_{n,n'}} \le \pi_{n'}^{m'},\,\forall n,n' \in \mathcal{N},\forall b\in\mathcal{\text{$\mathcal{P}$}}_{n,n'}, \\
&\forall m,m' \in \mathcal{M}_{t,d},\forall e_{m,m'} \in \mathcal{E}_{t,d},\forall{t}\in \mathcal{T}_\text{c},\forall{d}\in {\mathcal{D}_{t\_\text{c}}}.
\end{align}
Therefore, constraints (C4-1 to C4-3) should be applied instead of C4.
\subsection{Node Resources Limitations}
The sum of the allocated resources to VMs, placed on a node $n$ must not exceed its capacity ($R'_n$), thus, the following constraints prevail:
\begin{align}\label{Node Resources Limitations}
%% 3
&\text{C5: }\sum\limits_{t\in \mathcal{T}_\text{c}}\sum\limits_{d\in{\mathcal{D}_\text{$t\_\text{c}$}}}\sum\limits_{m \in \mathcal{M}_{t,d}} \pi_{n}^m\times{\nu_m} = {U_n},\forall n \in \mathcal{N}.\\
%% 4
&\text{C6: }{U_n} \le {{R'}_n}, \forall n\in \mathcal{N}.
\end{align}
Because each node and VM have three types of resources CPU, RAM, and storage, the new integer variable $U_n$ is defined to specify the sum of the required resources by the VMs, placed on node $n$. The $U_n$ is a vector containing the required CPU, RAM, and storage by the VMs.
\subsection{Link Bandwidth Limitation}
To determine the physical link $l_ {u, u'}$ as a member of the path $p_ {n, n'} ^ {b}$, the binary indicator $I_ {p_ {n, n'} ^ b} ^ {l_ {u , u'}}$ is defined, where, $\mathcal{L}_ {p_ {n, n'} ^ {b}}$ is the set of all physical links in path $p_ {n, n'} ^ {b}$. The following constraints specify the bandwidth of each link $l_{n,n'}$ applied by VLs on it considering the available bandwidth ($B'_{l_{u,u'}}$). The variable $U'_{l_{u,u'}}$ is defined to specify the sum of the required bandwidths by VLs on the physical link $l_{n,n'}$ and the variable ${\vartheta}_{l_{n,n'}}$ is defined to specify whether link $l_{n,n'}$ is applied in the current time slot or not, consequently, if link $l_{n,n'}$ is applied in the current time slot, this variable is set to 1, otherwise 0.
\begin{align}\label{Link Bandwidth Limitations}\nonumber
%% 5
&\text{C7: }\sum\limits_{t\in \mathcal{T}_\text{c}}\sum\limits_{d\in \mathcal{D}_{t\_\text{c}}} \sum\limits_{e_{m,m'} \in \mathcal{E}_{t,k}} \sum\limits_{n\in \mathcal{N}} \sum\limits_{n'\in \mathcal{N}}
\sum\limits_{b\in\mathcal{\mathcal{P}}_{n,n'}} I^{l_{u,u'}}_{p_{n,n'}^b}\times\\ &\xi^{e_{m,m'}}_{p^b_{n,n'}}\times\omega_{e_{m,m'}} = {U'}_{l_{u,u'}} ,\,\forall l_{u,u'}\in \mathcal{L}_{p_{n,n'}^b}.\\
%% 6
&\text{C8: }{{U'}_{l_{u,u'}}} \le {\vartheta}_{l_{u,u'}}\times{{B'}_{l_{u,u'}}},\forall l_{u,u'}\in \mathcal{L}_{p_{n,n'}^b}.
\end{align}

\subsection{Delay Model}
To assume the VLs’ maximum tolerable propagation delays, the following constraint is defined:
\begin{align}\nonumber
%% 19
&\text{C9: }\sum\limits_{l_{u,u'} \in \mathcal{L}_{p_{n,n'}^b}} I^{l_{u,u'}}_{p_{n,n'}^b}\times \xi^{e_{m,m'}}_{p^b_{n,n'}} \times\tau_{l_{u,u'}} \le \tau^{e_{m,m'}}_{\max} ,\,\\\nonumber
&\forall n,n' \in \mathcal{N},\forall b \in \mathcal{P}_{n,n'}, \forall e_{m,m'}\in\mathcal{E}_{t,d},\,
\forall{t}\in \mathcal{T}_\text{c},\,
\\
&\forall{d}\in \mathcal{D}_{t\_\text{c}}.
\end{align}
Because the tenant as a slice owner is responsible for slice management, the IP does not have any information about the VNFs that will be executed on VMs. Therefore, the execution delay cannot be computed. Also, the transmission delay is not considered, for simplicity \cite{ebrahimi2019joint}. 
\subsection{Node Power Consumption}
The linear node power consumption model proposed by \cite{dayarathna2015data} is modified as follows: 
\begin{align}\label{Node Power Consumption}\nonumber
%% 2
&\text{C10: }\sum_{n\in \mathcal{N}}(P^{\text{Max}}_n - P^{\text{Idle}}_n)\times \frac{{U}^{\text{CPU}}_n}{R^\text{CPU}_n} + (1-\alpha_n)\times \beta_{n}\times P^{\text{Idle}}_n \\
&={N}_{\text{c}}.\\\nonumber
%% 10
&\text{C11: }\pi_{n}^{m} \le \alpha_n,\forall n\in \mathcal{N}_\text{Used},\forall m \in \mathcal{M}_{t,d},\forall{t}\in\mathcal{T}_\text{c},\\
&\forall {d} \in \mathcal{D}_{t\_\text{c}}.\\\nonumber
%% 12
&\text{C12: }\pi_{n}^{m} \le \beta_{n},\forall n\in \mathcal{N}_\text{Unused},\forall m \in \mathcal{M}_{t,d},\forall{t}\in\mathcal{T}_\text{c},\\
&\forall{d}\in \mathcal{D}_{t\_\text{c}}.
\end{align}
In constraint C10, $P_{n} ^ \text{Idle}$ and $P_{n} ^ \text{Max}$ are equal to the power consumption volumes of the node $n$ at idle and maximum usage, respectively. The $\frac{U_{n} ^ \text{CPU}}{R_{n}^\text{CPU}}$ specifies the CPU utilization of node $n$ volume. 
To propose an online admission control and resource allocation with the aim of power consumption minimization, we need to divide nodes (set $\mathcal{N}$) into two sets $\mathcal{N}_{Used}$ and $\mathcal{N}_{Unused}$ to use nodes that have already been turned on as much as possible. The variable $\beta_{n}$ is set to 1 if node $n \in \mathcal{N}_{Unused}$ is turned on in the current time slot, while, $\alpha_n$ specifies the previously turned-on node $n \in \mathcal{N}_{Used}$. $\alpha_n$ is 0 if node $n$ is not previously applied.
%The set $\mathcal{N}$ is divided into the $\mathcal{N}_\text{Used}$ and $\mathcal{N}_\text{Unused}$ sets that specify nodes status, respectively. 
The circumstances that the nodes are turned-on and VMs are placed on the turned-on nodes is verified through constraints C11 and C12. By applying these constraints, a continuous variable, $N_{\text{c}}$, specifies the total nodes’ power consumption in the current time slot.
\subsection{Switch Power Consumption}
The volume of switch power consumption is computed through constraint C13. The switch power consumption model is proposed by \cite{farkiani2019fast}. For simplicity, a parameter $\zeta_{l_{u,u'}} $ is defined for each link $l_{n,n'}$, is the sum volume of: 1) the idle power consumption volume of switch $u$, 2) the idle power consumption volume of switch $u'$ , 3) the power consumption volume of switch $u$’s applied port for link $l_{u,u'}$, and 4) the power consumption volume of switch $u'$’s applied port for link $l_{u,u'}$. The formulation for computing the power of each link in its cumulative sense is as follows:
\begin{align}\label{link_power_calculation}
%% 1
&\zeta_{l_{u,u'}} = (2\times S^\text{Port}_u)+\frac{S_{u}}{S'_{u}} + \frac{S_{u'}}{S'_{u'}},\forall l_{u,u'} \in \mathcal{L}, u \neq u'. \\ 
& \label{link_power_calculation2} \zeta_{l_{u,u'}} = (S^\text{Port}_u)+\frac{S_{u}}{S'_{u}},\forall l_{u,u'} \in \mathcal{L}, u = u'.
\end{align}
In these two equations, $S_u$ and $S^\text{Port}_u$ are switch $u$ and its one port power consumption, respectively. Also, $S'_u$ specifies number of connected ports of switch $u$ to other switches. When a previously unused link $l_{u,u'}$ is applied in the current time slot for embedding VLs of newly accepted slices, first, the $\vartheta_{l_{u,u'}}$ is set to 1, and next, the sum of $\zeta_{l_{u,u'}}$ for all newly activated links in current time slot specifies the switches' power consumption. Here, a continuous variable, $S_{\text{c}}$, specifies the total switches' power consumption in the current time slot.
\begin{align}\label{Switch Power Consumption}
%% 7
&\text{C13: }\sum\limits_{u\in\mathcal{N}}\sum\limits_{u'\in\mathcal{N}}\vartheta_{l_{u,u'}}\times{\zeta}_{l_{u,u'}} = {S}_{\text{c}}, \vartheta_{l_{u,u'}} \neq \gamma_{l_{u,u'}}.
\end{align}

\subsection{The Admission Control and Resource Allocation Base Model}
\label{The Admission Control and Resource Allocation Base Model}
 The nodes' and switches' power consumption and the accepted slices’ count are considered as the main components of the objective function in this article to reduce the power consumption volume and the rejected slices’ count. Here, the objective function, named  C, is expressed as follows:
 \begin{align}\label{objective function}\nonumber
%% 1
&C =  (\frac{{N}_{\text{c}}}{{N}_{\text{Total}}}) +
(\frac{{S}_{\text{c}}}{{S}_{\text{Total}}}) + \eta+\sum\limits_{n\in\mathcal{N}}(\frac{U^\text{RAM}_n}{R^\text{RAM}_n}) + \\
&\sum\limits_{n\in\mathcal{N}}(\frac{U^\text{STOR}_n}{R^\text{STOR}_n}) + \sum\limits_{u\in\mathcal{N}}\sum\limits_{u'\in\mathcal{N}} (\frac{{U'}_{l_{u,u'}}}{{B}_\text{Total}}).
\end{align}
In the objective function C all components are normalized, to give priority to minimize $\eta$, the one that specifies the rejected slices’ count. Therefore, first $\eta$ and then the rest of the components are minimized. The nodes’ and switches’ power consumption are computed through $N_\text{c}$ and $S_\text{c}$, respectively. Because of the considered online approach in this article, the last three components in the objective function are defined to avoid the resources waste. The base model of admission control and resource allocation is presented as follows:
\begin{align}\label{joint_optimization}\nonumber
&\mathop{\min}~~C\\\nonumber
&\text{subject to:}\\\nonumber
&\text{(C1)-(C3),(C4-1)-(C4-3),(C5)-(C13)}\\\nonumber
&\text{C14: }\beta_n \in \{0,1\}, \forall n.\\\nonumber
&\text{C15: }\delta_{t,d} \in \{0,1\}, \forall t,d.\\\nonumber
&\text{C16: }\pi^m_n \in \{0,1\}, \forall t,d,m,n.\\
&\text{C17: }\xi^{e_{m,m'}}_{p^b_{n,n'}} \in \{0,1\}, \forall e_{m,m'},b,n,n',t,d.\\\nonumber
&\text{C18: }\vartheta^{l_{n,n'}} \in \{0,1\}, \forall l_{n,n'}.\\\nonumber
%&\text{C19: }\eta \geq 0.\\\nonumber
%&\text{C20: }U_n \geq 0, \forall n.\\\nonumber
%&\text{C21: }{U'}_{l_{n,n'}} \geq 0, \forall l_{n,n'}.\\\nonumber
%&\text{C22: }N_\text{c} \geq 0.\\\nonumber
%&\text{C23: }S_\text{c} \geq 0.
\end{align}
The rejected slices’ count is computed through constraint C1. Constraint C2 is applied in placing the accepted slices’ VMs on the nodes. By enforcing constraints C3 to C4-3, it is assumed that the VLs of accepted slices are located on only one path between nodes that include the associated VMs of the VLs.  The volume of applied resources in each node $n$ by VMs on it concerning the available resources is specified  by enforcing constraints C5 and C6. The applied bandwidth of each link $l_{u,u'}$ by VLs on it concerning the available bandwidth is specified by enforcing constraints C7 and C8. Constraint C9 is related to delay limit of each VL. The volume of nodes’ power consumption is computed through constraints C10, C11 and C12.  The volume of switches’ power consumption is computed through constraint C13. The binary decision variables of the problem are specified through constraints C14-C18.
\subsection{The Admission Control and Resource Allocation Robust Model}
%and reduces resource efficiency. For example, for a slice requested by a tenant, the amount of resources which  has considered for a virtual machine, as well as the data rate that orders for each of its requested slice links, are usually for peak conditions and it doesn't always use all the requested resources. So the assumption that the virtual links data rates and virtual machines resources are fixed and constant is
%physical links and nodes resources are not well used and wasted.
Considering the resources of each VM (CPU, RAM, and storage) and the data rate for each slice's link on constant basis, is a very ideal and unreal assumption. In this situation, if the demand  face unusual changes, the VLs and VMs become congested, leading to slice violation. To overcome this phenomenon and formulate the problem in conditions closer to reality, the uncertainties in the data rates of VLs and the workloads of VMs are considered as the stochastic variables. For this purpose, the requested resources of a slice are modeled in a sense that for slice $s_ {t, d}$, ${\nu _ {m}}=[\nu _ {m}^\text{CPU},\nu_ {m} ^ \text{RAM}, \nu _ {m} ^ \text{STOR}]$ and data rate $\omega_ {e_ {m, m'}}$ are considered as ${\widetilde{\nu}_{m}} = [{\widetilde{\nu} _ {m}^ \text{CPU}}, {\widetilde{\nu} _ {m} ^ \text{RAM}}, {\widetilde{\nu} _ {m} ^ \text{STOR}}]$ and ${\widetilde{\omega}_ {e_ {m, m'}}}$ rather than being fixed, where the volumes of ${\widetilde{\nu} _ {m}^ \text{CPU}}$ , ${\widetilde{\nu} _ {m} ^ \text{RAM}}$ , ${\widetilde{\nu} _ {m} ^ \text{STOR}}$, and ${\widetilde{\omega}_ {e_ {m, m'}}}$ have a uniform distribution \cite{wen2017robust} in the intervals $[{ \overline{\nu} _ {m} ^ \text{CPU}}- {\widehat{\nu} _ {m} ^ \text{CPU}}, {\overline{\nu} _ {m} ^ \text{CPU}} + {\widehat{\nu} _ {m} ^ \text{CPU}}]$, $[{\overline{\nu} _ {m} ^ \text{RAM}}- {\widehat{\nu} _ {m} ^ \text{RAM}}, { \overline{\nu} _ {m} ^ \text{RAM}} + {\widehat{\nu} _ {m} ^ \text{RAM}}]$, $[{\overline{\nu} _ {m} ^ \text{STOR}}- {\widehat{\nu} _ {m} ^ \text{STOR}}, {\overline{\nu} _ {m} ^ \text{STOR}} + {\widehat{\nu} _ {m} ^ \text{STOR}}]$, and $[{\overline{\omega}_ {e_ {m, m'}}}- {\widehat{\omega}_ {e_ {m, m'}}}, {\overline{\omega}_ {e_ {m, m'}}}+ {\widehat{\omega}_ {e_ {m, m'}}}]$, respectively. In the defined intervals, ${\overline{\nu} _ {m} ^ \text{CPU}}, {\overline{\nu} _ {m} ^ \text{RAM}}, {\overline{\nu} _ {m} ^ \text{STOR}}$, and ${\overline{\omega}_ {e_ {m, m'}}}$ are the centers of intervals or said otherwise, the nominal volumes, and ${\widehat{\nu} _ {m} ^ \text{CPU}} , {\widehat{\nu} _ {m} ^ \text{RAM}}, {\widehat{\nu} _ {m} ^ \text{STOR}}$, and ${\widehat{\omega}_ {e_ {m, m'}}}$ are the maximum deviation of demand. These intervals can be defined by the tenant, where, a tenant for a given slice, declares the least and the most slice’s required data rates, likewise, an interval for the requested resources’ volumes by the tenant can be of concern, which would overcome the fluctuations in traffics and workloads. An optimization model of admission control and resource allocation with the defined stochastic variables is presented, where, the changes that is considered in slice requests affect the constraints C5 and C7 of the base model (\ref{joint_optimization}). The stated constraints will be changed as follows:
\begin{align}\label{constraints_ with_stochastic_variables}
%% 3
&\sum\limits_{t\in \mathcal{T}_\text{c}}\sum\limits_{d\in{\mathcal{D}_\text{$t\_\text{c}$}}}\sum\limits_{m \in \mathcal{M}_{t,d}} \pi_{n}^m\times{{\widetilde{\nu_m}}} = {U_n}, \forall n \in \mathcal{N}.\\\nonumber
%% 5
&\sum\limits_{t\in \mathcal{T}_\text{c}}\sum\limits_{d\in \mathcal{D}_{t\_\text{c}}} \sum\limits_{e_{m,m'} \in \mathcal{E}_{t,k}} \sum\limits_{n\in\mathcal{N}} \sum\limits_{n'\in\mathcal{N}}
\sum\limits_{b\in\mathcal{\mathcal{P}}_{n,n'}} I^{l_{u,u'}}_{p_{n,n'}^b}\times\text{${\xi}$}^{e_{m,m'}}_{p^b_{n,n'}}\\
&\times{\widetilde{\omega}_{e_{m,m'}}} = {U'}_{l_{u,u'}} ,\,\forall l_{u,u'}\in \mathcal{L}_{p_{n,n'}^b}.
\end{align}

Because the formulation with these modified constraints that includes stochastic variables cannot be solved directly, for its worst-case robust counterpart formulation, the $\Gamma$-Robustness is applied to encounter uncertainties. In this robust formulation, the stochastic variables are converted into deterministic ones, using the conversion process presented in \cite{marotta2017energy}. 
%\begin{align}\label{robust_conterpart}
%\nonumber
%%% 3
%&\sum\limits_{t}\sum\limits_{d}\sum\limits_{m} \pi_{n}^m\times{\overline{\nu_m}} +  \mathop{\max}\limits_{|\epsilon_m| \le 1}\{\sum\limits_{t} \sum\limits_{d} \sum\limits_{m}
%\pi_{n}^{m} \\
%&\times\epsilon_{m}\times\widehat{{\nu}_{m}}\} = {U_n}, \forall n \in \mathcal{N}.\\\nonumber
%% 5
%&\sum\limits_{t}\sum\limits_{d} \sum\limits_{e_{m,m'}} \sum\limits_{n} \sum\limits_{n'}\sum\limits_{b} I^{l_{u,u'}}_{p_{n,n'}^b}\times \xi^{e_{m,m'}}_{p^b_{n,n'}}\times\overline{ \omega_{e_{m,m'}}}+ \\\nonumber
%&+\mathop{\max}\limits_{|\epsilon'_{e_{m,m'}}| \le 1}\{\sum\limits_{t}\sum\limits_{d} \sum\limits_{e_{m,m'}} \sum\limits_{n} \sum\limits_{n'}\sum\limits_{b} I^{l_{u,u'}}_{p_{n,n'}^b}\times \xi^{e_{m,m'}}_{p^b_{n,n'}}\\
%&\times\epsilon'_{e_{m,m'}}\times\widehat{ \omega_{e_{m,m'}}}\} = {U'}_{l_{u,u'}} ,\,\forall l_{u,u'}\in \mathcal{L}_{p_{n,n'}^b}.
%\end{align}
Therefore, the linear $\Gamma$-Robustness based model with 2 parameters $\Gamma_1$ and $\Gamma_2$, named \textbf{ROBINS}, is as follows:\\ \\
\textbf{(ROBINS)}
\begin{align}\label{robust_joint_optimization}\nonumber
%% 1
&\mathop{\min}~~C\\\nonumber
&\text{subject to:}\\\nonumber
&\text{(C1)-(C3),(C4-1)-(C4-3)}\\\nonumber
&\text{C5-1: }\sum\limits_{t}\sum\limits_{d}\sum\limits_{m} {\pi}_{n}^m\times{\overline{\nu}_m} + \sum\limits_{t}\sum\limits_{d}\sum\limits_{m}{\rho_{1}}^{m}_{n} + \Gamma_1\times{z_{1}}_{n}\\\nonumber 
&= {U_n},\forall n \in \mathcal{N}.\\\nonumber
%% 3-1
&\text{C5-2: }\pi_{n}^m\times{\widehat{\nu_m}} \le {\rho_{1}}^{m}_{n} + {z_{1}}_{n},\quad \forall n,m,t,d.\\\nonumber
&\text{(C6)}\\\nonumber
&\text{C7-1: }\sum\limits_{t}\sum\limits_{d} \sum\limits_{e_{m,m'}} \sum\limits_{n} \sum\limits_{n'}
\sum\limits_{b} I^{l_{u,u'}}_{p_{n,n'}^b}\times \xi^{e_{m,m'}}_{p^b_{n,n'}}\times\overline{\omega}_{e_{m,m'}}+\\\nonumber
&\sum\limits_{t}\sum\limits_{d}\sum\limits_{e_{m,m'}}\sum\limits_{n}\sum\limits_{n'}
\sum\limits_{b}I^{l_{u,u'}}_{p_{n,n'}^b}\times{\rho}^{e_{m,m'}}_{p^b_{n,n'}} + \Gamma_2\times{z_2}_{l_{u,u'}}  =\\ 
&{U'}_{l_{u,u'}} ,\forall l_{u,u'}.\\\nonumber
&\text{C7-2: }I^{l_{u,u'}}_{p_{n,n'}^b} \times\xi^{e_{m,m'}}_{p^b_{n,n'}}\times{\widehat{\omega}_{e_{m,m'}}} \le {\rho_2}^{e_{m,m'}}_{p^b_{n,n'}} + {z_2}_{l_{u,u'}},\\\nonumber
&\forall e_{m,m'},b,n,n',l_{u,u'}.\\\nonumber
&\text{(C8)-(C18)}\\\nonumber
%&\text{C24: }{\rho_{1}}^{m}_{n} \geq 0, \forall m,n.\\\nonumber
%&\text{C25: }{z_{1}}_{n} \geq 0, \forall n.\\\nonumber
%&\text{C26: }{\rho_2}^{e_{m,m'}}_{p^b_{n,n'}} \geq 0,\forall e_{m,m'},b,n,n'.\\\nonumber
%&\text{C27: }{z_2}_{l_{u,u'}} \geq 0, \forall l_{u,u'}.
\end{align}
In ROBINS, the robustness auxiliary variables ${\rho_{1}}^{m}_{n}$, ${z_{1}}_{n}$, ${\rho_2}^{e_{m,m'}}_{p^b_{n,n'}}$, and ${z_2}_{l_{u,u'}}$ are integers. To specify the relative deviations, the ${\widehat{\nu} _ {m}} = \Delta_1 \times {\overline{\nu} _ {m}}$ and ${\widehat{\omega}_ {e_ {m, m'}}} = \Delta_2 \times {\overline{\omega}_ {e_ {m, m'}}}$ such that $\Delta_1$ and $\Delta_2$ are considered within 0 to 1 range. By applying $\Gamma_1$ and $\Gamma_2$ parameters, the count of the VMs and VLs that their demand can fluctuate, are specified, respectively. It is worth noting that this worst-case model that considers the resource requests according to the worst conditions, can be applied to any resource request distribution \cite{jiang2016data}, including uniform, normal, etc. On the other hand, the advantage of $\Gamma$-Robustness compared to the general worst-case form \cite{wen2017robust}, is that, the parameter $\Gamma$ can be used to control the number of random variables whose values can fluctuate. Therefore, the IP can use this parameter to control the robustness alongside the $\Delta$ parameter that specifies the fluctuation interval.
\begin{center}
  \begin{figure}[t!h]
	\centering
	\includegraphics[width=\columnwidth]{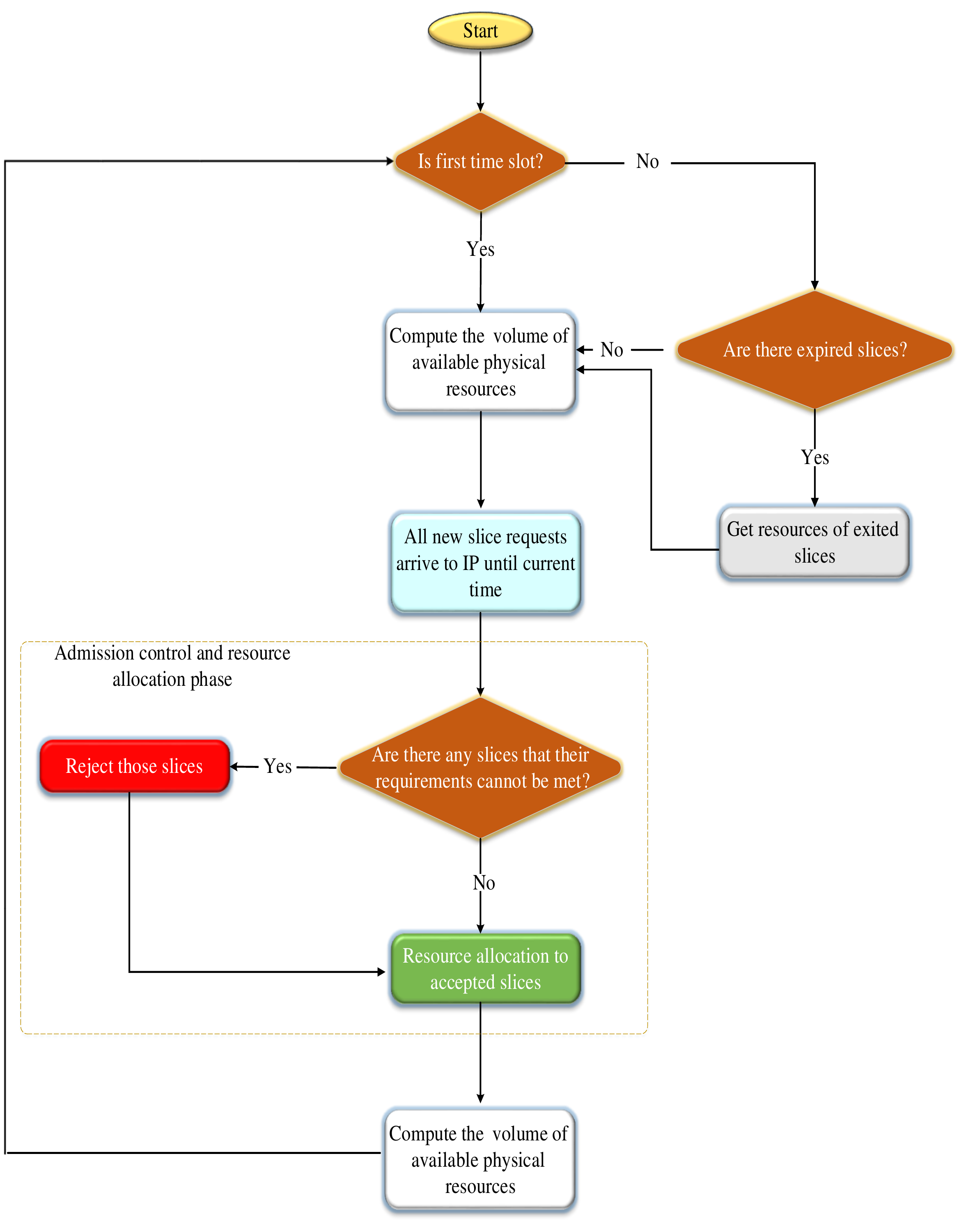}
	\caption{Online Admission Control and Resource Allocation Process}\label{flowchart}
\end{figure}
%\vspace{-1em}
\end{center}
\section{online robust admission control and resource allocation: Solution Methodology}
\label{online robust admission control and resource allocation: Solution Methodology}
An admission control and resource allocation algorithm is proposed through the ROBINS. ROBINS is based on BLP that can be solved by using any commercial optimization solver, such as IBM ILOG CPLEX. However, finding the optimal solution in reasonable amount of time may have difficulties for an advanced state-of-the-art slover like CPLEX when the size of the problem increase (e.g. real-world and large-scale networks)\cite{marotta2017energy}. Therefore, to find a near-optimal solution in a reasonable amount of time, a heuristic algorithm is presented. The general process of admission control and resource allocation of the newly arrived slices applied in this article is shown in Fig. (\ref{flowchart}), where, first, in each time slot except the first, the allocated resources to the slices that are expired are taken back, next, the possibility of acceptance of arrived slices are checked and if there exist slices which their requirements cannot be met, they are rejected and then the requested resources’ are allocated to the accepted slices.
\subsection{Optimal Admission Control and Resource Allocation Algorithm}\label{Optimal Admission control and Resource Allocation Algorithm}
An algorithm is proposed for optimal admission control and resource allocation, according to process shown in Fig. (\ref{flowchart}) and ROBINS, named OEA-ONSU. Algorithm \ref{algorithm_OEA_ONSU} shows the OEA-ONSU. In this algorithm, each time slot, consist of three steps: 1) the existence of expired slices are checked (line 2). If the slice lifespan is expired, both the allocated resources to its VMs (line 4-8), and VLs (line 9-15) will be taken back and the sets $\mathcal{T}$, $\mathcal{D}_t$,and $\mathcal{S}$ must be updated, otherwise, only the  $\phi_{t,d}$' volume is reduced one unit (lines 19-21). If there exist no VMs and VLs on a previously applied server and link, respectively, these two will be turned OFF (lines 24-25 and 34-35), otherwise, if the lifespans of the slices in a previous time slot are expired, the resources considered for robustness will be taken back to the servers and links (lines 26-31 and 36-41). After the expired slices’ and the available network resources’ status are determined, it is time to apply the admission control and resource allocation to the newly arrived slices, that is, entering step 2), where, the ROBINS model must be solved (line 43) and 3) where, the sets $\mathcal{T}$, $\mathcal{D}_t$, and $\mathcal{S}$, and the volume of applied resources, and consumed power must be updated. \\

%\begin{figure}[!h]
%    \begin{minipage}{0.98\columnwidth}
%        \begin{algorithm}[H]
\begin{algorithm}[t]
    \caption{OEA-ONSU}
    \footnotesize
    \tiny
    \SetAlgoLined
    \label{algorithm_OEA_ONSU}
    %\DontPrintSemicolon
    %\normalsize
    \KwInput{$G,\mathcal{S},\mathcal{S}_\text{c},\mathcal{T},\mathcal{D}_t,\mathcal{T}_\text{c},\mathcal{D}_{t\_\text{c}},N_\text{Used},S_\text{Used},O^{t'}_n,O'^{t'}_{l_{n,n'}}, \Gamma_1,\Gamma_2,\Delta_1,\Delta_2$}
    \KwOutput{Rejected slices of $\mathcal{S}_\text{c}$, Servers and paths for embedding accepted slices of $\mathcal{S}_\text{c}$}
    \For {each time slot}
    {
        \tcc{STEP 1: Get resources of expired slices}
        \For {each tenant $t$ in $\mathcal{T}$ and each slice $d$ in $\mathcal{D}_t$}
        {
            \eIf {$\phi_{t,d} == 0$}
            {
                \For{each $m$ in $\mathcal{M}_{t,d}$ and each $n$ in $\mathcal{N}_\text{Used}$, if $\pi^m_n == 1$}{

                    $\pi^m_n = 0$\;
                    ${R'}_n += \nu_m$\;
                    Update $N_\text{Used}$\;

                }
                \For{each $e_{m,m'}$ in $\mathcal{E}_{t,d}$ and each Path $b$ in $\mathcal{P}_{n,n'}$, if $\xi^{e_{m,m'}}_{p^b_{n,n'}} == 1$}
                {
                    $\xi^{e_{m,m'}}_{p^b_{n,n'}} = 0$\;
                    \For{each used link $l_{u,u'}$, if  $l_{u,u'}$ in $p^b_{n,n'}$}
                    {
                        ${B'}_{l_{u,u'}} += \omega_{e_{m,m'}}$\;
                        Update $S_\text{Used}$\;
                    }
                
                }
                Update $\mathcal{T}$\;
                Update $\mathcal{D}_t$\;
                Update $\mathcal{S}$\;
            }{
            $\phi_{t,d} -= 1$\;
            }
            
        }
        % az aval for ta injast khata
        \For {each $n$ in $\mathcal{N}_\text{Used}$}
        {
            \eIf {number of VMs on $n == 0$}
            {
                %$NR^{Remaining}_n = NR_n$\\
                $\mathcal{N}_\text{Used}$.remove$(n)$\;
            }{
                \For {each time slot $t'$ in previous time slots, if in current time slot,$\phi$ of all slices in $t' == 0$}
                {
                    $R'_n += O^{t'}_{n}$\;
                    Update $N_\text{Used}$\;
                }
            }
        }
        \For{each used link $l_{u,u'}$}
        {
            \eIf {number of VLs on $l_{u,u'} == 0$}
            {
                $\gamma_{l_{u,u'}} = 0$\;
            }{
                \For{each time slot $t'$ in previous time slots, if in current time slot, $\phi$ of all slices in $t' == 0$}
                {
                    ${B'}_{l_{u,u'}} + = O'^{t'}_{l_{u,u'}}$\;
                    Update $S_\text{Used}$\;
                }
            }
        }
        \tcc{STEP 2: Admission control and Resource allocation}
        \textbf{Solve ROBINS}\;
        \tcc{STEP 3: Update $\mathcal{T}$, $\mathcal{D}_t$, $\mathcal{S}$, and the volume of available resources,  applied resources for robustness, and the power consumption}
        \For {each accepted slice in current time slot}
        {
            Update $\mathcal{T}$\;
            Update $\mathcal{D}_t$\;
            Update $\mathcal{S}$\;
        }
        \For {each $n$ in $\mathcal{N}_\text{Used}$}
        {
            Update ${R'}_n$\;
            Update $O^t_n$ for current time slot $t$\;
        }
        \For{each used link ${l_{u,u'}}$}
        {
            Update ${B'}_{l_{u,u'}}$\;
            Update $O'^t_{l_{u,u'}}$ for current time slot $t$\;
        }
        Update $N_\text{Used}$\;
        Update $S_\text{Used}$\;
    }
\end{algorithm}

\subsection{Near Optimal Admission control and Resource Allocation Algorithm}\label{Near Optimal Admission control and Resource Allocation Algorithm}
To find a near-optimal solution in a reasonable amount of time in the step 2 of Algorithm \ref{algorithm_OEA_ONSU}, a greedy algorithm is devised to use instead of ROBINS. In this context, the new three-step algorithm, named NEA-ONSU, is devised,  and Algorithm \ref{algorithm_NEA_ONSU} shows its details. Step 1 of this algorithm resembles that of Algorithm \ref{algorithm_OEA_ONSU}. In step 2, first, all the servers are sorted in the descending order of their available capacities (line 3), next, all slices are sorted in a descending order based on the total capacity required for their VMs (line 4) and for each slice, the requested VMs are sorted in descending order of their capacities (line 6). For each VM, first, the first proper server should be found according to the resources required for that VM and robustness, and next, for each VM that is connected to this VM, the same process is run, where the required bandwidth capacity and robustness and the maximum tolerable propagation delay of the VL between the two current VMs are of concern. Consequently, first, the existing paths between the two candidate servers for placing the two VMs based on their propagation delays in an ascending order must be sorted and next, the first path that meets the requirements (lines 7-30) must be selected. If the required servers and links cannot be found for a slice, that slice will be rejected (lines 31-32), otherwise, the sets $\mathcal{T}$, $\mathcal{D}_t$, and $\mathcal{S}$ must be updated (lines 33-37). In step 3, the applied servers and links must be specified, and the volume of applied resources for robustness and the power consumption must be computed.
\begin{itemize}
    \item \textbf{Computational complexity:} Consider $t$ as the current time slot count, and $|\mathcal{T}|$ and $|\mathcal{D}_t|$ as the total accepted tenants’ and slices’ count up to the current time slot, respectively; $|\mathcal{T}_\text{c}|$ and $|\mathcal{D}_{t\_\text{c}}|$ as the current time slot tenants’ and slices’ count, respectively; $\mathcal{|N|}$ as the nodes’ count; $|\mathcal{M}|$ as the VMs’ count; $|\mathcal{E}|$ as the VLs’ set length; $b$ is the count of the paths between two nodes, and $|\mathcal{L}|$ as the physical links’ set length. The computational complexity of step 1 is O($|\mathcal{T}|.|\mathcal{D}_t|.({\mathcal{|N|}}.|\mathcal{M}|+{\mathcal{|N|}}.t+|\mathcal{E}|.\mathcal{|N|}^2b.|\mathcal{L}|)$), step  2 is O($(|\mathcal{T}_\text{c}|.|\mathcal{D}_{t\_\text{c}}|).\log(|\mathcal{T}_\text{c}|.|\mathcal{D}_{t\_\text{c}}|)+\mathcal{|N|}.\log \mathcal{|N|} + (|\mathcal{T}_\text{c}|.|\mathcal{D}_{t\_\text{c}}|).(|\mathcal{M}|^2.\mathcal{|N|}^4.b.\log(\mathcal{|N|}^2.b) + \mathcal{|N|}^4.b^2.|\mathcal{L}|^2.|\mathcal{E}| + |\mathcal{M}|.\mathcal{|N|}^3.b.|\mathcal{L}|)$), and step 3 is O($|\mathcal{T}_\text{c}|.|\mathcal{D}_{t\_\text{c}}|.({\mathcal{|N|}}.|\mathcal{M}|+|\mathcal{E}|.\mathcal{|N|}^2b.|\mathcal{L}|)$). Therefore, the total computational complexity of Algorithm \ref{algorithm_NEA_ONSU} is the sum of the step 1 and step 2 complexities, because, the complexity of step 3 is negligible compared to the steps 1 and 2.
\end{itemize}
\begin{algorithm}[t]
   \caption{NEA-ONSU}
    \footnotesize
    \tiny
    %\DontPrintSemicolon
    \SetAlgoLined
    \label{algorithm_NEA_ONSU}
    \KwInput{$G,\mathcal{S},\mathcal{S}_\text{c},\mathcal{T},\mathcal{D}_t,\mathcal{T}_\text{c},\mathcal{D}_{t\_\text{c}},N_\text{Used},S_\text{Used},O^{t'}_n,O'^{t'}_{l_{n,n'}},\Gamma_1,\Gamma_2,\Delta_1,\Delta_2$}
    \KwOutput{Rejected slices of $\mathcal{S}_\text{c}$, Servers and paths for embedding accepted slices of $\mathcal{S}_\text{c}$}
    \For {each time slot}
    {
        \tcc{STEP 1: Get resources of expired slices}
        Same~as~OEA-ONSU \\
        \tcc{STEP 2: Admission control and Resource allocation}
        
        $\textbf{sorted\_N}$: sort~$\mathcal{N}$~descending~according~to~available~capacity~of~each~node\;
        $\textbf{sorted\_slices}$: sort~$\mathcal{S}_\text{c}$~descending~according~to~requested~VMs'~resources~of~slices\;
        \For {each slice $s$ in sorted\_slices}
        {
            $\textbf{sorted\_VMs}$: sort~VMs~descending according to their requested capacities\;
            \For{each $m$ in sorted\_VMs and each node $n$ in sorted\_N, if node~$n$~is~proper~for~$m$~by~considering~robustness}
            {
                \For{each $m'!=m$ in sorted\_VMs, if ${e}_{m,m'}$~exists}
                {
                    \For{each node $n'$ in sorted\_N, if node $n'$ is proper for $m'$ by considering robustness}
                    {
                        $\textbf{sorted\_paths}$: sort~paths~between~$n$ ,~$n'$~ascending according~to~propagation~delay\;
                        \For{each $p$ in ${{sorted\_paths}}$, if $p.prop\_delay$ $\le$ ${{\tau}^{e_{m,m'}}_{max}}$}
                        {
                            \For{each link ${l_{u,u'}}\in p$, if ${{B'}_{l_{u,u'}}}$ is enough for ${{e}_{m,m'}}$
                            by considering robustness}
                            {
                                \textbf{ }\\
                                ${\pi}^m_n = 1$\;
                                \textbf{ }\\
                                ${R'}_n -= \nu_m$\;
                                \textbf{ }\\
                                ${\pi}^{m'}_{n'} = 1$\;
                                \textbf{ }\\
                                ${R'}_{n'} -= \nu_{m'}$\;
                                \textbf{ }\\
                                ${\xi}^{e_{m,m'}}_{p} = 1$\;
                                \For{each link $l_{u,u'} \in p$}
                                {
                                    ${B'}_{l_{u,u'}}-=\omega_{e_{m,m'}}$\;
                                }
                            }
                        }
                    }
                }

            }
            \eIf{suitable servers and paths not found}
            {
                $\delta_{t,d} = 0$\;
            }{
                Update $\mathcal{T}$\;
                Update $\mathcal{D}_t$\;
                Update $\mathcal{S}$\;
            }
        }
        \tcc{STEP 3: Specify the applied servers and links, and compute the volume of applied resources for robustness and the power consumption}
        
        Update $\mathcal{N}_{Used}$\;
        \For{each $l_{n,n'}$}
        {
            Update $\vartheta_{l_{n,n'}}$\;
        }
        \For {each $n$ in ${\mathcal{N}_\text{Used}}$}
        {
            Update $O^t_n$ for current time slot $t$\;
            ${R'}_n -= O^t_n$\;
            
        }
        \For{each used link ${l_{u,u'}}$}
        {
            Update $O'^t_{l_{u,u'}}$ for current time slot $t$\;
            ${B'}_{l_{u,u'}} -= O'^t_{l_{u,u'}}$\;
        }
        Update $N_\text{Used}$\;
        Update $S_\text{Used}$\;
    
    } 
\end{algorithm}
\section{Numerical Results}
\label{Numerical Results}\subsection{Simulation Environment}
\begin{figure}
     \centering
        \includegraphics[width=0.65\columnwidth]{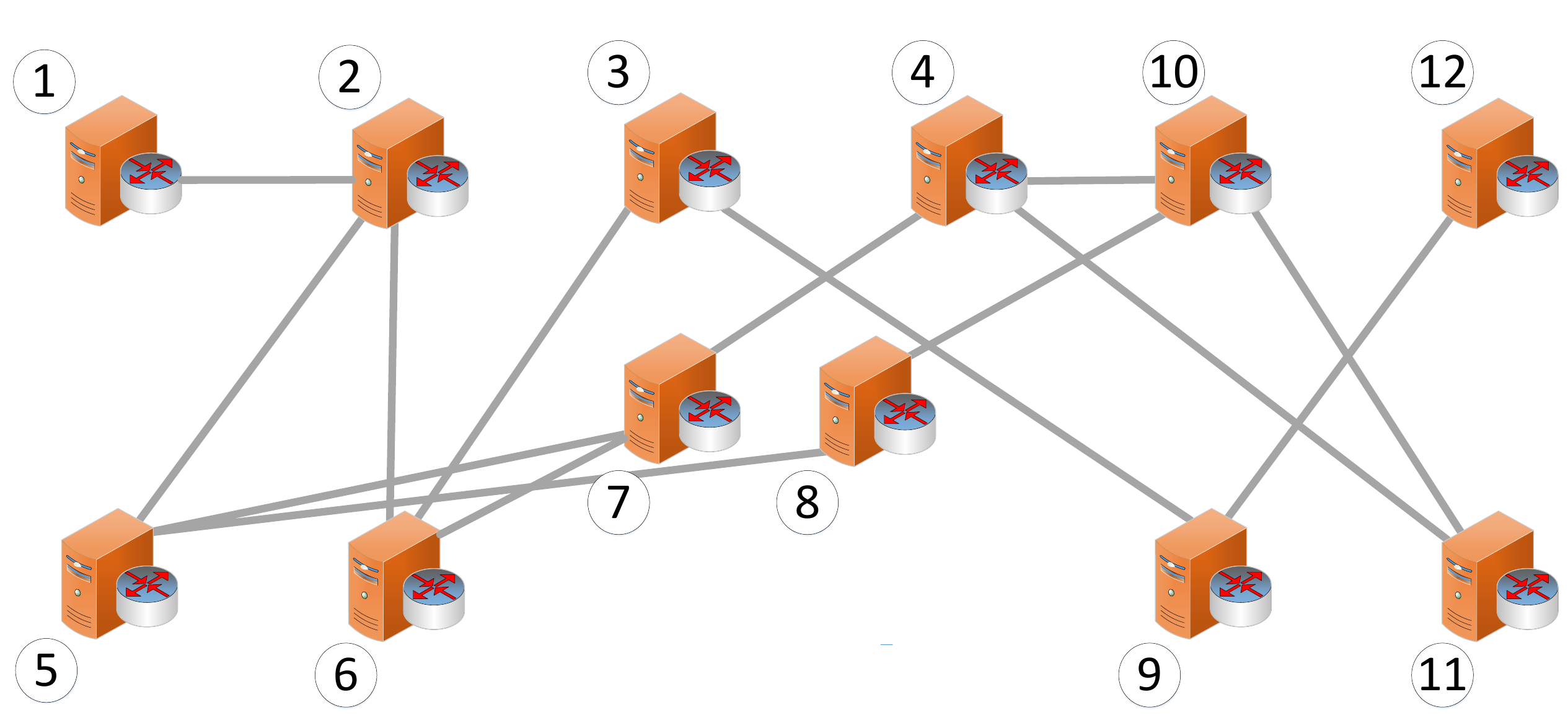}
        \caption{Abilene Network Topology}
        \label{fig:Abilene-Network}
\end{figure}

The simulation is run to evaluate the efficiency of the proposed algorithms. To implement the infrastructure network, the Abilene network \cite{orlowski2010sndlib} with 12 nodes and 27 links is involved, Fig. (\ref{fig:Abilene-Network}). The main Abilene network has 15 links, but here, each node has a link to itself to allow the VMs to interconnect on the common server, therefore, the link capacity for all links that interconnect the switches is set to $BW(i,j)=10$ Gbps, while, the capacity of each node’s local link is set to 40 Gbps, thus, no bottleneck is formed in the local communications.
The algorithms are implemented in Python 3.6. The docplex Python library 2.4.61 and IBM CPLEX Optimizer 12.8 are applied to implement and solve the ROBINS. Simulations are run on a machine with 20 cores 2.3GHz Intel Xeon E5-2695 v3 CPU and 128 GB RAM. The Poisson distribution is applied for new slice requests’ arrivals with the mean arrival rate $\lambda$ = 2 and max arrival rate which is set to 5 per time slot. The slice lifespan ($\mu$) is drawn from an exponential distribution with a mean of 10 time slots (40 time slots of the new and expired slice requests are simulated). The simulation parameters are tabulated in Table~\ref{simulation-parameters}. It is assumed that each tenant has only one slice request. 
Specifications of the physical servers proposed by \cite{farkiani2019fast} are modified and tabulated in Table \ref{servers-specifications}. The VMs' specifications are tabulated in Table \ref{VMs-specifications}. The Barabasi-Albert model \cite{hosseini2019probabilistic,barabasi1999emergence} is used to generate the VNs topologies. As observed in Table \ref{servers-specifications}, two types of servers are of concern and in each simulation iteration, for each server of the infrastructure network, the specifications are randomly selected from these two types. As observed in Table \ref{VMs-specifications}, three types of VMs are of concern and in any type of VM, the resources are designed to be compatible. The switch power data are extracted from \cite{marotta2017fast,bari2019esso} and are modified into: the power of each switch is 184 Watts, the power of each 10 Gbps port is 4.3 Watts and the power of each 40 Gbps port is 13.6 Watts. Consequently, the power of each link is computed through Eqs. (\ref{link_power_calculation} and \ref{link_power_calculation2}). The source code of simulation is available in \cite{ecv7-nz24-21}.

%\newcolumntype{P}[1]{>{\centering\arraybackslash}p{#1}}
\begin{table}[htpb]
\centering
\caption{Simulation Parameters}
\label{simulation-parameters}
\small

\begin{center}
 \begin{tabular}{||m{5cm}|m{3cm}||}
 \hline
 {Number of time slots}& {40} \\  
 \hline
 {$\Gamma_1,\Gamma_2$} & {$[0,4],[0,4]$} \\
 \hline
 {$\Delta_1,\Delta_2$} & {$(0\%,10\%,30\%)$}\\
 \hline
 {Max number of $|\mathcal{T}_\text{c}|$} & 5 \\
 \hline
 {$|\mathcal{D}_{t_\text{c}}|$} & 1 \\
 \hline
 {Number of VMs of each slice request} & U(2-4) \\
 \hline
 {Bandwidth of each VL (Mbps)} & U(100-1500) \\
 \hline
 {Propagation delay of each VL (ms)} & U(4-13) \\
 \hline
 \end{tabular}
 \end{center}

\end{table}

\begin{table}
\centering
\caption{Servers Specifications}
\label{servers-specifications}
\small
\begin{center}
 \begin{tabular}{||m{2cm}|m{2cm}|m{2cm}||}
 \hline
 Type & 1 & 2\\
 \hline
 {CPU (core)}& 32 & 48\\  
 \hline
 {RAM (GB)} & 192 & 768 \\
 \hline
 {Storage (GB)} & 4000 & 4000\\
 \hline
 {$P^\text{Max}$ (Watts)} & 540 & 700 \\
 \hline
 {$P^\text{Idle}$ (Watts)} & 170 & 180 \\
 \hline
 \end{tabular}
 \end{center}
\end{table}

\begin{table}
\centering
\caption{VMs Specifications}
\label{VMs-specifications}
\small
\begin{center}
 \begin{tabular}{||m{2cm}|m{1cm}|m{1cm}|m{1cm}||}
 \hline
 {Type}& {1} & {2}& {3} \\  
 \hline
 {CPU (core)} & 1 & 2 & 4 \\
 \hline
 {RAM (GB)} & 2 & 4 & 16\\
 \hline
 {Storage (GB)} & 120 & 120 & 120 \\
 \hline
 \end{tabular}
 \end{center}
\end{table}
\begin{figure}
     \centering
     \begin{subfigure}[b]{0.48\columnwidth}
         \centering
         \includegraphics[width=\columnwidth]{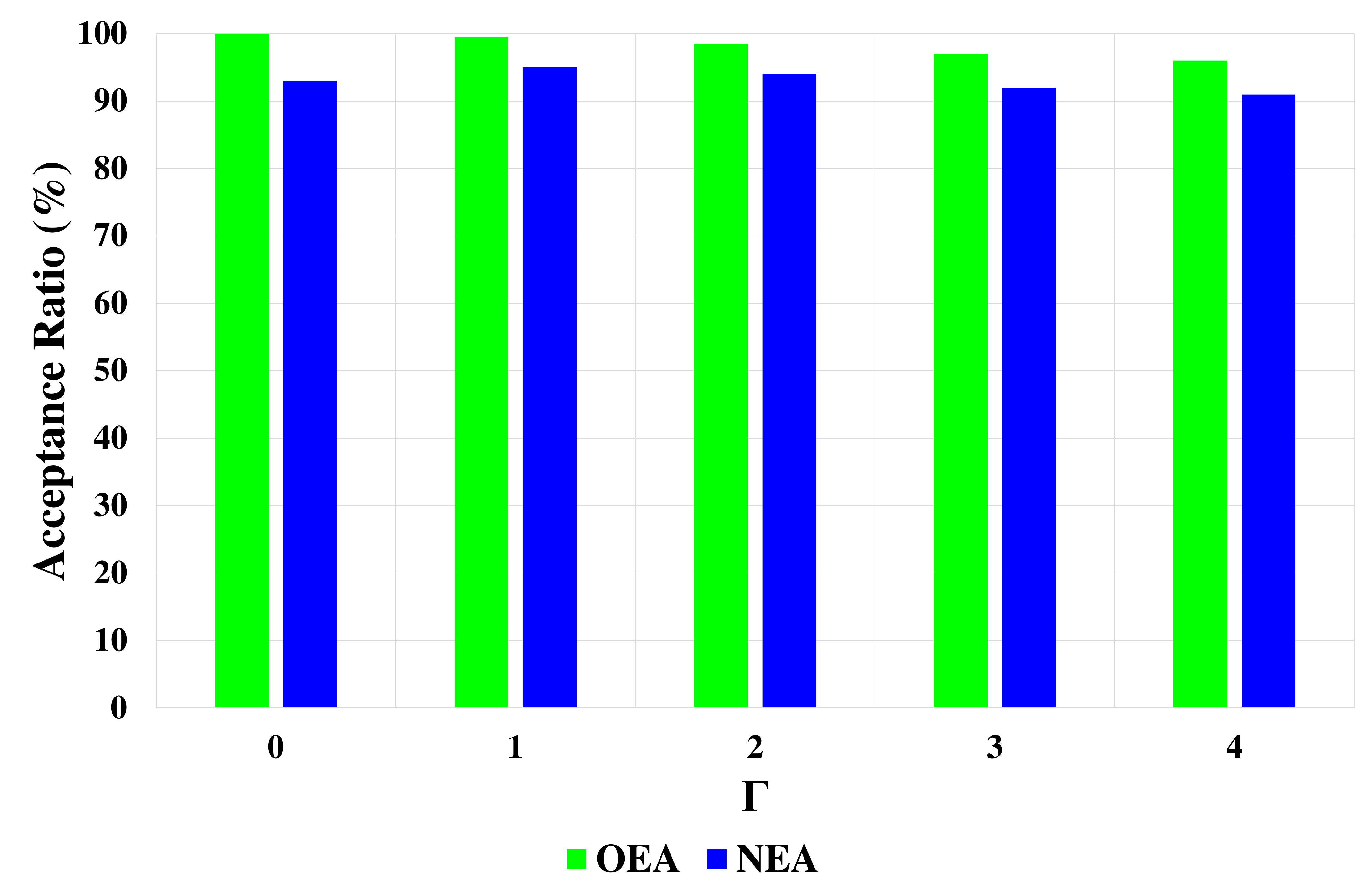}
         \caption{Different Protection Levels ($\Gamma_1$=$\Gamma_2$, $\Delta_1$=$\Delta_2$ and Their Values Are Equal to 10\%)}
         \label{fig:comp-AccRatio-Different-protection-levels}
     \end{subfigure}
     \hfill
     \begin{subfigure}[b]{0.48\columnwidth}
         \centering
         \includegraphics[width=\columnwidth]{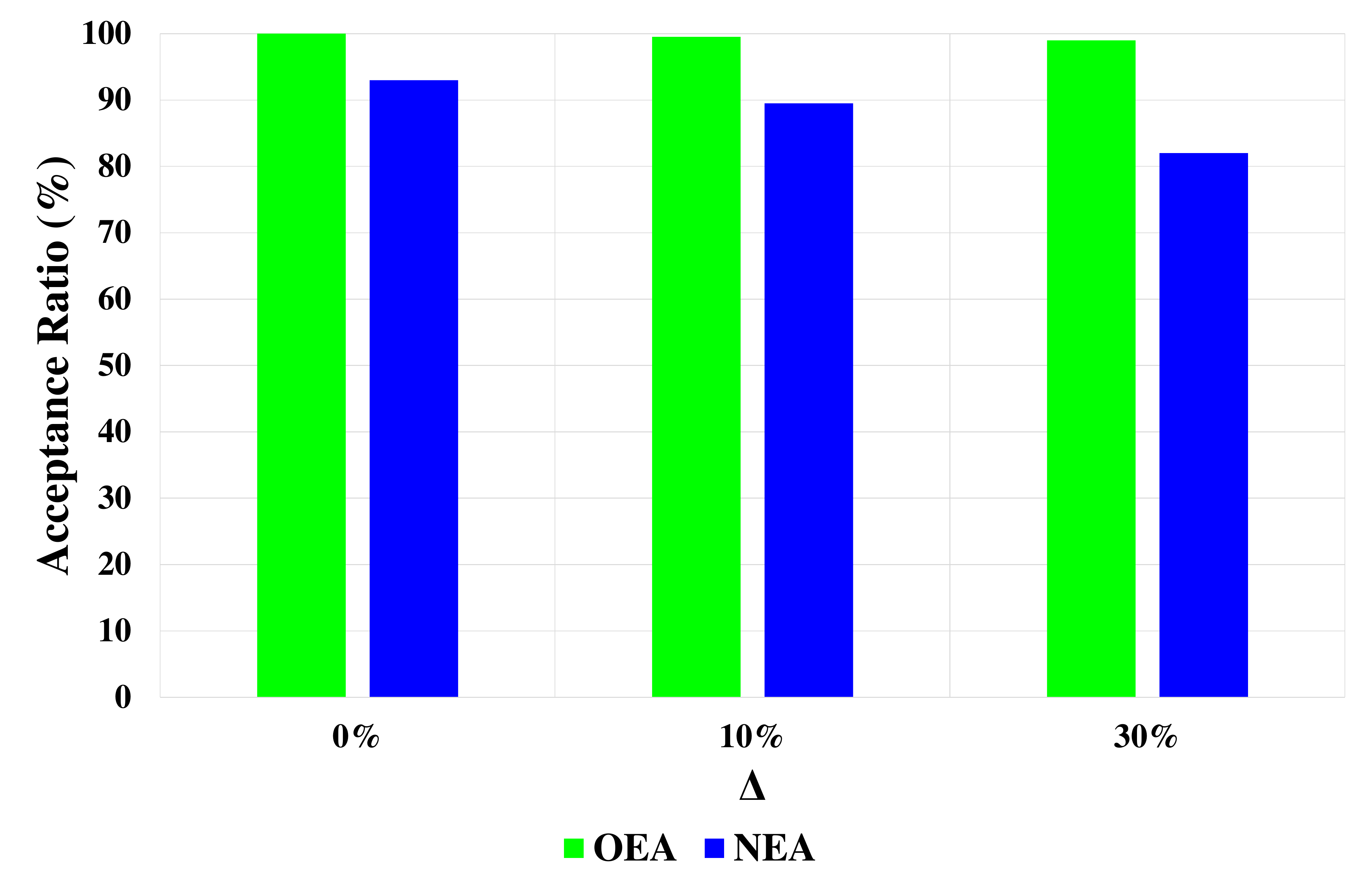}
         \caption{Different Relative Deviations ($\Delta_1$=$\Delta_2$, $\Gamma_1$=$\Gamma_2$ and Their Values Are Equal to 1)}
         \label{fig:comp-AccRatio-Different-relative-deviations}
     \end{subfigure}
     \hfill
        \caption{Acceptance Ratio}
        \label{fig:comp-AccRatio}
    \end{figure}
\subsection{Performance Metrics and Results}
The results are presented for two scenarios: in the first, the relative deviations $\Delta_1$=$\Delta_2$\footnote{Considering $\Delta_1$=$\Delta_2$ is for simplicity and the algorithms have the ability to take different values for these two parameters as input.} for all requested VMs’ resources and VLs’ data rates are set to $10\%$, and results are presented for different protection levels that are specified with $\Gamma_1$=$\Gamma_2$\footnote{Considering $\Gamma_1$=$\Gamma_2$ is for simplicity and the algorithms have the ability to take different values for these two parameters as input.}, with 0 to 4 volumes; in the second, the $\Gamma_1$=$\Gamma_2$ and their volumes are set to 1 and the results are presented for different $\Delta_1$=$\Delta_2$, with 0\%, 10\%, and 30\% volumes. The reason for running these two scenarios is to compare the applied robustness using $\Gamma$ and $\Delta$, which are the two general parameters in providing robustness. The baseline is resource allocation without robustness \cite{marotta2017fast,wen2017robust}. Therefore, we consider the resource allocation without robustness ($\Delta$=0\% or $\Gamma$=0) as the baseline and compare the results of our algorithms in 1) $\Delta$=0\% or $\Gamma$=0 (as the baseline) with 2) $\Delta$ or $\Gamma$ with different values except 0, in two stated scenarios, based on the following performance metrics. The simulation results constitute the 20 simulations’ average. 
    \begin{figure*}
     \centering
     \begin{subfigure}[b]{0.32\textwidth}
         \centering
         \includegraphics[width=\textwidth]{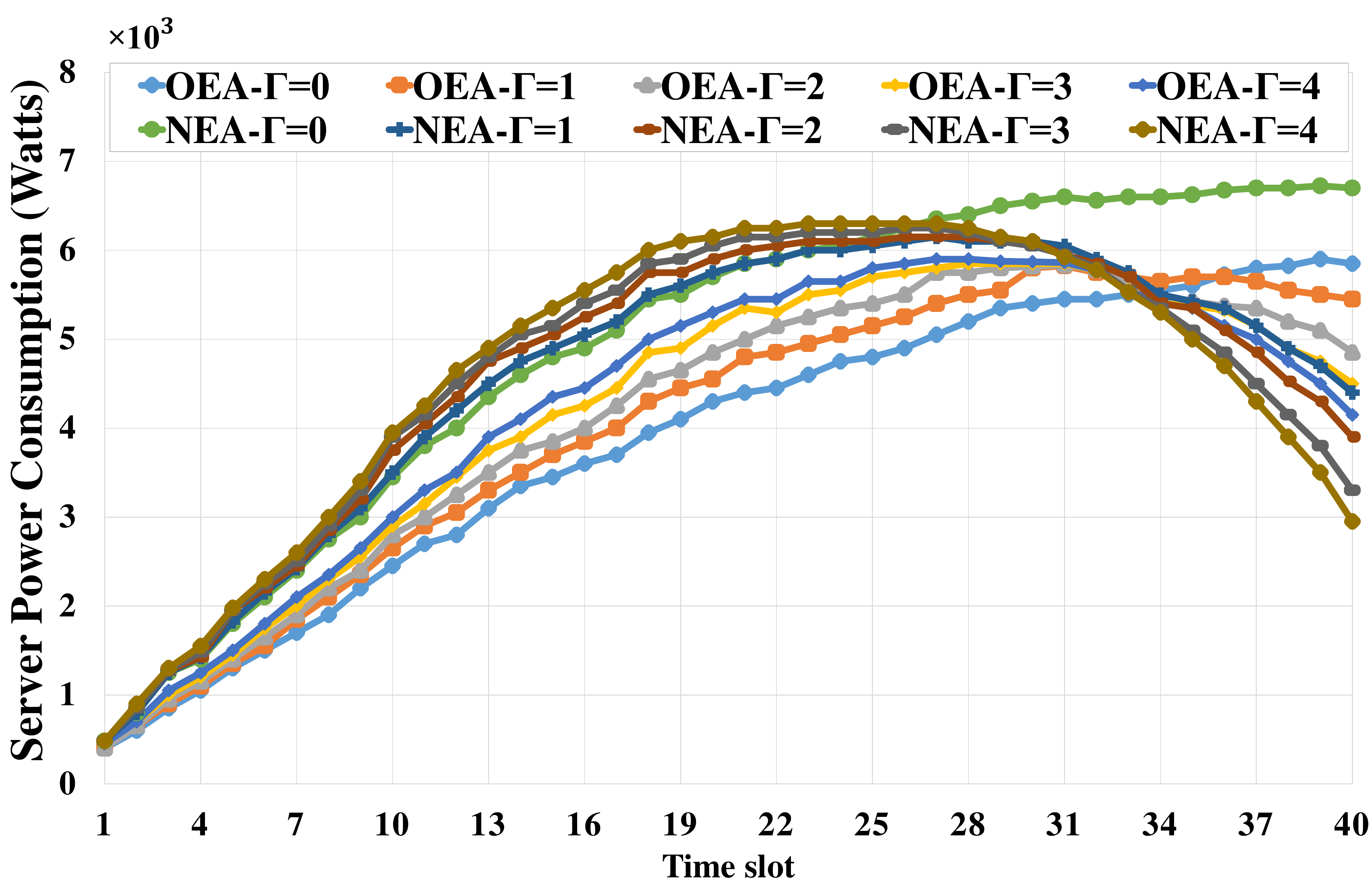}
         \caption{Server Power Consumption in Different Protection Levels ($\Gamma_1$=$\Gamma_2$, $\Delta_1$=$\Delta_2$ and Their Values Are Equal to 10\%)}
         \label{fig:comp-NodePower-timeslot-Different-protection-levels}
     \end{subfigure}
     \hfill
       \begin{subfigure}[b]{0.32\textwidth}
         \centering
         \includegraphics[width=\textwidth]{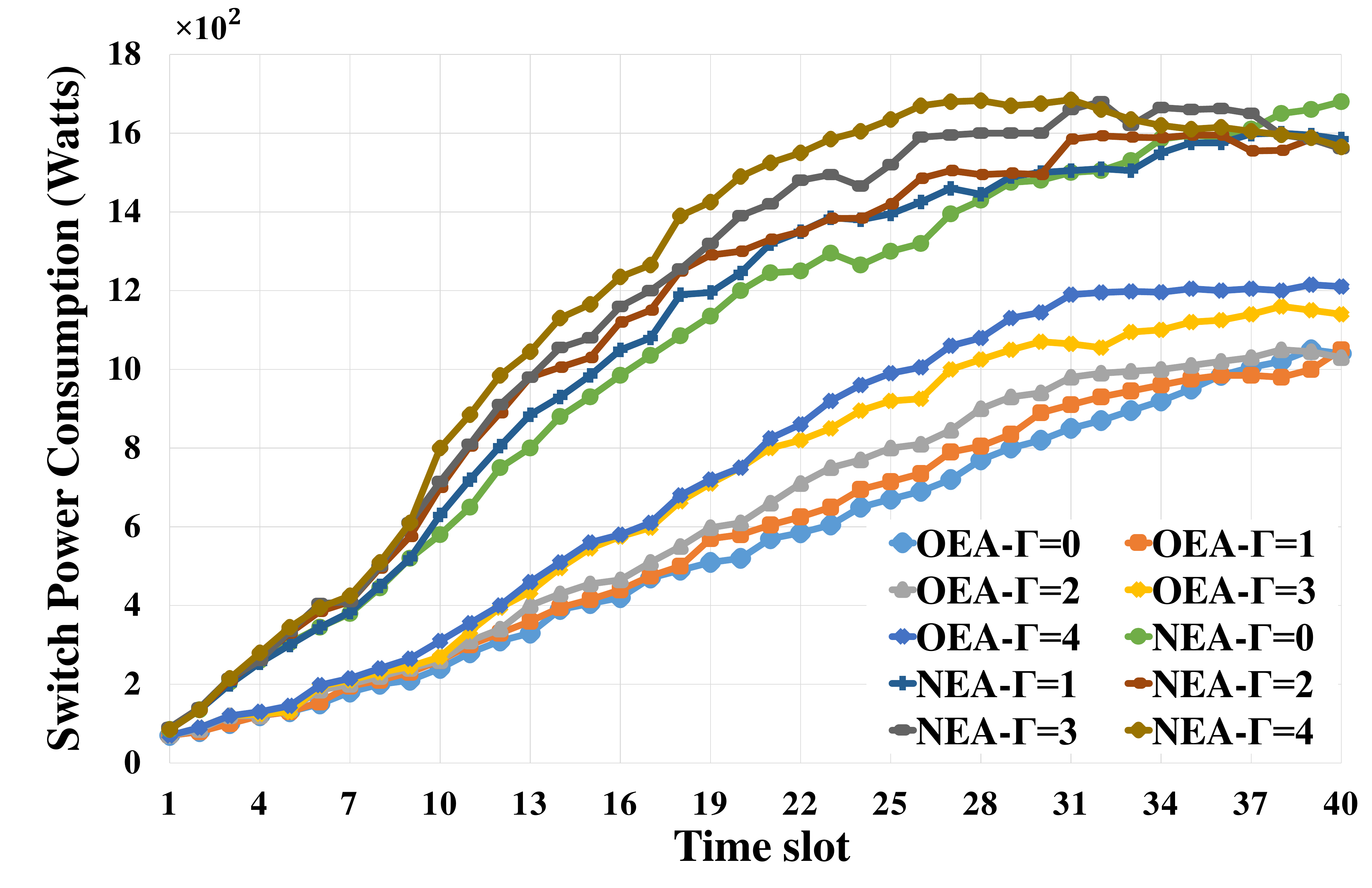}
         \caption{Switch Power Consumption in Different Protection Levels ($\Gamma_1$=$\Gamma_2$, $\Delta_1$=$\Delta_2$ and Their Values Are Equal to 10\%)}
         \label{fig:comp-SwitchPower-timeslot-Different-protection-levels}
     \end{subfigure}
     \hfill
     \begin{subfigure}[b]{0.32\textwidth}
         \centering
         \includegraphics[width=\textwidth]{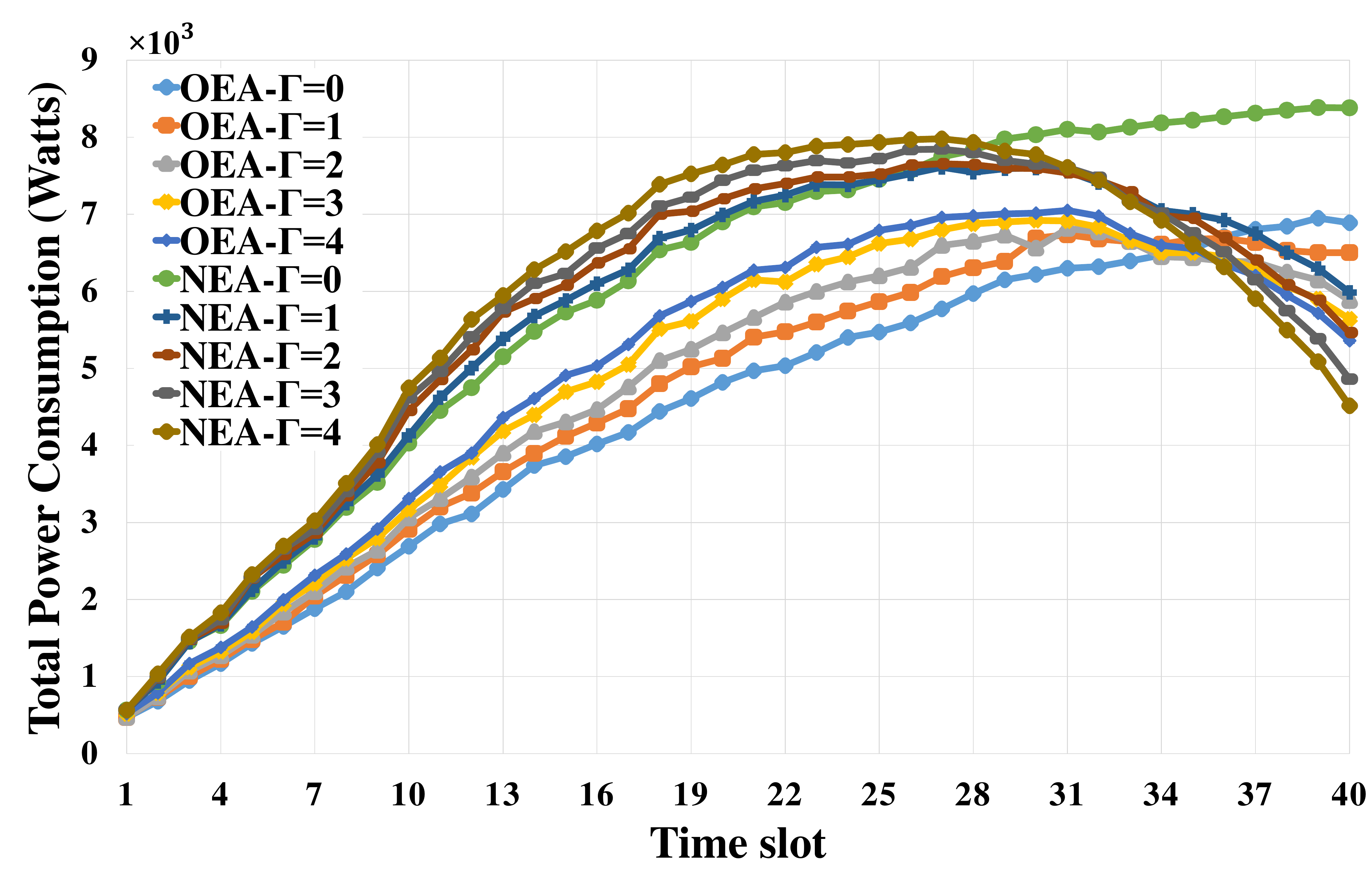}
         \caption{Total Power Consumption in Different Protection Levels ($\Gamma_1$=$\Gamma_2$, $\Delta_1$=$\Delta_2$ and Their Values Are Equal to 10\%)}
         \label{fig:comp-TotalPower-timeslot-Different-protection-levels}
     \end{subfigure}
     \hfill
     \begin{subfigure}[b]{0.32\textwidth}
         \centering
         \includegraphics[width=\textwidth]{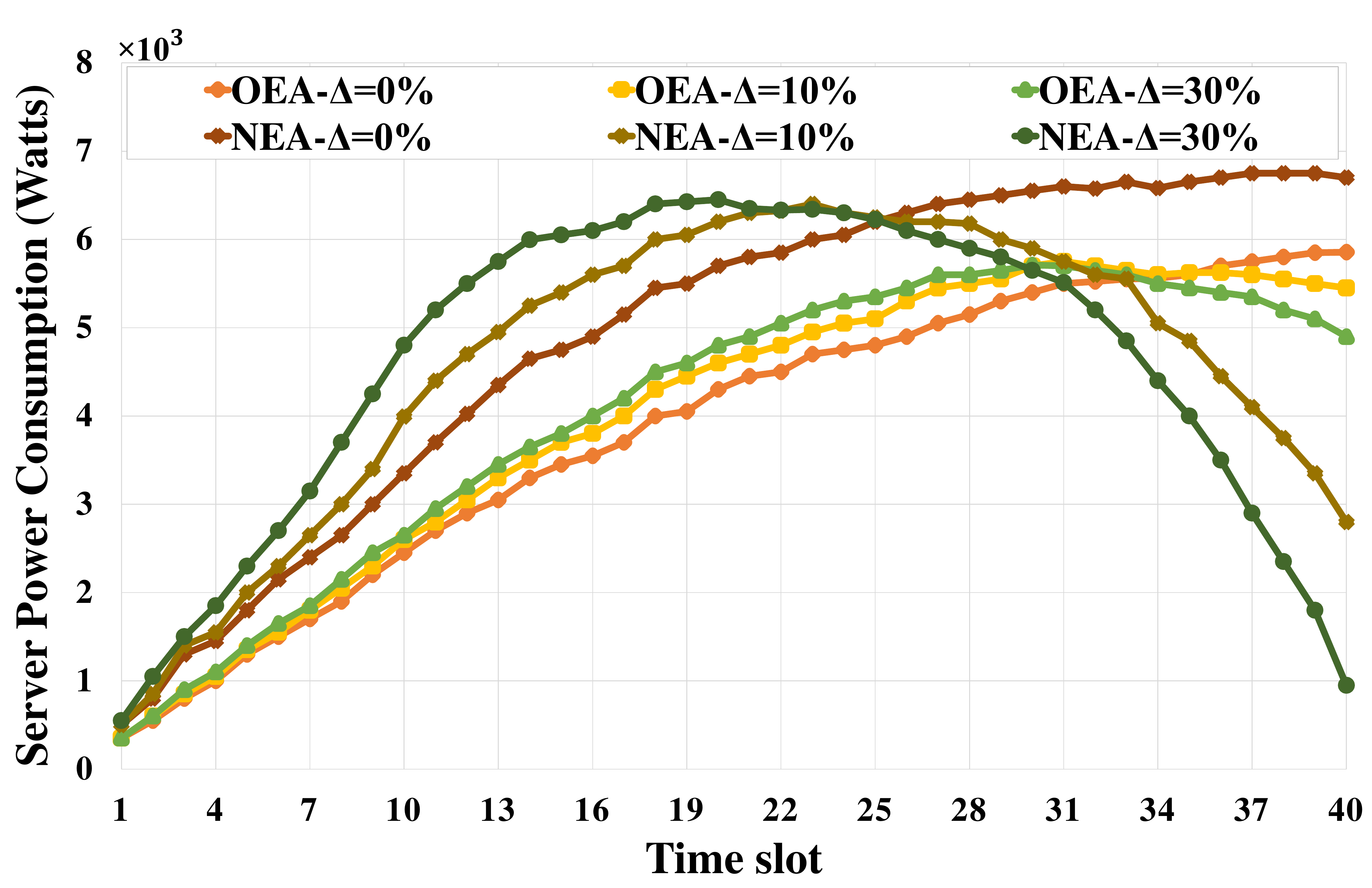}
         \caption{Server Power Consumption in Different Relative Deviations ($\Delta_1$=$\Delta_2$, $\Gamma_1$=$\Gamma_2$ and Their Values Are Equal to 1)}
         \label{fig:comp-NodePower-timeslot-Different-intervals}
     \end{subfigure}
     \hfill
     \begin{subfigure}[b]{0.32\textwidth}
         \centering
         \includegraphics[width=\textwidth]{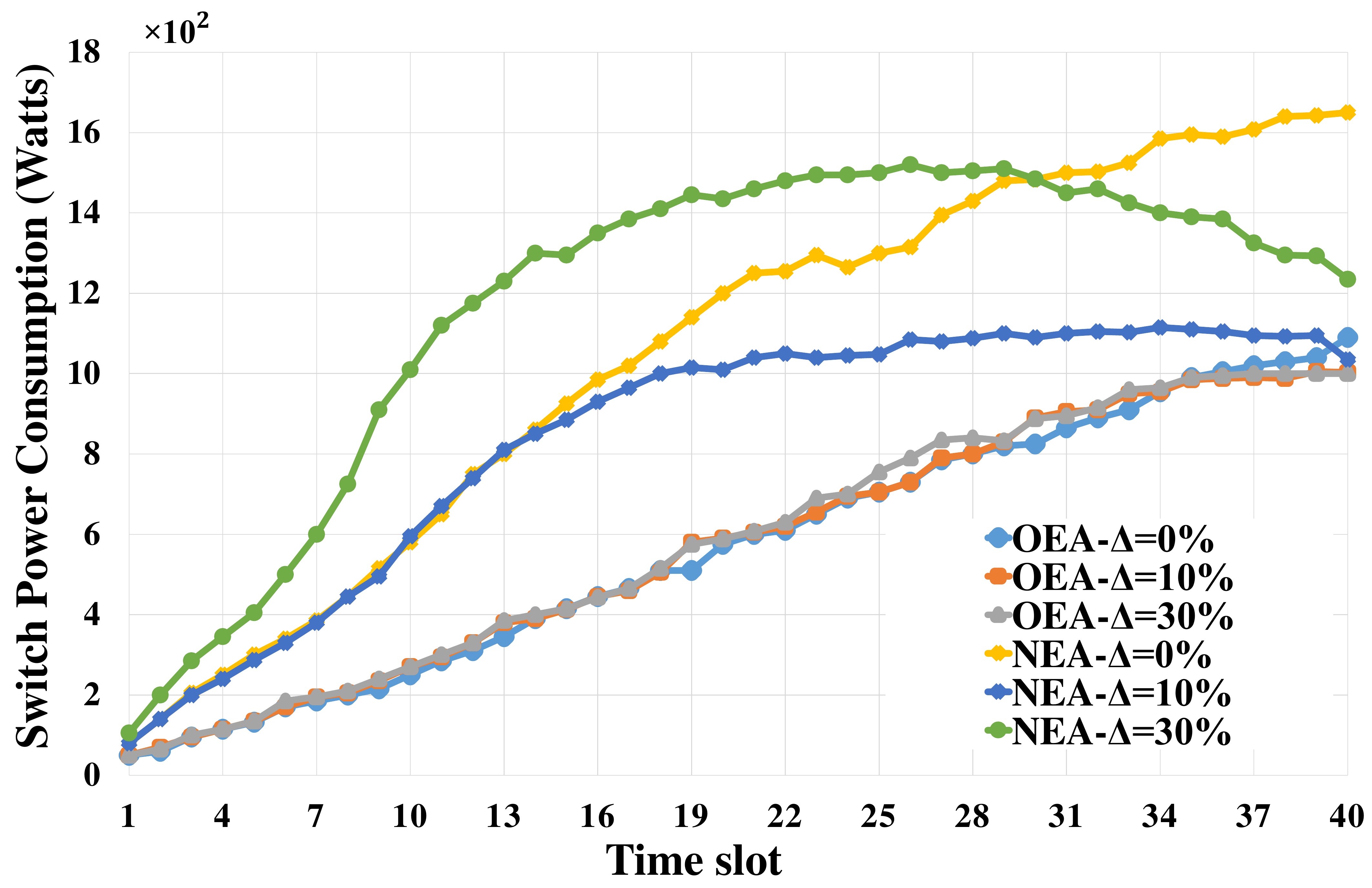}
         \caption{Switch Power Consumption in Different Relative Deviations ($\Delta_1$=$\Delta_2$, $\Gamma_1$=$\Gamma_2$ and Their Values Are Equal to 1)}
         \label{fig:comp-SwitchPower-timeslot-Different-intervals}
     \end{subfigure}
     \hfill
     \begin{subfigure}[b]{0.32\textwidth}
         \centering
         \includegraphics[width=\textwidth]{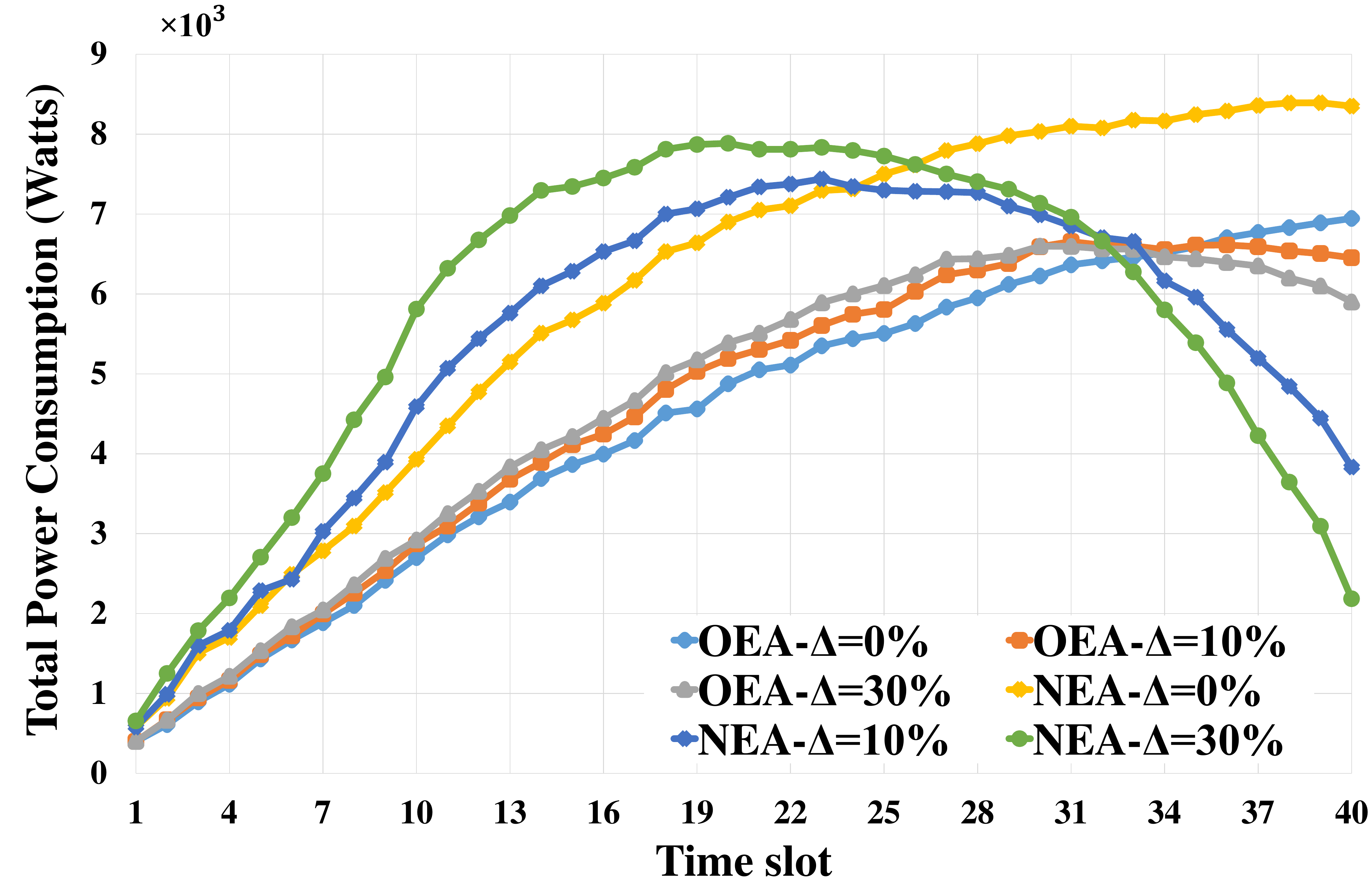}
         \caption{Total Power Consumption in Different Relative Deviations ($\Delta_1$=$\Delta_2$, $\Gamma_1$=$\Gamma_2$ and Their Values Are Equal to 1)}
         \label{fig:comp-TotalPower-timeslot-Different-intervals}
     \end{subfigure}

     %\hfill
        \caption{Power Consumption Changes in Time Slots}
        \label{fig:comp-PowerConsTimeslots}
    \end{figure*}

    \begin{figure*}
     \centering
     \begin{subfigure}[b]{0.32\textwidth}
         \centering
         \includegraphics[width=\textwidth]{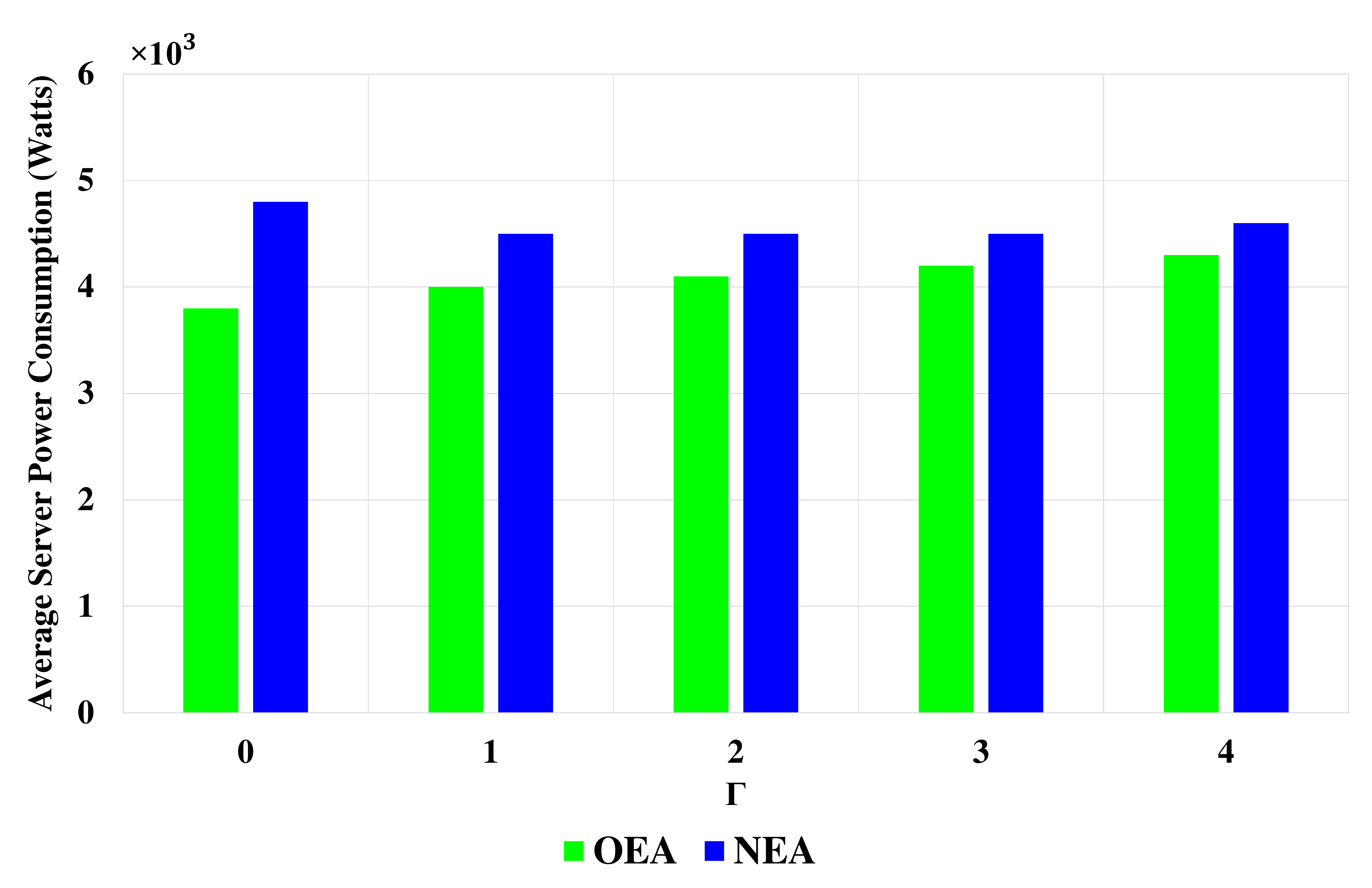}
         \caption{Server Power Consumption in Different Protection Levels ($\Gamma_1$=$\Gamma_2$, $\Delta_1$=$\Delta_2$ and Their Values Are Equal to 10\%)}
         \label{fig:comp-NodePower-Different-protection-levels}
     \end{subfigure}
     \hfill
     \begin{subfigure}[b]{0.32\textwidth}
         \centering
         \includegraphics[width=\textwidth]{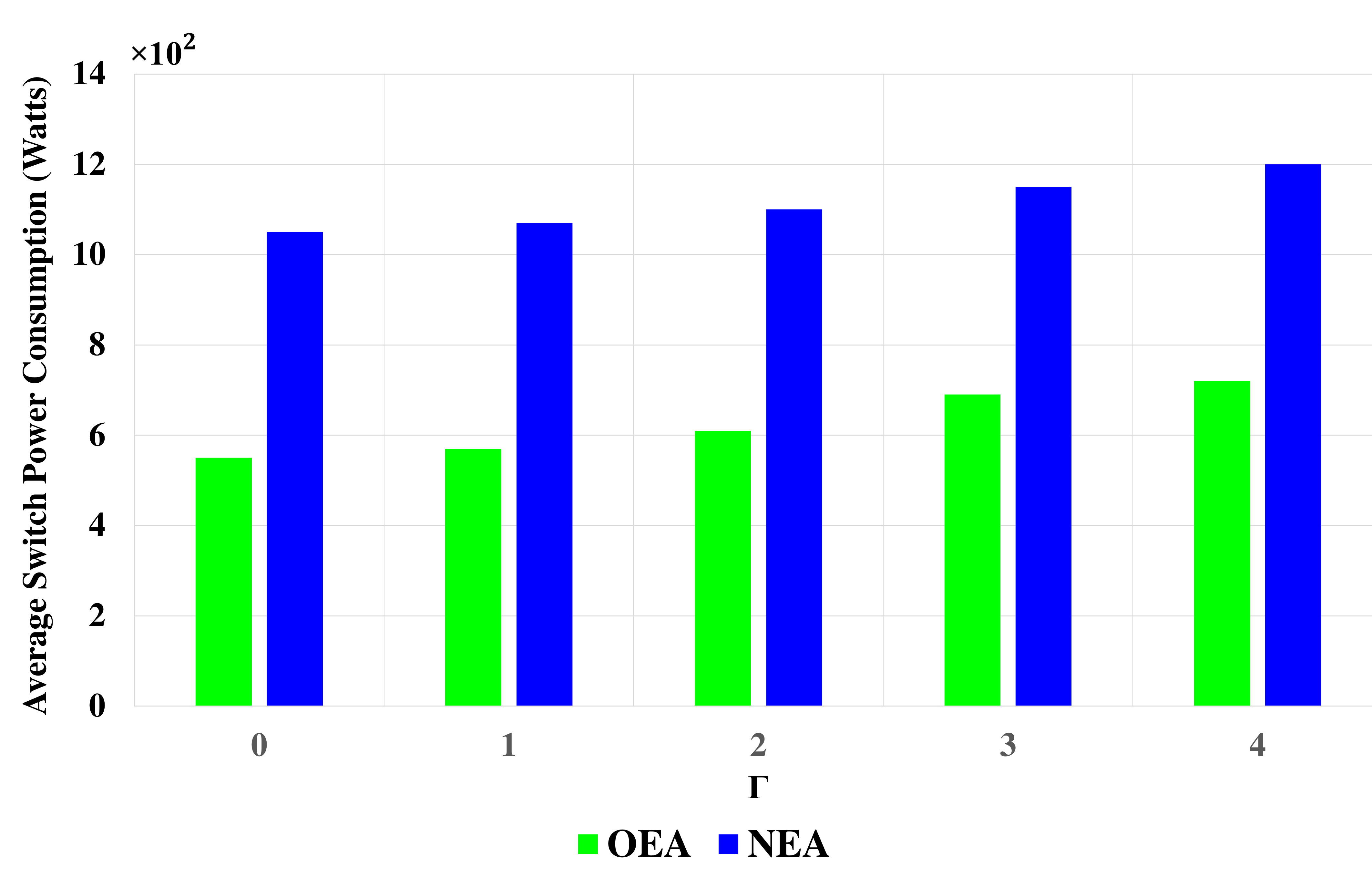}
         \caption{Switch Power Consumption in Different Protection Levels ($\Gamma_1$=$\Gamma_2$, $\Delta_1$=$\Delta_2$ and Their Values Are Equal to 10\%)}
         \label{fig:comp-SwitchPower-Different-protection-levels}
     \end{subfigure}
     \hfill
     \begin{subfigure}[b]{0.32\textwidth}
         \centering
         \includegraphics[width=\textwidth]{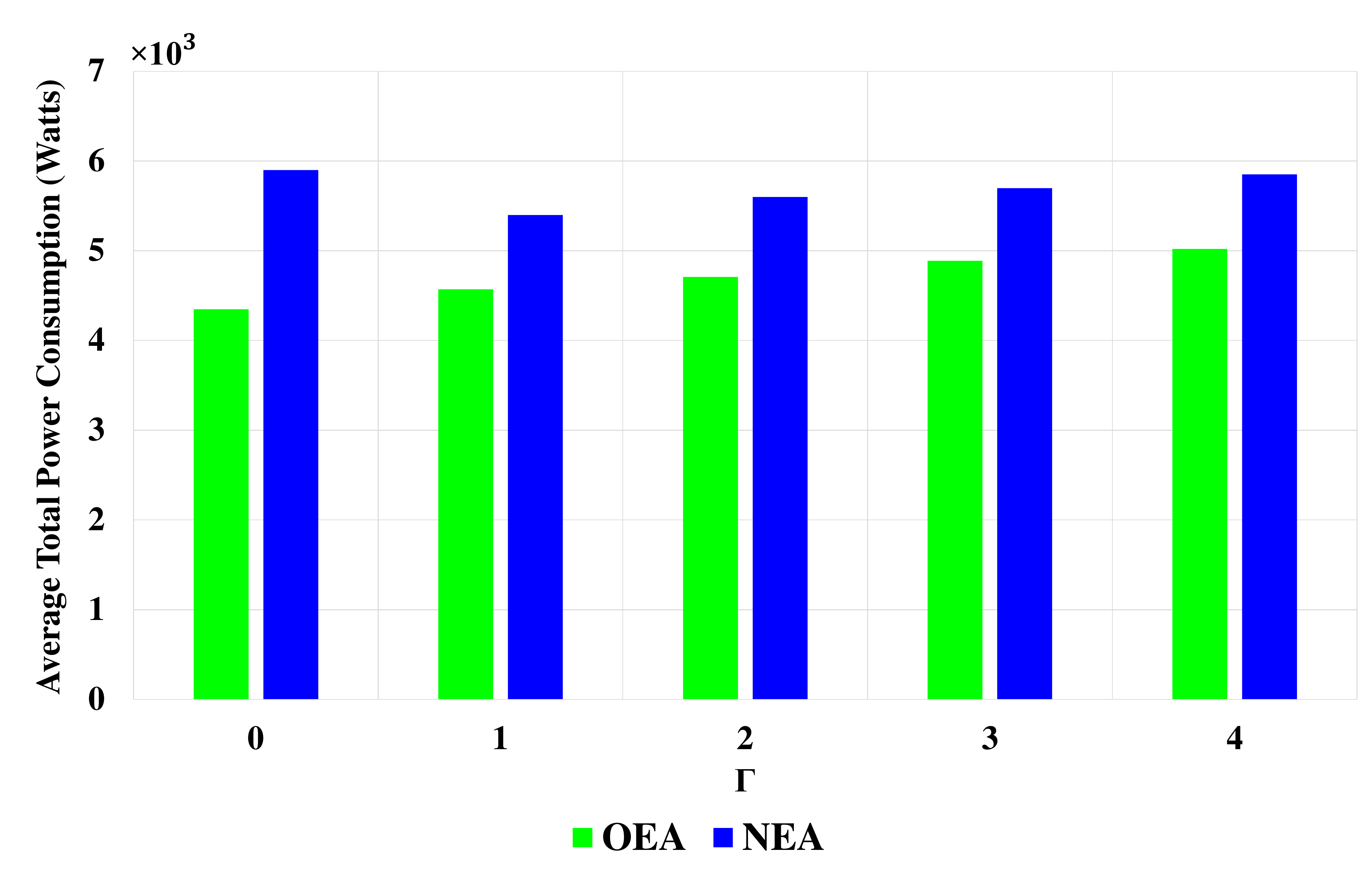}
         \caption{Total Power Consumption in Different Protection Levels ($\Gamma_1$=$\Gamma_2$, $\Delta_1$=$\Delta_2$ and Their Values Are Equal to 10\%)}
         \label{fig:comp-TotalPower-Different-protection-levels}
     \end{subfigure}
     \hfill
     \begin{subfigure}[b]{0.32\textwidth}
         \centering
         \includegraphics[width=\textwidth]{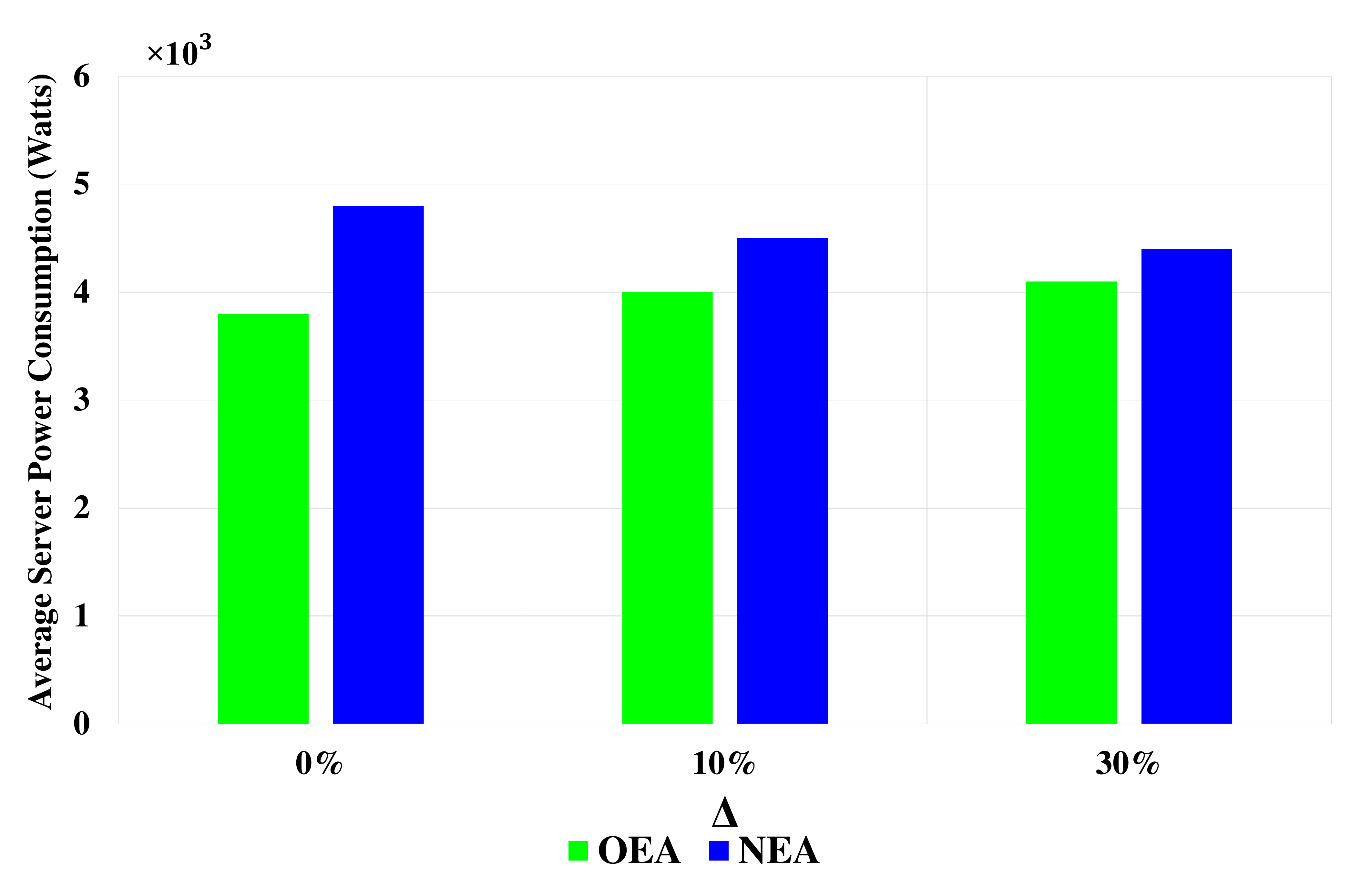}
         \caption{Server Power Consumption in Different Relative Deviations ($\Delta_1$=$\Delta_2$, $\Gamma_1$=$\Gamma_2$ and Their Values Are Equal to 1)}
         \label{fig:comp-NodePower-Different-intervals}
     \end{subfigure}
     \hfill
     \begin{subfigure}[b]{0.32\textwidth}
         \centering
         \includegraphics[width=\textwidth]{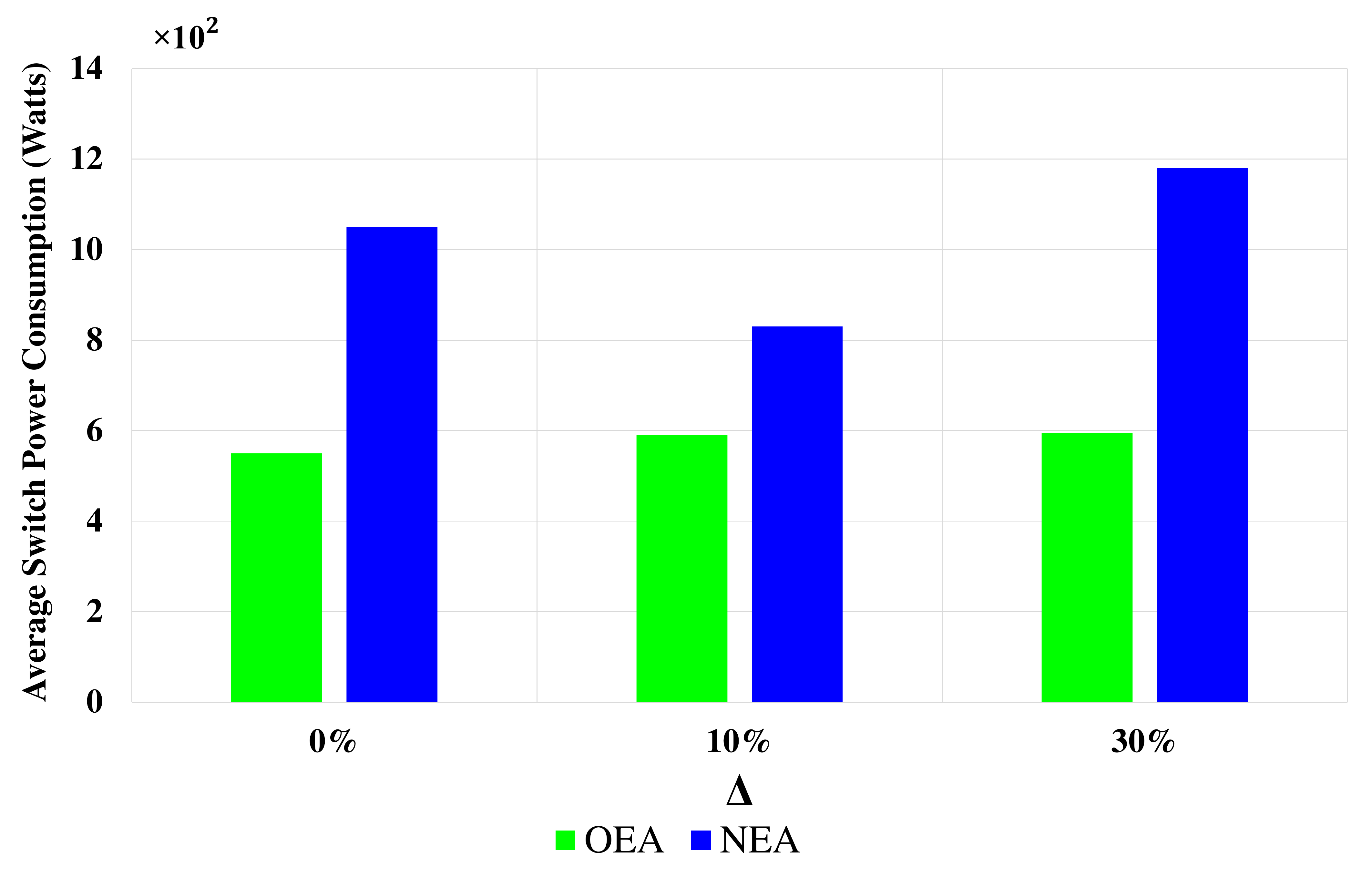}
         \caption{Switch Power Consumption in Different Relative Deviations ($\Delta_1$=$\Delta_2$, $\Gamma_1$=$\Gamma_2$ and Their Values Are Equal to 1)}
         \label{fig:comp-SwitchPower-Different-intervals}
     \end{subfigure}
     \hfill
     \begin{subfigure}[b]{0.32\textwidth}
         \centering
         \includegraphics[width=\textwidth]{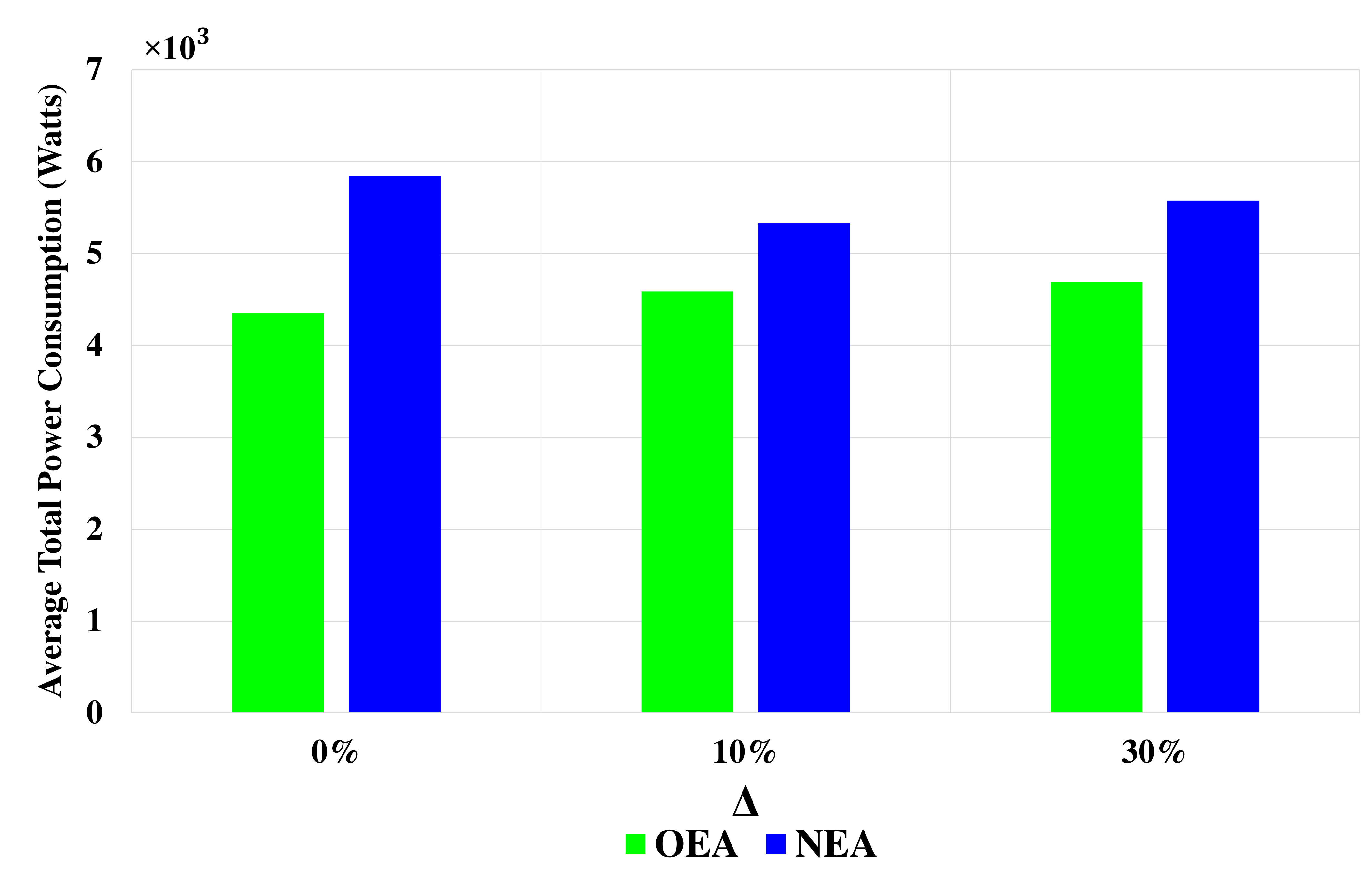}
         \caption{Total Power Consumption in Different Relative Deviations ($\Delta_1$=$\Delta_2$, $\Gamma_1$=$\Gamma_2$ and Their Values Are Equal to 1)}
         \label{fig:comp-TotalPower-Different-intervals}
     \end{subfigure}

        \caption{Average Power Consumption Over Time Slots}
        \label{fig:comp-AvgPowerCons}
    \end{figure*}
\begin{enumerate}
 \begin{figure*}
     \centering
     \begin{subfigure}[b]{0.49\columnwidth}
         \centering
         \includegraphics[width=\columnwidth]{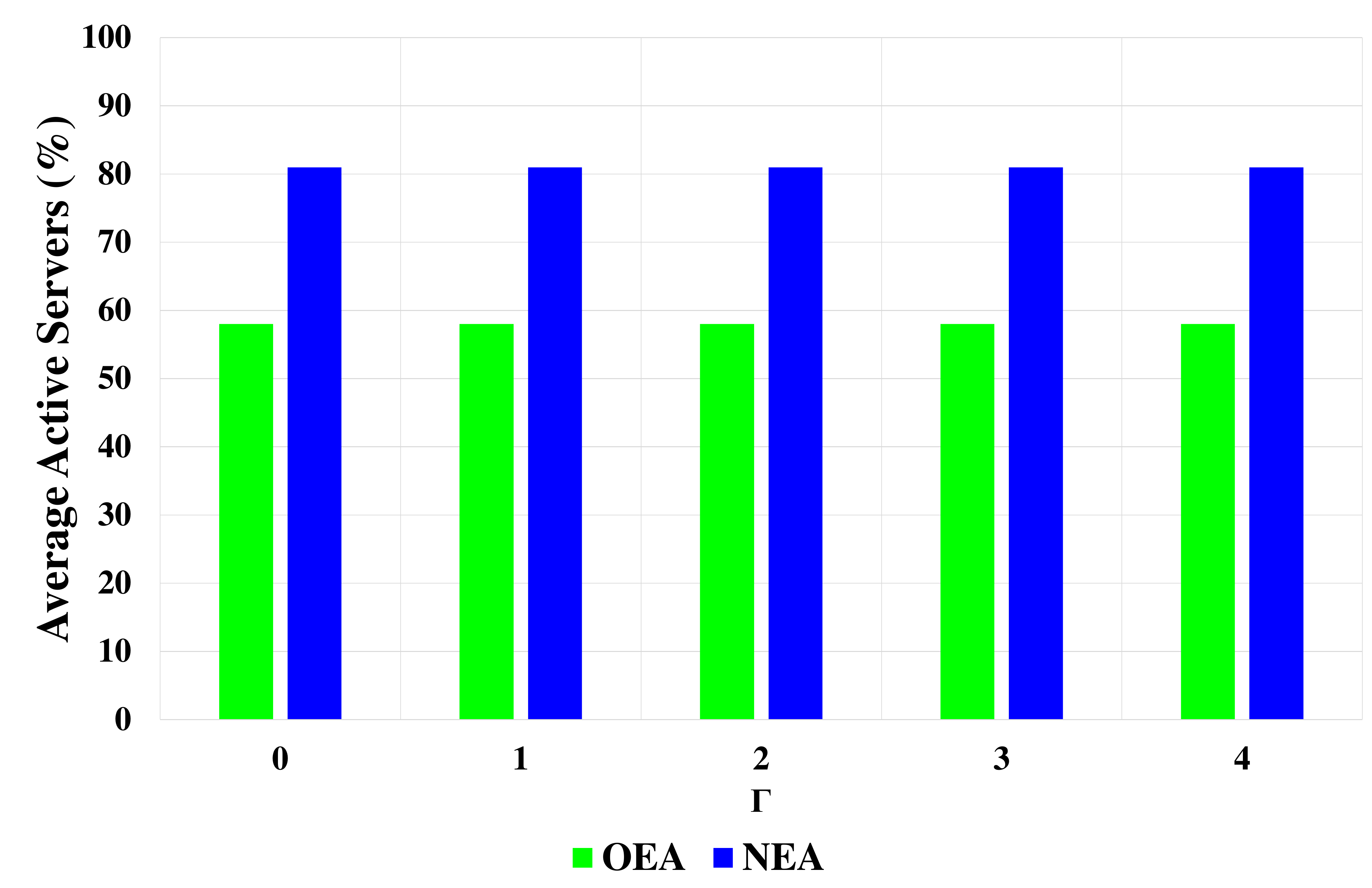}
         \caption{Active Servers in Different Protection Levels ($\Gamma_1$=$\Gamma_2$, $\Delta_1$=$\Delta_2$ and Their Values Are Equal to 10\%)}
         \label{fig:comp-ActiveServers-Different-protection-levels}
     \end{subfigure}
     \hfill
     \begin{subfigure}[b]{0.49\columnwidth}
         \centering
         \includegraphics[width=\columnwidth]{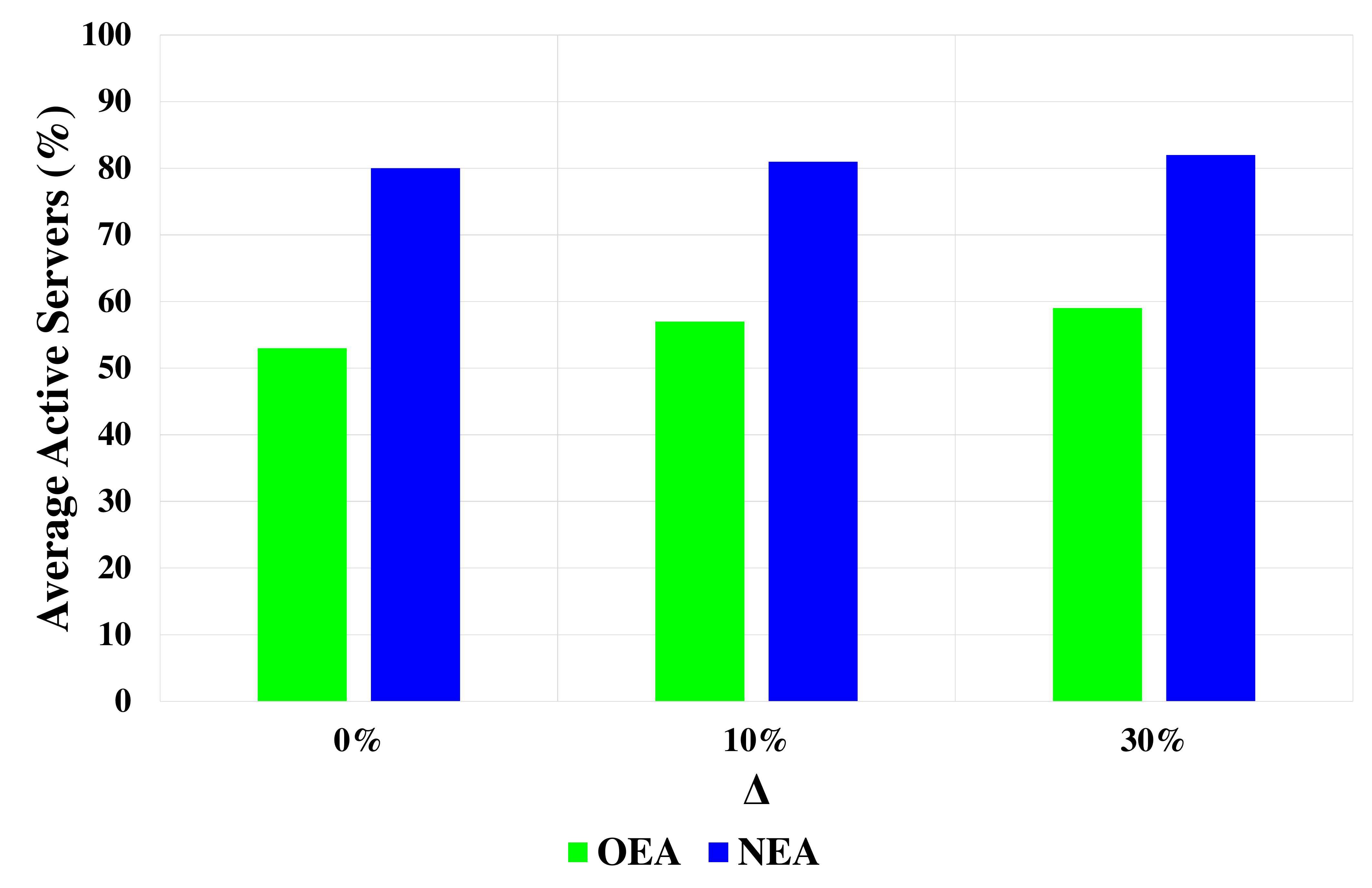}
         \caption{Active Servers in Different Relative Deviations ($\Delta_1$=$\Delta_2$, $\Gamma_1$=$\Gamma_2$ and Their Values Are Equal to 1)}
         \label{fig:comp-ActiveServers-Different-relative-deviations}
     \end{subfigure}
     \hfill
     \begin{subfigure}[b]{0.49\columnwidth}
         \centering
         \includegraphics[width=\columnwidth]{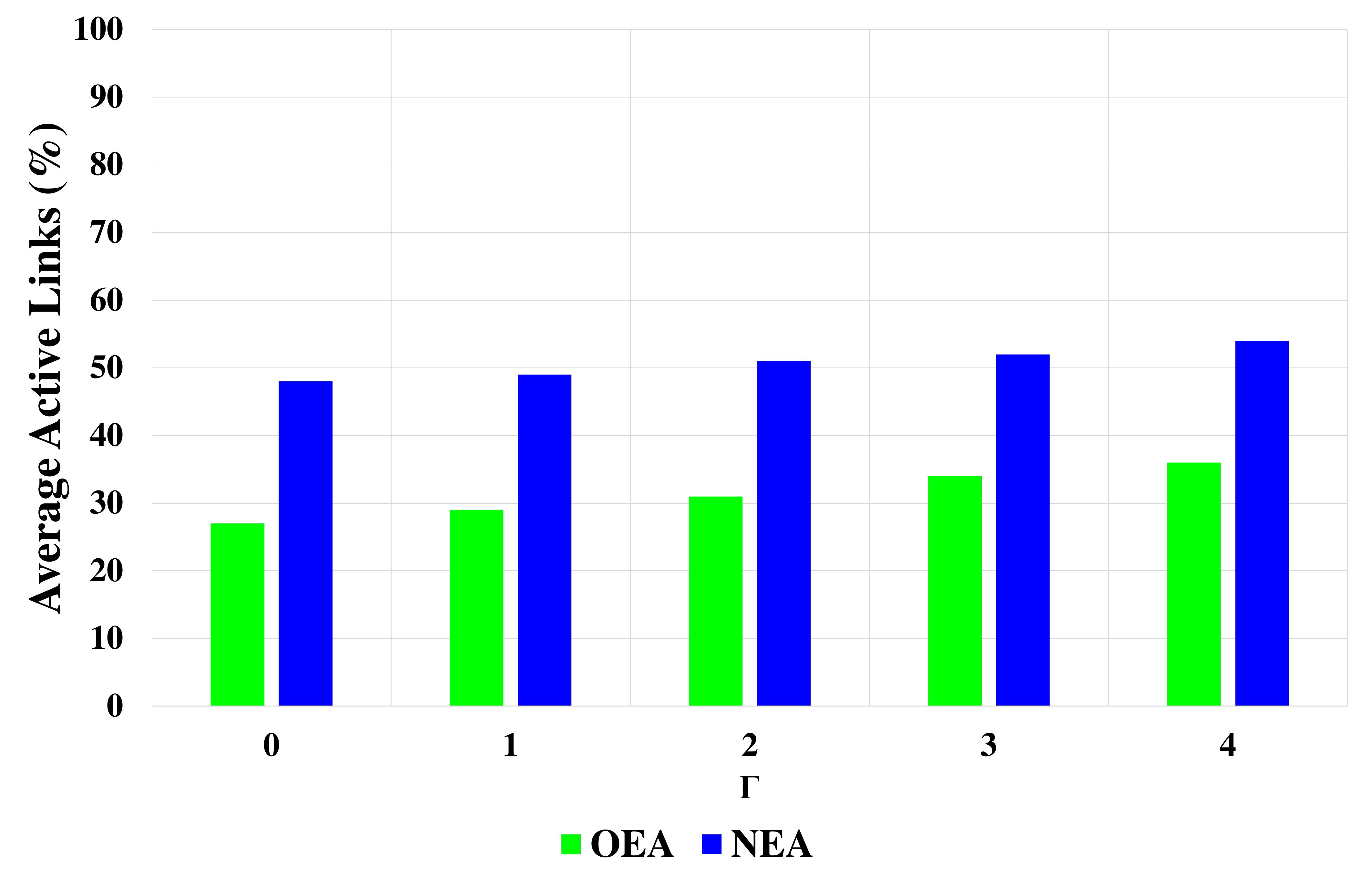}
         \caption{Active Links in Different Protection Levels ($\Gamma_1$=$\Gamma_2$, $\Delta_1$=$\Delta_2$ and Their Values Are Equal to 10\%)}
         \label{fig:comp-ActiveLinks-Different-protection-levels}
     \end{subfigure}
     \hfill
     \begin{subfigure}[b]{0.49\columnwidth}
         \centering
         \includegraphics[width=\columnwidth]{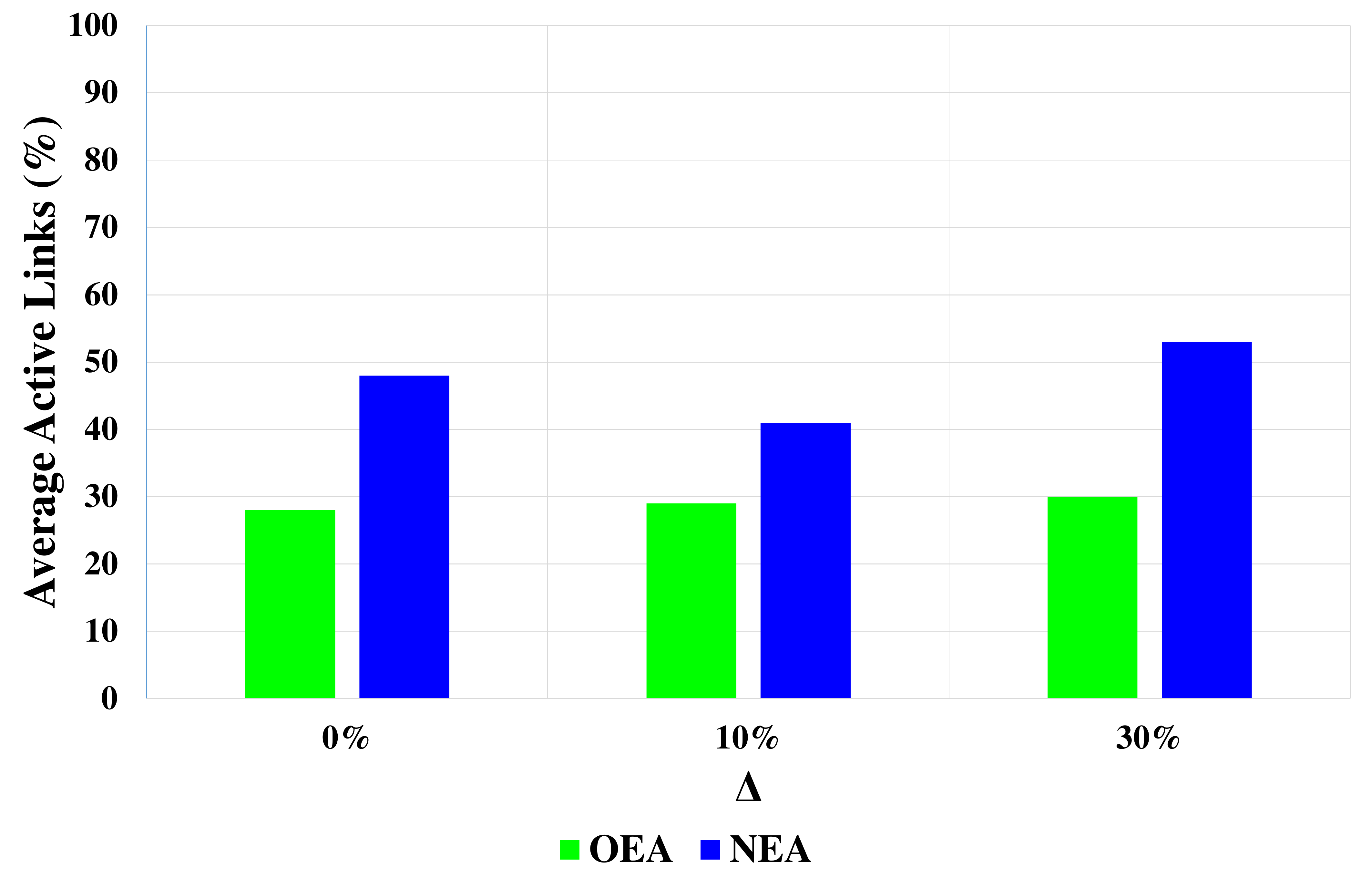}
         \caption{Active Links in Different Relative Deviations ($\Delta_1$=$\Delta_2$, $\Gamma_1$=$\Gamma_2$ and Their Values Are Equal to 1)}
         \label{fig:comp-ActiveLinks-Different-relative-deviations}
     \end{subfigure}
     \hfill
        \caption{Average Active Servers and Links}
        \label{fig:comp-ActiveServersLinks}
    \end{figure*}
	\item \textbf{Acceptance ratio:}
	The first metric is the accepted slices’ count to the total arrived slices’ count ratio, expressed through Eq. (\ref{Acc_ratio_calculation}): 
	\begin{align}
	\label{Acc_ratio_calculation}
    &{Acceptance\_ratio} = {\frac{\#Accepted\_slices}{\#Arrived\_slices}}\times100.
    \end{align}
    The ratios of OEA-ONSU and NEA-ONSU are compared in two stated scenarios, Fig. (\ref{fig:comp-AccRatio}). The average of arrived slices in 20 simulations is 100. As observed in scenario 1, Fig. (\ref{fig:comp-AccRatio-Different-protection-levels})), an increase in $\Gamma_1$ and $\Gamma_2$, decreases the OEA-ONSU’s gained acceptance ratio, because the capacity of resources are limited and this algorithm cannot accept all slices with their requirements, while, in scenario 2 in Fig. (\ref{fig:comp-AccRatio-Different-relative-deviations}), the OEA-ONSU acceptance ratio is almost fixed and usually, higher than that of the scenario 1. As to NEA-ONSU, in scenario 1, compared to 2, its acceptance ratios are closer to the optimal state, and in scenario 2, acceptance ratios decrease by about 10\%. Unlike OEA-ONSU’s behavior in scenario 1, at $\Gamma$=0, the NEA-ONSU’s acceptance ratio is less than the acceptance ratio at $\Gamma$=1 and 2, and this is due to the greedy behavior of NEA-ONSU in choosing nodes to place VMs. On average, in terms of acceptance ratio, the OEA-ONSU has about 2\% gap with baseline and the NEA-ONSU has about 7\% optimality gap with OEA-ONSU. 
    \item \textbf{Power consumption:}
    One of other metrics applied in assessing OEA-ONSU and NEA-ONSU is the power consumption metric, consisting of the servers’ power consumption, switches’ power consumption, and total power consumption, the sum of the two prior powers. The changes in this metric during time slots, are in two batches, described as follows: 1) Figs. (\ref{fig:comp-NodePower-timeslot-Different-protection-levels}, \ref{fig:comp-SwitchPower-timeslot-Different-protection-levels}, and \ref{fig:comp-TotalPower-timeslot-Different-protection-levels}), scenario 1, 2) Figs. (\ref{fig:comp-NodePower-timeslot-Different-intervals}, \ref{fig:comp-SwitchPower-timeslot-Different-intervals}, and \ref{fig:comp-TotalPower-timeslot-Different-intervals}), scenario 2.
    The average total power of servers and switches, when they are turned-on under maximum load,  is 7450 and 2500 Watts, respectively. The average power consumption over time slots includes two batches of figures: 1) Figs. (\ref{fig:comp-NodePower-Different-protection-levels}, \ref{fig:comp-SwitchPower-Different-protection-levels}, and \ref{fig:comp-TotalPower-Different-protection-levels}), scenario 1, 2) Figs. (\ref{fig:comp-NodePower-Different-intervals}, \ref{fig:comp-SwitchPower-Different-intervals}, and \ref{fig:comp-TotalPower-Different-intervals}), scenario 2. As observed in Figs. (\ref{fig:comp-NodePower-timeslot-Different-protection-levels} and \ref{fig:comp-NodePower-timeslot-Different-intervals}), and considering the comparison of OEA-ONSU and NEA-ONSU in terms of acceptance ratio, the power consumption of the servers in scenario 1 with an increase $\Gamma$, from time slot 23 onwards, and in scenario 2 with an increase $\Delta$, from time slot 19 onwards, as approaches The average volume of total power of servers under maximum load, it begins to decrease due to the inability to accept the new slice requests and the expired slices exit. The server power consumption reduction in scenario 2 is higher than that of the scenario 1, which is due to a higher reduction in acceptance ratio in scenario 2. These conditions in switch and total power consumption, respectively, are evident in the figures. Because the effect of servers’ power consumption in total power consumption is more than the switches’ power consumption, the total power consumption changes are more affected by servers’ power consumption. In Fig. (\ref{fig:comp-AvgPowerCons}), where as observed, in general, by increasing $\Gamma$ and $\Delta$, the average power consumption increase. As observed in Fig. (\ref{fig:comp-NodePower-Different-protection-levels}), although, acceptance ratio of $\Gamma$=0,  Fig. (\ref{fig:comp-AccRatio-Different-protection-levels}), is less than $\Gamma$=1 in NEA-ONSU it is expected that power consumption to be low, because the slice requests need more capacity for robustness and it is possible that the order of choosing servers in NEA-ONSU for $\Gamma$=1 vary from $\Gamma$=0, the server power consumption is decreased, and because the applied links’ count increase, the switch power consumption in $\Gamma$=1 becomes more than $\Gamma$=0. On average, as to power consumption, the OEA-ONSU has about 4\% gap with baseline and the NEA-ONSU has about 10\% optimality gap with OEA-ONSU.

 	\item \textbf{Active servers and links:}
	The third evaluation metric includes the active servers' and links' count, separately. As observed in Fig. (\ref{fig:comp-ActiveServersLinks}), by increasing $\Gamma$, the active servers’ count, Fig. (\ref{fig:comp-ActiveServers-Different-protection-levels}), remains almost fixed, while the active links’ count, Fig. (\ref{fig:comp-ActiveLinks-Different-protection-levels}), increases. In scenario 1, increasing the servers’ power consumption, Fig. (\ref{fig:comp-NodePower-Different-protection-levels}), is due to increasing the applied resources’ volume instead of increasing applied servers’ counts. In Figs. (\ref{fig:comp-ActiveServers-Different-relative-deviations} and \ref{fig:comp-ActiveLinks-Different-relative-deviations}), by increasing $\Delta$, active servers’ count in OEA-ONSU increases more compared to the active links’ count, and as to NEA-ONSU, the active servers’ count increase. In Fig. (\ref{fig:comp-ActiveLinks-Different-relative-deviations}), for NEA-ONSU, the active links’ count in $\Delta$=10\% is less than $\Delta$=0\%, duo to low acceptance ratio in $\Delta$=10\%. At $\Delta$=30\%, although, the acceptance ratio is less than $\Delta$=0\%, the active links’ count is high, that is, the slices’ requests require more resources, that is, the active links’ count affects switch power consumption, Fig. (\ref{fig:comp-SwitchPower-Different-intervals}). In general, the percentage of active servers and links in scenario 1 are more than that of the scenario 2. In this metric, computing the gap of OEA-ONSU with baseline, and optimality gap of NEA-ONSU are not essential because, according to the objective function, here, the gaps are essential only in the two previous metrics.
	\item \textbf{Execution time:}
	Because finding the optimal solution for the ROBINS BLP problem may be unreachable in reasonable amount of time for real-world and large-scale networks, and, the near-optimal solution in a short time is sought, consequently, the algorithm NEA-ONSU is devised as an alternative to OEA-ONSU. In this context, execution time is a proper metric for comparing these two algorithms. The execution time of these algorithms is subject to the execution time of the three steps therein. The average execution time of these algorithms in each time slot is shown in Fig. (\ref{fig:comp-Executiontime}), where, as observed, NEA-ONSU is about 30X faster than OEA-ONSU.
	\begin{figure}
     \centering
        \includegraphics[width=0.8\columnwidth]{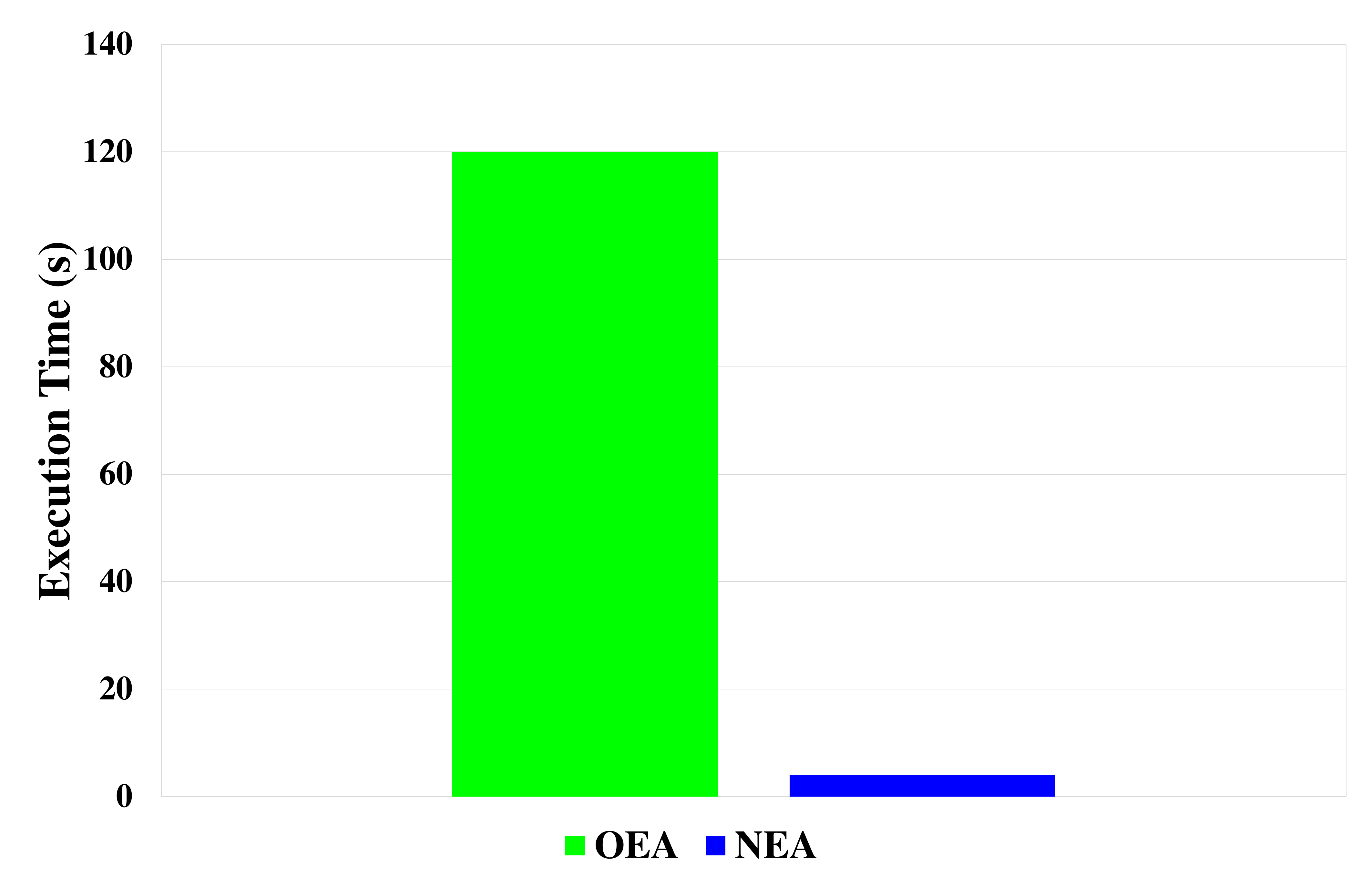}
        \caption{Average Execution Time}
        \label{fig:comp-Executiontime}
    \end{figure}
    %\item \textbf{Robustness price:}
    %To be robust, more resources of servers and links are required and there exist fluctuative relation between robustness and 1) power consumption, 2) acceptance ratio, all named robustness price. By applying this metric, the effect of increasing robustness on the acceptance ratio and power consumption is assessed, in simultaneous manner. As observed in Fig. (\ref{fig:comp-Robustness price}), in general, an increase in robustness may decrease the acceptance ratio (e.g. NEA-ONSU results in Fig. (\ref{fig:comp-Robustness price-RD10}) and Fig. (\ref{fig:comp-Robustness price-RD30}) compared to Fig. (\ref{fig:comp-Robustness price-RD0})) and may increase power consumption (e.g. OEA-ONSU results in Figs. (\ref{fig:comp-Robustness price-gamma1}-\ref{fig:comp-Robustness price-gamma4}) compared to Fig. (\ref{fig:comp-Robustness price-gamma0})). In some situations, because of a decrease in acceptance ratio, the power consumption decrease (e.g. NEA-ONSU result in Fig. (\ref{fig:comp-Robustness price-RD10}) compared to Fig. (\ref{fig:comp-Robustness price-RD0})).
\end{enumerate}
Generally, the exact gain of optimal algorithm compared to the stated baseline is providing reliable slices with a little gap with the baseline.
\section{Conclusion}
\label{Conclusion}
We proposed two online admission control and resource allocation algorithms for network slicing under bandwidth and workload uncertainties. These include robustness vs. high fluctuations on both requested VLs' bandwidths and VMs' resources. Besides, because the slices' arrive to the network in different times and their lifespans vary, the solution dynamically react to the online slice requests. The joint problem of online admission control and resource allocation considering the energy consumption is formulated mathematically. The formulation is a BLP, where the $\Gamma$-Robustness concept is exploited to overcome VLs bandwidths' and VNFs workloads' uncertainties. Then, an optimal algorithm, named OEA-ONSU, that adopts this mathematical model is proposed. To find near-optimal solution in reasonable amount of time, a new heuristic algorithm, named NEA-ONSU, is proposed. The assessments’ results indicate that the efficiency of NEA-ONSU is vital in increasing the accepted requests’ count, decreasing power consumption and providing adjustable tolerance vs. the VNFs workloads’ and VLs traffics’ uncertainties, separately. Considering the acceptance ratio and power consumption that constitute the two important components of the objective function, NEA-ONSU has about 7\% and 10\% optimality gaps, respectively, while being about 30X faster than that of OEA-ONSU. A prospective extension to this work is to let the IP allocates more resources to the slices dynamically, if they need more resources than the provisioned resources for robustness. In other words, we need to add resource reallocation process to this work. This allows trade-off between proactive and reactive approaches.

\ifCLASSOPTIONcaptionsoff
  \newpage
\fi

\bibliographystyle{IEEEtran}
\small{\bibliography{references.bib}}

\begin{thebibliography}{10}
\providecommand{\url}[1]{#1}
\csname url@samestyle\endcsname
\providecommand{\newblock}{\relax}
\providecommand{\bibinfo}[2]{#2}
\providecommand{\BIBentrySTDinterwordspacing}{\spaceskip=0pt\relax}
\providecommand{\BIBentryALTinterwordstretchfactor}{4}
\providecommand{\BIBentryALTinterwordspacing}{\spaceskip=\fontdimen2\font plus
\BIBentryALTinterwordstretchfactor\fontdimen3\font minus
  \fontdimen4\font\relax}
\providecommand{\BIBforeignlanguage}[2]{{%
\expandafter\ifx\csname l@#1\endcsname\relax
\typeout{** WARNING: IEEEtran.bst: No hyphenation pattern has been}%
\typeout{** loaded for the language `#1'. Using the pattern for}%
\typeout{** the default language instead.}%
\else
\language=\csname l@#1\endcsname
\fi
#2}}
\providecommand{\BIBdecl}{\relax}
\BIBdecl

\bibitem{MSU-CSE-06-2}
N.~Alliance, ``\uppercase{NGMN} 5\uppercase{G} white paper,'' Tech. Rep.,
  February 2015.

\bibitem{alliance2016description}
------, ``Description of network slicing concept,'' \emph{NGMN 5G P}, vol.~1,
  no.~1, 2016.

\bibitem{MSU-CSE-06-1}
5G-PPP, ``View on 5\uppercase{G} architecture,'' Tech. Rep., July 2016.

\bibitem{ebrahimi2019joint}
S.~Ebrahimi, A.~Zakeri, B.~Akbari, and N.~Mokari, ``Joint resource and
  admission management for slice-enabled networks,'' \emph{arXiv}, pp.
  arXiv--1912, 2019.

\bibitem{fendt2018network}
A.~Fendt, S.~Lohmuller, L.~C. Schmelz, and B.~Bauer, ``A network slice resource
  allocation and optimization model for end-to-end mobile networks,'' in
  \emph{IEEE 5G World Forum (5GWF)}, California, USA, 2018.

\bibitem{farkiani2019fast}
B.~Farkiani, B.~Bakhshi, and S.~A. Mirhassani, ``A fast near-optimal approach
  for energy-aware sfc deployment,'' \emph{IEEE Transactions on Network and
  Service Management}, vol.~16, no.~4, pp. 1360--1373, 2019.

\bibitem{chen2020network}
W.-K. Chen, Y.-F. Liu, A.~De~Domenico, and Z.-Q. Luo, ``Network slicing for
  service-oriented networks with flexible routing and guaranteed e2e latency,''
  \emph{arXiv preprint arXiv:2002.07380}, 2020.

\bibitem{bhamare2017optimal}
D.~Bhamare, M.~Samaka, A.~Erbad, R.~Jain, L.~Gupta, and H.~A. Chan, ``Optimal
  virtual network function placement in multi-cloud service function chaining
  architecture,'' \emph{Computer Communications}, vol. 102, pp. 1--16, 2017.

\bibitem{halabian2019distributed}
H.~Halabian, ``Distributed resource allocation optimization in 5\uppercase{G}
  virtualized networks,'' \emph{IEEE Journal on Selected Areas in
  Communications}, vol.~37, no.~3, pp. 627--642, 2019.

\bibitem{marotta2017energy}
A.~Marotta, F.~D’Andreagiovanni, A.~Kassler, and E.~Zola, ``On the energy
  cost of robustness for green virtual network function placement in
  5\uppercase{G} virtualized infrastructures,'' \emph{Computer Networks}, vol.
  125, pp. 64--75, 2017.

\bibitem{bauschert2014network}
T.~Bauschert, C.~B{\"u}sing, F.~D'Andreagiovanni, A.~M. Koster, M.~Kutschka,
  and U.~Steglich, ``Network planning under demand uncertainty with robust
  optimization,'' \emph{IEEE Communications Magazine}, vol.~52, no.~2, pp.
  178--185, 2014.

\bibitem{orlowski2010sndlib}
S.~Orlowski, R.~Wess{\"a}ly, M.~Pi{\'o}ro, and A.~Tomaszewski, ``Sndlib
  1.0—survivable network design library,'' \emph{Networks: An International
  Journal}, vol.~55, no.~3, pp. 276--286, 2010.

\bibitem{sun2019energy}
G.~Sun, Y.~Li, H.~Yu, A.~V. Vasilakos, X.~Du, and M.~Guizani,
  ``Energy-efficient and traffic-aware service function chaining orchestration
  in multi-domain networks,'' \emph{Future Generation Computer Systems},
  vol.~91, pp. 347--360, 2019.

\bibitem{soualah2019online}
O.~Soualah, M.~Mechtri, C.~Ghribi, and D.~Zeghlache, ``Online and batch
  algorithms for \uppercase{VNF}s placement and chaining,'' \emph{Computer
  Networks}, vol. 158, pp. 98--113, 2019.

\bibitem{ghazizadeh2019joint}
A.~Ghazizadeh, B.~Akbari, and M.~M. Tajiki, ``Joint reliability-aware and cost
  efficient path allocation and \uppercase{VNF} placement using sharing
  scheme,'' \emph{arXiv preprint arXiv:1912.06742}, 2019.

\bibitem{varasteh2021holu}
A.~Varasteh, B.~Madiwalar, A.~Van~Bemten, W.~Kellerer, and C.~Mas-Machuca,
  ``Holu: Power-aware and delay-constrained vnf placement and chaining,''
  \emph{IEEE Transactions on Network and Service Management}, vol.~18, no.~2,
  pp. 1524--1539, 2021.

\bibitem{chen2021optimal}
W.-K. Chen, Y.-F. Liu, A.~De~Domenico, Z.-Q. Luo, and Y.-H. Dai, ``Optimal
  network slicing for service-oriented networks with flexible routing and
  guaranteed e2e latency,'' \emph{IEEE Transactions on Network and Service
  Management}, vol.~18, no.~4, 2021.

\bibitem{hosseini2019probabilistic}
F.~Hosseini, A.~James, and M.~Ghaderi, ``Probabilistic virtual link embedding
  under demand uncertainty,'' \emph{IEEE Transactions on Network and Service
  Management}, vol.~16, no.~4, pp. 1552--1566, 2019.

\bibitem{marotta2017fast}
A.~Marotta, E.~Zola, F.~d'Andreagiovanni, and A.~Kassler, ``A fast robust
  optimization-based heuristic for the deployment of green virtual network
  functions,'' \emph{Journal of Network and Computer Applications}, vol.~95,
  pp. 42--53, 2017.

\bibitem{nguyen2019proactive}
M.~Nguyen, M.~Dolati, and M.~Ghaderi, ``Proactive service orchestration with
  deadline,'' in \emph{IEEE Conference on Network Softwarization (NetSoft)},
  Paris, France, 2019.

\bibitem{wen2018robustness}
R.~Wen, G.~Feng, J.~Tang, T.~Q. Quek, G.~Wang, W.~Tan, and S.~Qin, ``On
  robustness of network slicing for next-generation mobile networks,''
  \emph{IEEE Transactions on Communications}, vol.~67, no.~1, pp. 430--444,
  2018.

\bibitem{reddy2016robust}
V.~S. Reddy, A.~Baumgartner, and T.~Bauschert, ``Robust embedding of
  \uppercase{VNF}/service chains with delay bounds,'' in \emph{IEEE Conference
  on Network Function Virtualization and Software Defined Networks (NFV-SDN)},
  California, USA, 2016.

\bibitem{baumgartner2017network}
A.~Baumgartner, T.~Bauschert, A.~A. Blzarour, and V.~S. Reddy, ``Network slice
  embedding under traffic uncertainties—a light robust approach,'' in
  \emph{IEEE 13th International Conference on Network and Service Management
  (CNSM)}, Tokyo, Japan, 2017.

\bibitem{wen2017robust}
R.~Wen, J.~Tang, T.~Q. Quek, G.~Feng, G.~Wang, and W.~Tan, ``Robust network
  slicing in software-defined 5\uppercase{G} networks,'' in \emph{IEEE Global
  Communications Conference}, Singapour, 2017.

\bibitem{bauschert2020fast}
T.~Bauschert and V.~S. Reddy, ``A fast, scalable meta-heuristic for network
  slicing under traffic uncertainty,'' in \emph{International Conference on the
  Applications of Evolutionary Computation (Part of EvoStar)}, Seville, Spain,
  2020.

\bibitem{nguyen2020deadline}
M.~Nguyen, M.~Dolati, and M.~Ghaderi, ``Deadline-aware sfc orchestration under
  demand uncertainty,'' \emph{IEEE Transactions on Network and Service
  Management}, vol.~17, no.~4, pp. 2275--2290, 2020.

\bibitem{luu2021uncertainty}
Q.-T. Luu, S.~Kerboeuf, and M.~Kieffer, ``Uncertainty-aware resource
  provisioning for network slicing,'' \emph{IEEE Transactions on Network and
  Service Management}, vol.~18, no.~1, pp. 79--93, 2021.

\bibitem{luu2022admission}
------, ``Admission control and resource reservation for prioritized slice
  requests with guaranted sla under uncertainties,'' \emph{IEEE Transactions on
  Network and Service Management}, 2022.

\bibitem{queseth20175g}
O.~Queseth, {\"O}.~Bulakci, P.~Spapis, P.~Bisson, P.~Marsch, P.~Arnold,
  P.~Rost, Q.~Wang, R.~Blom, S.~Salsano \emph{et~al.}, ``5\uppercase{G} ppp
  architecture working group: View on 5\uppercase{G} architecture (version 2.0,
  december 2017),'' 2017.

\bibitem{wang2019reconfiguration}
G.~Wang, G.~Feng, T.~Q. Quek, S.~Qin, R.~Wen, and W.~Tan, ``Reconfiguration in
  network slicing—optimizing the profit and performance,'' \emph{IEEE
  Transactions on Network and Service Management}, vol.~16, no.~2, 2019.

\bibitem{dayarathna2015data}
M.~Dayarathna, Y.~Wen, and R.~Fan, ``Data center energy consumption modeling: A
  survey,'' \emph{IEEE Communications Surveys \& Tutorials}, vol.~18, no.~1,
  pp. 732--794, 2015.

\bibitem{jiang2016data}
R.~Jiang and Y.~Guan, ``Data-driven chance constrained stochastic program,''
  \emph{Mathematical Programming}, vol. 158, no.~1, pp. 291--327, 2016.

\bibitem{barabasi1999emergence}
A.-L. Barab{\'a}si and R.~Albert, ``Emergence of scaling in random networks,''
  \emph{science}, vol. 286, no. 5439, pp. 509--512, 1999.

\bibitem{bari2019esso}
M.~F. Bari, S.~R. Chowdhury, and R.~Boutaba, ``Esso: An energy smart service
  function chain orchestrator,'' \emph{IEEE Transactions on Network and Service
  Management}, vol.~16, no.~4, pp. 1345--1359, 2019.

\bibitem{ecv7-nz24-21}
\BIBentryALTinterwordspacing
S.~Gholamipour, ``Online admission control and resource allocation in network
  slicing subject to demand uncertainties,'' 2021. [Online]. Available:
  \url{https://dx.doi.org/10.21227/ecv7-nz24}
\BIBentrySTDinterwordspacing

\end{thebibliography}
%\vspace{-1.0cm}
\begin{IEEEbiography}[{\includegraphics[width=0.9in,height=1.1in,clip,keepaspectratio]{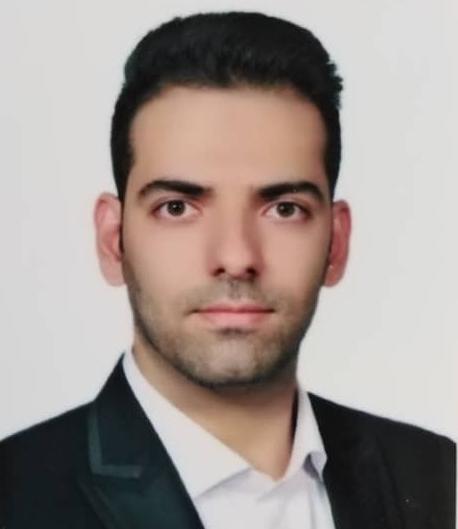}}]{Sajjad Gholamipour} obtained his M.Sc  in computer engineering from Tehran University, Tehran, Iran, in 2018, and is a PhD candidate at Tarbiat Modares University, Tehran, Iran. His main research interests are in Next Generation Network Slicing, SDN, NFV, Network Management, Cloud Computing, Resource Management in Next Generation Networks like 5G/6G, and AI.
\end{IEEEbiography}
%\vspace{-1.0cm}
\begin{IEEEbiography}[{\includegraphics[width=0.9in,height=1.1in,clip,keepaspectratio]{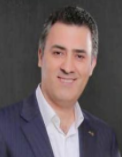}}]{Behzad Akbari} received the B.S., M.S., and Ph.D. degree in computer engineering from the Sharif University of Technology, Tehran, Iran, in 1999, 2002, and 2008 respectively. He is currently an Associate Professor of Computer Engineering in Tarbiat Modares University. His research interest includes Computer Networking, Multimedia Networking, Cloud Computing and Networking, SDN, Network Virtualization and Resource Management in 5G Networking, AI-Based Networking, Network QoS, Network Performance Modeling and Analysis, Network Security Events Analysis and Correlation.
\end{IEEEbiography}
%\vspace{-1.25cm}
\begin{IEEEbiography}[{\includegraphics[width=0.9in,height=1.1in,clip,keepaspectratio]{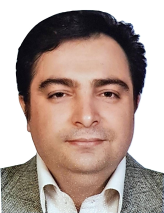}}]{Neder Mokari}(Senior Member, IEEE) obtained his PhD in electrical Engineering at Tarbiat Modares University, Tehran, Iran in 2014. His thesis received the IEEE outstanding PhD thesis award. He joined the Department of Electrical and Computer Engineering, Tarbiat Modares University as an assistant professor in October 2015. He has been elected as an IEEE exemplary reviewer in 2016 by IEEE Communications Society. At present, he is an Associated Professor at the Department of Electrical and Computer Engineering, Tarbiat Modares University, Tehran, Iran. His research interests cover the many aspects of wireless technologies with a narrow focus on wireless networks. In recent years, his research has been funded by Iranian Mobile Telecommunication Companies, Iranian National Science Foundation (INSF). He is involved in a number of large scale network design and consulting projects in the telecom industry.
\end{IEEEbiography}
%\vspace{-1.25cm}
\begin{IEEEbiography}[{\includegraphics[width=0.9in,height=1.1in,clip,keepaspectratio]{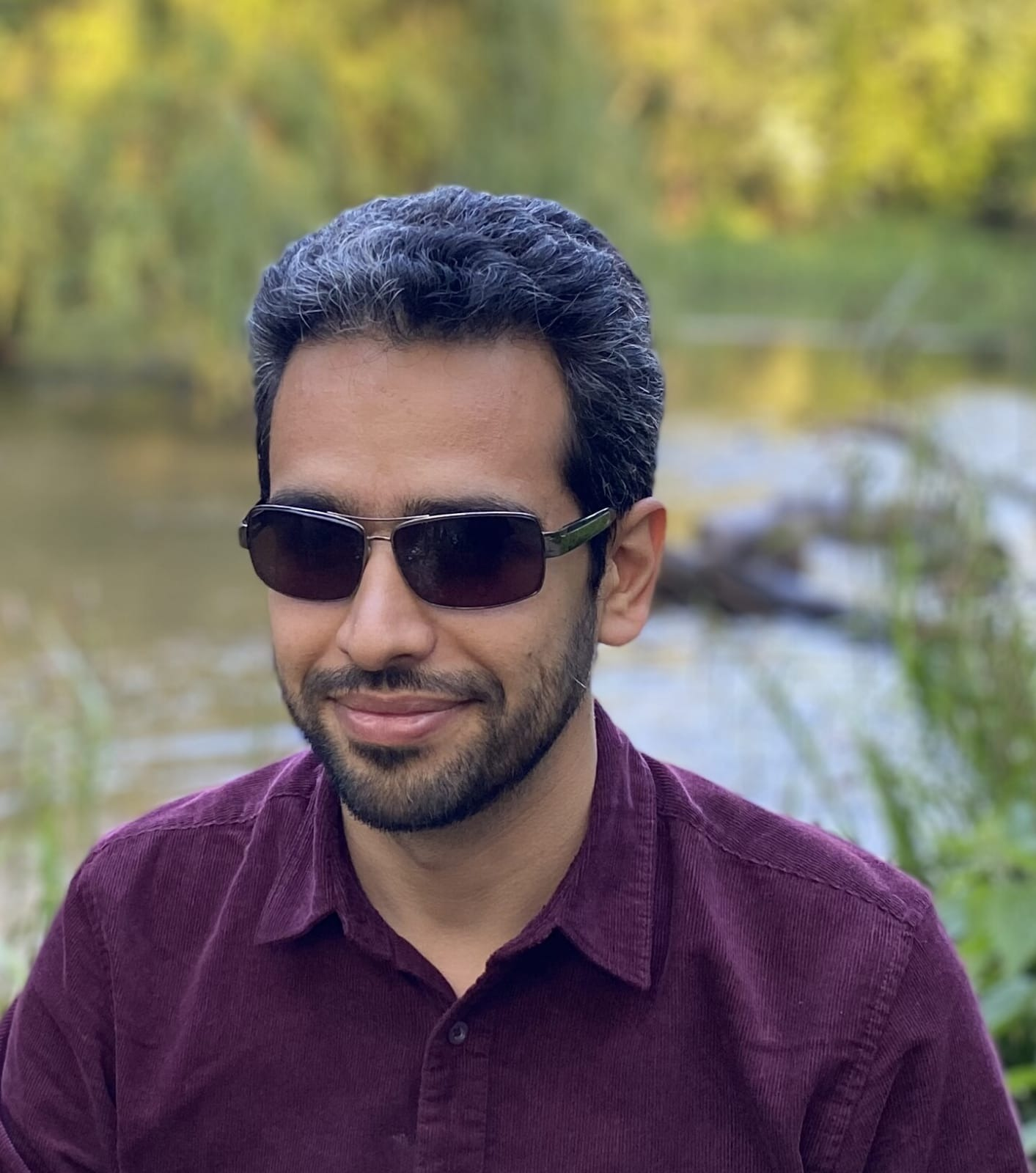}}]{Mohammad M. Tajiki}
is currently a research assistant at the Queen Mary University of London. He holds a double PhD degree from University of Rome Tor Vergata and Tarbiat Modares University in Computer Engineering and Electrical Engineering, respectively. His research interests are in Network Monitoring, AI, Service Function Chaining, IPv6 Segment Routing, SDN-based Networking, and Network Analysis.
\end{IEEEbiography}
%\vspace{-1.25cm}
\begin{IEEEbiography}[{\includegraphics[width=0.9in,height=1.1in,clip,keepaspectratio]{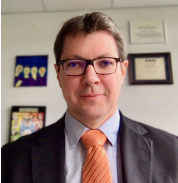}}]{Eduard A. Jorswieck}(S’01–M’03–SM’08–F’20) is the Chair of Communications Systems and a Full Professor at Technical University Braunschweig, Germany, since August 2019. From 2008 until 2019, he was the Chair of Communications Theory and Full Professor at Dresden University of Technology (TUD), Germany. His main research interests are in the wide span of communication sciences. He has published more than 125 articles, 13 book chapters, 3 monographs, and presented about 290 conference papers on these topics. Dr. Jorswieck is an IEEE Fellow. He is a member of the IEEE SAM Technical Committee since 2015. Since 2017, he serves as Editor-in-Chief of the EURASIP Journal on Wireless Communications and Networking. He currently serves on the editorial board for IEEE Transactions on Information Forensics and Security. In 2006, he received the IEEE Signal Processing Society Best Paper Award.
\end{IEEEbiography}
\end{document}